\RequirePackage[T1]{fontenc}
\documentclass[11pt, urlcolor=blue, linkcolor=blue]{article} 
\usepackage{cite}
\usepackage{float}
\usepackage[utf8]{inputenc}

\usepackage{multicol}
\usepackage{multirow}
\usepackage{tablefootnote}
\usepackage{amsmath, amsthm, amssymb,slashed,mathtools}
\usepackage{ifpdf}
\ifpdf
  \usepackage[pdftex]{graphicx}
  \usepackage{epstopdf}
\else
  \usepackage[dvips]{graphicx}
\fi
\parskip=\baselineskip

\usepackage[usenames, dvipsnames]{color}
\usepackage[svgnames]{xcolor}
\usepackage[colorlinks,citecolor=RoyalBlue, urlcolor=RoyalBlue, linkcolor=RoyalBlue ]{hyperref} 

\allowdisplaybreaks[1]

\numberwithin{equation}{section}

\usepackage{booktabs}

\theoremstyle{definition}

\newcommand{\cblue}[1]{\textcolor{blue}{#1}}

\definecolor{mygray}{gray}{0.6}

\usepackage{upgreek}
\usepackage{bbm}

\newcommand{\ABK}{\text{ABK}}

\newenvironment{myfont}[2][]{\csname#2\endcsname[#1]}{}

\usepackage{slashed}
\usepackage[makeroom]{cancel}
\usepackage[normalem]{ulem}
\usepackage{soul}
\newcommand{\stkout}[1]{\ifmmode\text{\sout{\ensuremath{#1}}}\else\sout{#1}\fi}

\usepackage{sseq}
\usepackage[all,cmtip]{xy}
\usepackage{tikz-cd}
\usepackage{tikz}
\usetikzlibrary{matrix}

\newcommand{\bea}{\begin{eqnarray}}
\newcommand{\eea}{\end{eqnarray}}
\def\be{\begin{equation}}
\def\ee{\end{equation}}

\newcommand{\e}{\hspace{1pt}\mathrm{e}}

\newcommand{\im}{\hspace{1pt}\mathrm{i}\hspace{1pt}}
\newcommand{\ii}{\hspace{1pt}\mathrm{i}\hspace{1pt}}

\def\RP{{\mathbb{RP}}}

\newcommand{\nn}{\nonumber}

\definecolor{red}{rgb}{1,0,0}
\definecolor{blue}{rgb}{0,0,1}
\definecolor{dblue}{rgb}{0,0,0.4}
\definecolor{green}{rgb}{0,1,0}
\definecolor{black}{rgb}{0,0,0}
\definecolor{white}{rgb}{1,1,1}

\definecolor{brn}{rgb}{.8,.4,.0}
\definecolor{redo}{rgb}{1,.5,.0}
\definecolor{ddgrn}{rgb}{0,0.4,0}
\definecolor{dgrn}{rgb}{0,0.55,0}
\definecolor{dbl}{rgb}{0,0,0.5}

\usepackage[bbgreekl]{mathbbol}
\usepackage{amscd}

\newcommand{\Z}{\mathbb{Z}}
\newcommand{\C}{\mathbb{C}}
\newcommand{\R}{\mathbb{R}}

\newcommand{\dd}{\hspace{1pt}\mathrm{d}}
\newcommand{\<}{\langle} 
\renewcommand{\>}{\rangle}

\newcommand{\Refe}[1]{Ref.~\cite{#1}}
\newcommand{\Eq}[1]{(\ref{#1})} 
\newcommand{\eq}[1]{(\ref{#1})} 
\newcommand{\eqn}[1]{eqn.~(\ref{#1})}

\newcommand{\Tr}{{\rm Tr}}

\newcommand{\prt}{\partial}

\newcommand{\bpm}{\begin{pmatrix}}
\newcommand{\epm}{\end{pmatrix}}
\newcommand{\bmm}{\begin{matrix}}
\newcommand{\emm}{\end{matrix}}

\newcommand{\cA}{ {\cal A} }

\newcommand{\cD}{ {\cal D} }

\newcommand{\al}{\alpha} 
\newcommand{\bt}{\beta}

\def\CA{{\cal A}}
\def\CB{{\cal B}}

\def\CJ{{\cal J}}

\def\CN{{\cal N}}

\def\Z{{\mathbb{Z}}}

\def\R{{\mathbb{R}}}
\def\C{{\mathbb{C}}}

\def\Tr{{\mathrm{Tr}}}

\def \H{\operatorname{H}}

\def \Z{\mathbb{Z}}
\def \Pin{\mathrm{Pin}}

\def \RP{\mathbb{RP}}

\newcommand{\Sec}[1]{Sec.~\ref{#1}} 
\newcommand{\Table}[1]{Table \ref{#1}}

\usetikzlibrary{decorations.markings}
\usetikzlibrary{positioning}
\usetikzlibrary{shadings}

\newcommand{\SO}{{\rm SO}}
\newcommand{\Spin}{{\rm Spin}}
\newcommand{\U}{{\rm U}}
\newcommand{\SU}{{\rm SU}}

\newcommand{\rS}{{\rm S}}

\newcommand{\rF}{{\rm F}}

\def\Sq{\mathrm{Sq}}

\def\B{\mathrm{B}}

\def\bfB{\mathbf{B}} 
\def\bfL{\mathbf{L}}

\def \EPin{\mathrm{EPin}}

\def\TP{\mathrm{TP}}

\usepackage{datetime}
\usepackage{enumitem} 
\usepackage{moreenum}

\newcommand{\sharpfootnote}[1]{%
\let\oldthefootnote=\thefootnote%
\stepcounter{mpfootnote}%
\addtocounter{footnote}{-1}%
\renewcommand{\thefootnote}{{W$^+$}} 
\footnote{#1}%
\let\thefootnote=\oldthefootnote%
}

\newcommand{\naturalfootnote}[1]{%
\let\oldthefootnote=\thefootnote%
\stepcounter{mpfootnote}%
\addtocounter{footnote}{-1}%
\renewcommand{\thefootnote}{{W$^-\natural$}}
\footnote{#1}%
\let\thefootnote=\oldthefootnote%
}

\newcommand{\flatfootnote}[1]{%
\let\oldthefootnote=\thefootnote%
\stepcounter{mpfootnote}%
\addtocounter{footnote}{-1}%
\renewcommand{\thefootnote}{{W$^-\flat$}}
\footnote{#1}%
\let\thefootnote=\oldthefootnote%
}

\DeclareRobustCommand\sWang
{\cblue{\includegraphics[height=3.2ex]{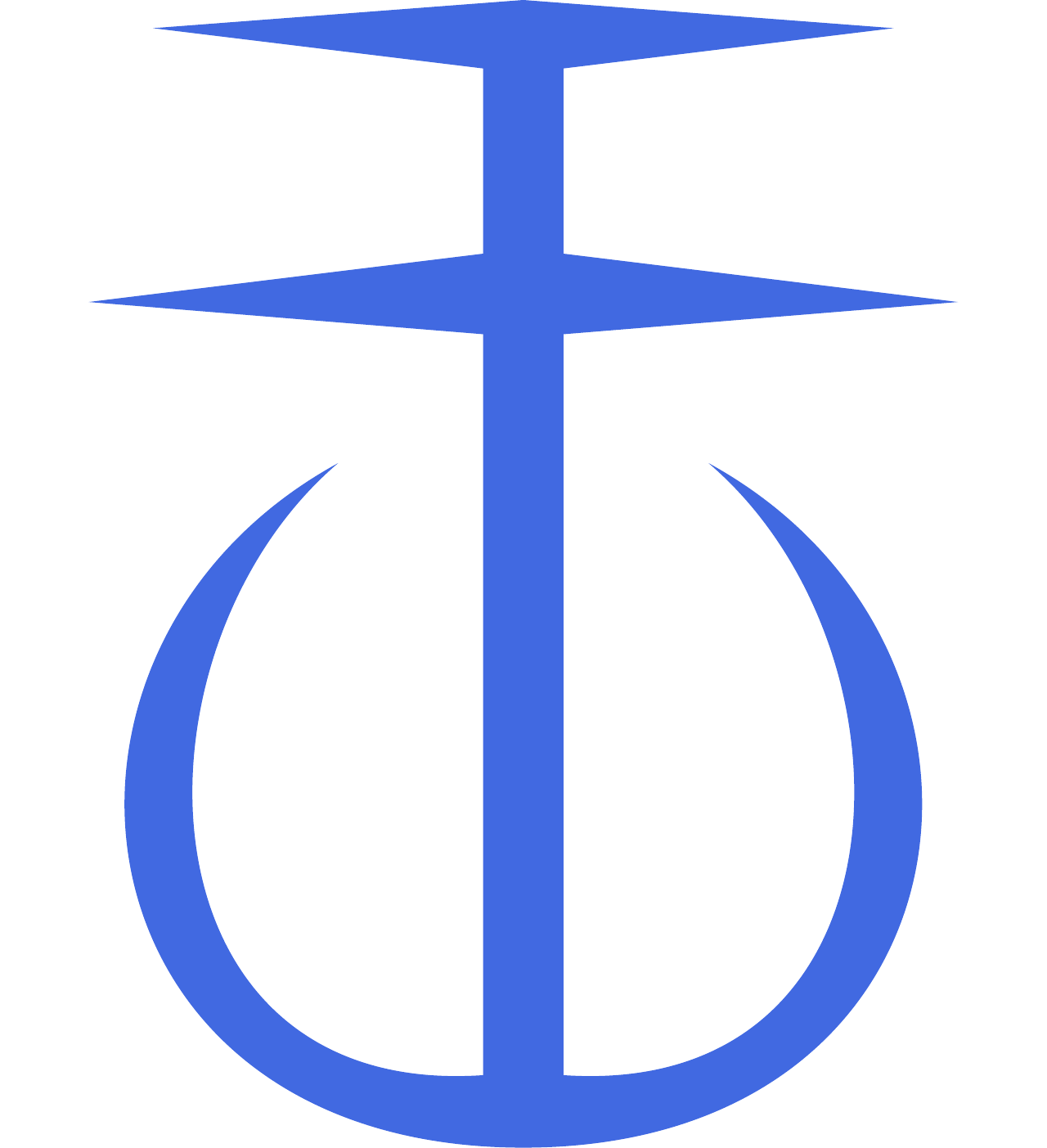}}}

\newcommand{\Wangfootnote}[1]{%
\let\oldthefootnote=\thefootnote%
\stepcounter{mpfootnote}%
\addtocounter{footnote}{-1}%
\renewcommand{\thefootnote}{\sWang}
\footnote{#1}%
\let\thefootnote=\oldthefootnote%
}

\usepackage{comment}

\usepackage{upgreek}
\newcommand{\Fig}[1]{Fig.~\ref{#1}}

\def\ra{\mathrm{a}}
\def\rb{\mathrm{b}}

\def\rq{\mathrm{q}}

\usepackage{mathrsfs}

\newcommand{\dal}{\dot{\al}} 
\newcommand{\dbt}{\dot{\beta}}

\def\SL{\rm SL}
\usepackage{esint} 

\begin{document}
\begin{titlepage}
\vskip1.25in
\begin{center}


{\bf\LARGE{ 
Anomaly and Cobordism Constraints
\\[8mm]
Beyond the Standard Model:  
Topological Force 
}}

\vskip0.5cm 
\Large{\quad\quad Juven Wang\Wangfootnote{
{\tt jw@cmsa.fas.harvard.edu} \quad\; 
\includegraphics[width=3.0in]{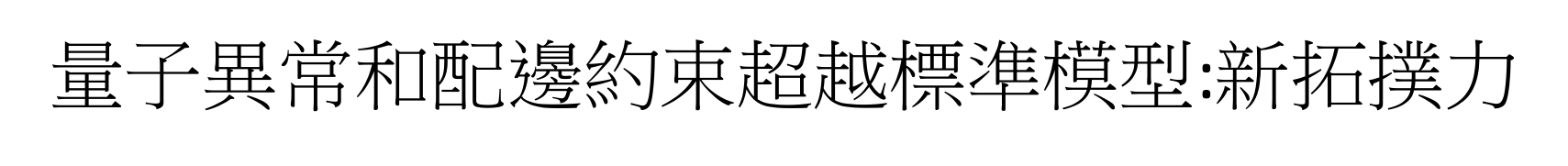}
\quad\quad \href{http://sns.ias.edu/~juven/}{\bf http://sns.ias.edu/$\sim$juven/} \\[2mm]
} 
\\[2.75mm]  
} 
\vskip.5cm
{ {\small{\textit{{Center of Mathematical Sciences and Applications, Harvard University,  Cambridge, MA 02138, USA}}}}
}

\end{center}

\vskip 0.5cm
\baselineskip 16pt
\begin{abstract}

Standard lore uses perturbative local anomalies to check the kinematic consistency of
 gauge theories coupled to chiral fermions, such as the Standard Models (SM) of particle physics. 
In this work, 
based on a systematic cobordism classification, 
we examine the constraints from invertible quantum anomalies
(including all perturbative local anomalies and all nonperturbative global anomalies) for
SM and chiral gauge theories.
We also clarify the different uses of these anomalies: including (1) anomaly cancellations of dynamical gauge fields,
(2) 't Hooft anomaly matching conditions of background fields of global symmetries. 
We apply several 
4d anomaly constraints
of $\mathbb{Z}_{16}$, $\mathbb{Z}_{4}$, and $\mathbb{Z}_{2}$ classes, beyond
 the familiar Feynman-graph perturbative $\mathbb{Z}$ class local anomalies. 
 %
 As an application, for 
$\frac{{\SU(3)\times \SU(2)\times \U(1)}}{\Z_q}$ SM (with $q=1,2,3,6$) 
and SU(5) Grand Unification with 15n chiral Weyl fermions and with a discrete baryon minus lepton number 
$X=5({\bf B}- {\bf L})-4Y$ preserved,
we discover a new hidden gapped sector previously unknown to the SM and Georgi-Glashow model. 
The 
gapped sector 
at low energy contains 
either (1) 4d \emph{non-invertible} topological quantum field theory (TQFT, 
{above the energy gap 
with heavy fractionalized anyon excitations from 1d particle worldline and 2d string worldsheet,
inaccessible directly from Dirac \emph{or} Majorana mass gap of the 16th Weyl fermions [i.e., right-handed neutrinos],
but accessible via a topological quantum phase transition}),
or (2) 5d \emph{invertible} 
TQFT in \emph{extra dimensions}.
Above a higher energy scale, the discrete $X$ becomes dynamically gauged,
the entangled Universe in 4d and 5d 
is 
mediated by    
Topological Force. 
Our model potentially 
resolves 
puzzles, 
surmounting 
sterile neutrinos and dark matter, in fundamental physics. 





\flushright

\end{abstract}

\end{titlepage}

  \pagenumbering{arabic}
    \setcounter{page}{2}
    

\tableofcontents



~\newline ~\newline ~\newline ~\newline ~\newline ~\newline
~\newline ~\newline ~\newline

\hfill ``Der schwer gefa{\ss}te Entschlu{\ss}. Mu{\ss} es sein?  Es mu{\ss} sein!'' \\[2mm] 
\hspace*{\fill}``The heavy decision. Must it be? It must be!'' \\[2mm] 
\hspace*{\fill}String Quartet No.$\;$16 in F major, op. 135\\[2mm] 
\hspace*{\fill} Ludwig van Beethoven in 1826


\section{Introduction}
\label{sec:Intro}

The Universe where we reside, to our contemporary knowledge, 
is governed by the laws of quantum theory, the information and long-range entanglement, and gravity theory. 
Quantum field theory (QFT), especially 
gauge field theory, under the name of \emph{Gauge Principle} following Maxwell \cite{Maxwell1865zz}, Hilbert, Weyl \cite{Weyl1929ZPhy}, 
Pauli \cite{RevModPhys13.203Pauli}, and many  pioneers,
forms a foundation of fundamental physics. 
Yang-Mills (YM) gauge theory \cite{PhysRev96191YM1954}, generalizing the U(1) abelian gauge group to a non-abelian Lie group, 
has been given credits for theoretically and experimentally essential to 
describe the Standard Model (SM) physics 
\cite{Glashow1961trPartialSymmetriesofWeakInteractions, Salam1964ryElectromagneticWeakInteractions, Weinberg1967tqSMAModelofLeptons}.


The SM of particle physics is a chiral gauge theory in 4d\footnote{{We denote 
$d$d for the $d$ spacetime dimensions.
The $d+1$d means the $d$ spatial and 1 time dimensions.
The $n$D means the $n$ space dimensions.}}
encoding three of the four known fundamental forces or interactions 
(the electromagnetic, weak, and strong forces, but without gravity).
The SM also classifies all experimentally known elementary particles so far: 
Fermions include three generations of quarks and leptons. (See Table \ref{table:SMfermion} for $15 + 1 = 16$  Weyl fermions
in SM for each one of the three generations, where the additional 1 Weyl fermion is the sterile right-handed neutrino).  
Bosons include the vector gauge bosons: one electromagnetic force mediator 
photon $\gamma$, 
the eight strong force mediator 
gluon g (denoted the gauge field $A^{\ra}$), 
the three weak force mediator 
$W^\pm$ and $Z^0$
gauge bosons; and the scalar Higgs particle $\phi_H$.  
(See Table \ref{table:SMboson} for 12 gauge boson generators for SM.)
While the spin-2 boson, the graviton, has not yet been experimentally verified, and it is not within SM.
Physics experiments had confirmed that at a higher energy of SM, 
the electromagnetic and weak forces are unified into an electroweak interaction sector:
Glashow-Salam-Weinberg (GSW) SM
\cite{Glashow1961trPartialSymmetriesofWeakInteractions, Salam1964ryElectromagneticWeakInteractions, Weinberg1967tqSMAModelofLeptons}.  

Grand Unifications and 
Grand Unified Theories (GUT) 
hypothesize that at a further higher energy, 
the strong and the electroweak interactions will be unified into an electroweak-nuclear GUT interaction sector.
The GUT interaction is characterized by one larger gauge group 
and its force mediator gauge bosons 
with a single unified 
coupling constant \cite{Georgi1974syUnityofAllElementaryParticleForces, Fritzsch1974nnMinkowskiUnifiedInteractionsofLeptonsandHadrons}.
\footnote{Other than the three known fundamental forces, we also have other GUT forces predicted by GUT.
However, in our present work, we discover a possible new force based on the global anomaly matching, 
which we name \text{Topological Force.}
This {Topological Force} 
may be a possible new force not included in the known four fundamental forces nor included in GUT. 
The {Topological Force} is not fully explored in the prior 
particle physics phenomenology 
literature, at least not rigorously explored based on the cobordism and nonperturbative global anomaly matching framework. 
On the other hand, when we consider the constraints from the anomaly matching,
the gravity mostly plays the role of the gravitational \emph{background probed} fields (instead of \emph{dynamical} gravity), 
such as in  the gravitational anomaly or the mixed gauge-gravitational anomaly.  
We mostly leave out any dynamical gravity outside our model, so the present work is far from a Theory of Everything (TOE). 
The only exception that we discuss
the influences of dynamical gravity (or the hypothetical particle: graviton, 
such as in Quantum Gravity or Topological Gravity) 
will be in \Sec{sec:gravity}.
}

In this article, we revisit the
Glashow-Salam-Weinberg SM
\cite{Glashow1961trPartialSymmetriesofWeakInteractions, Salam1964ryElectromagneticWeakInteractions, Weinberg1967tqSMAModelofLeptons},
with four possible gauge groups:
\bea
G_{\text{SM}_q} \equiv \frac{\SU(3)\times \SU(2)\times \U(1)}{\Z_q}, \quad q=1,2,3,6.
\eea
We revisit the embedding of this special $q=6$ SM (with a gauge group $\frac{\SU(3)\times \SU(2)\times \U(1)}{\Z_q}$)
into
Georgi-Glashow SU(5) GUT with an SU(5) gauge group \cite{Georgi1974syUnityofAllElementaryParticleForces} (with 24 Lie algebra generators for SU(5) gauge bosons shown in Table \ref{table:SMboson}),
and Fritzsch-Minkowski SO(10) GUT with a Spin(10) gauge group \cite{Fritzsch1974nnMinkowskiUnifiedInteractionsofLeptonsandHadrons}.
Our main motivation to revisit these well-known SM and GUT models, following our prior work \cite{WanWang2018bns1812.11967, WW2019fxh1910.14668}, 
is that the recent systematic Freed-Hopkins' cobordism classification of topological invariants \cite{Freed2016}
can be applied to classify
all the invertible quantum anomalies,
including 
\begin{itemize}
\item all \emph{perturbative local anomalies}, and 
\item all \emph{nonperturbative global anomalies}, 
\end{itemize}
which
can further mathematically and rigorously constrain SM and GUT models.
In fact, many earlier works and recent works suggest the cobordism theory is the underlying math structure of invertible quantum anomalies,
see a list of selective References \cite{Atiyah1975jfAPS, Atiyah1976APS, Atiyah1980APS, Freed0607134, Freed2013bordism, Kapustin2014tfa1403.1467, 1405.7689, Kapustin1406.7329,  Kitaev2015, Witten:2015aba, 1711.11587GPW,
2019CMaPh.368.1121Y, Freed2019scoFreedHopkins1908.09916, Witten2019bou1909.08775} and an overview therein.
By the completion of all invertible anomalies, we mean that 
it is subject to a given symmetry group 
\bea
G\equiv ({\frac{{G_{\text{spacetime} }} \ltimes  {{G}_{\text{internal}} }}{{N_{\text{shared}}}}}),
\eea 
where the ${G_{\text{spacetime} }}$ is the spacetime symmetry,\footnote{For example, ${G_{\text{spacetime} }}$ can be the Spin($d$), the double cover of Euclidean rotational symmetry SO($d$) for the QFT of a spacetime dimension $d$.}
the ${{G}_{\text{internal}} }$ is the internal symmetry,\footnote{For example, ${{G}_{\text{internal}} }$ can be the 
$G_{\text{SM}_q} \equiv \frac{\SU(3)\times \SU(2)\times \U(1)}{\Z_q}, \quad q=1,2,3,6.$} 
the $\ltimes$ is a semi-direct product from a ``twisted'' 
extension,\footnote{The  ``twisted'' 
extension is due to the symmetry extension from ${{G}_{\text{internal}} }$ by ${G_{\text{spacetime} }}$. For a trivial extension,
the semi-direct product $\ltimes$ becomes a direct product $\times$.}
and the ${N_{\text{shared}}}$ is the shared common normal subgroup symmetry between ${G_{\text{spacetime} }}$ 
and ${{G}_{\text{internal}} }$.
 \Refe{WanWang2018bns1812.11967, WW2019fxh1910.14668, WangWen2018cai1809.11171} applies
Thom-Madsen-Tillmann spectra \cite{thom1954quelques,MadsenTillmann4}, Adams spectral sequence (ASS) \cite{Adams1958}, 
and Freed-Hopkins theorem \cite{Freed2016} in order
to compute, relevant for SM and GUT, the bordism group
\bea
\Omega_{d}^{G},
\eea
and a specific version of cobordism group (firstly defined to classify Topological Phases [TP] in \cite{Freed2016})
\bea\label{eq:TPG}
\Omega^{d}_{G} &\equiv&
\Omega^{d}_{({\frac{{G_{\text{spacetime} }} \ltimes  {{G}_{\text{internal}} }}{{N_{\text{shared}}}}})} 
\equiv
\TP_d(G).
\eea
For a given $G$, 
\Refe{WanWang2018bns1812.11967, WW2019fxh1910.14668, WangWen2018cai1809.11171} 
find out corresponding all possible topological terms and all possible anomalies relevant for SM and GUT.
See more mathematical definitions, details, and References therein our prior work \cite{WanWang2018bns1812.11967, WW2019fxh1910.14668}.

Along this research direction, other closely related pioneer and beautiful works, by
{Garcia-Etxebarria-Montero} \cite{GarciaEtxebarriaMontero2018ajm1808.00009} and Davighi-Gripaios-Lohitsiri \cite{2019arXiv191011277D}, 
use a different mathematical tool, based on Atiyah-Hirzebruch spectral sequence (AHSS) \cite{AtiyahHirzebruch1961} and Dai-Freed theorem  \cite{Dai1994kqFreed9405012},
also compute the bordism groups $\Omega_{d}^{G}$ and classify possible global anomalies in SM and GUT.\footnote{A practical comment is that
Adams spectral sequence (ASS) used in
 \Refe{Freed2016, WanWang2018bns1812.11967, WW2019fxh1910.14668}
 turns out to be  more powerful than the 
 Atiyah-Hirzebruch spectral sequence (AHSS) used in \cite{GarciaEtxebarriaMontero2018ajm1808.00009, 2019arXiv191011277D}.
 \begin{itemize}
 \item
\Refe{WanWang2018bns1812.11967,  WW2019fxh1910.14668, Wan2019sooWWZHAHSII1912.13504, Wan2019oaxWWHAHSIII1912.13514, WanWangv2 
} based on Adams spectral sequence (ASS) and 
Freed-Hopkins theorem \cite{Freed2016}
includes the more refined data, containing both \emph{module} and \emph{group} structure, thus with the advantages 
of having less differentials.
In addition,  
we can conveniently read and extract the topological terms and co/bordism invariants from the Adams chart and 
ASS \cite{WanWang2018bns1812.11967}.


\item
\Refe{GarciaEtxebarriaMontero2018ajm1808.00009, 2019arXiv191011277D} is based on  Atiyah-Hirzebruch spectral sequence (AHSS), 
which includes more differentials
and some undetermined extensions.
{It is also not straightforward  to extract the  topological terms and co/bordism invariants directly from the AHSS.}
\end{itemize}
}
As we will see, many SM and GUT with extra discrete symmetries require the twisted $G$,
whose $\Omega_{d}^{G}$ is very difficult, if not simply impossible, 
to be determined via AHSS alone \cite{GarciaEtxebarriaMontero2018ajm1808.00009, 2019arXiv191011277D}.
But the twisted $\Omega_{d}^{G}$ can still be relatively easily computed via ASS \cite{WanWang2018bns1812.11967, WanWangv2}.
Therefore, we will focus on the results obtained in \cite{WanWang2018bns1812.11967, WW2019fxh1910.14668}.

In this work, we have the plans and outline as follows, with topics by sections:
\begin{enumerate}[leftmargin=4.mm, label=Sec.~\textcolor{blue}{[\arabic*]}., ref={[\arabic*]}]
\setcounter{enumi}{1}
\item {\bf Overview on SM, GUT, and Anomalies}  in \Sec{sec:2Overview}:

We first overview the ingredients of various SM and GUT to set up the stage in \Sec{sec:2OverviewSMGUT}.
Then we comment on the different types of anomalies,
and we clarify the different uses of these anomalies: including (1) anomaly cancellations of dynamical 
gauge fields,\footnote{The perturbative parts of calculations are standard on the QFT textbook \cite{Weinberg1996Vol2, PeskinSchroeder1995book, Zee2003book, Srednicki2007book}}
(2) 't Hooft anomaly matching conditions \cite{tHooft1979ratanomaly} of background fields of global symmetries, and others, in \Sec{sec:2OverviewSMGUTdifferentanomalies} and \Fig{fig:triangle}.
We also describe the
perturbative local anomalies vs 
nonperturbative global anomalies;
also bosonic vs fermionic anomalies, etc.
These are presented in \Sec{sec:2Overview}.

\item {\bf Dynamical Gauge Anomaly Cancellation}  in
\Sec{sec:DynamicalGaugeAnomalyCancellation}:

We explicitly show the classification of all possible (invertible) anomalies of SM gauge groups by cobordism data in Table \ref{table:SU3SU2U1},
and explicitly check all anomaly cancellations for perturbative local anomalies vs 
nonperturbative global anomalies in
\Sec{sec:DynamicalGaugeAnomalyCancellation}.

\item {\bf Anomaly Matching for SM and GUT with Extra Symmetries}  in
\Sec{Sec:discretesymmetries}: 

It is natural to include 
additional extra symmetries (such as the ${ \mathbf{B}-  \mathbf{L}}$ baryon minus lepton numbers or $X=5({\bf B}- {\bf L})-4Y$ symmetry with the hypercharge $Y$ \cite{Wilczek1979hcZee}, 
motivated in \cite{GarciaEtxebarriaMontero2018ajm1808.00009} and \cite{WW2019fxh1910.14668}) 
into SM and SU(5) GUT. 
We show the classification of all possible (invertible) anomalies of some SM and GUT with extra symmetries
by cobordism data in Table \ref{table:SU3SU2U1-discrete}.
We explicitly check their anomaly cancellations for perturbative local anomalies vs 
nonperturbative global anomalies in
\Sec{Sec:discretesymmetries}.

\item {\bf Beyond Three ``Fundamental'' Forces: Hidden New Topological Force} in \Sec{sec:HiddenTopologicalSectors}:

We recall that to match the 't Hooft anomalies of global symmetries, there are several ways:
\begin{enumerate} 
\item Symmetry-breaking:\\ 
$\bullet$ (say, discrete or continuous $G$-symmetry breaking. Explicitly breaking or spontaneously breaking [may give Nambu-Goldstone modes]).\\
\item Symmetry-preserving: \\
$\bullet$ Degenerate ground states (like the ``Lieb-Schultz-Mattis theorem \cite{Lieb1961frLiebSchultzAOP, Hastings2003zx},'' may host intrinsic topological orders.\footnote{Topological order is in the sense of Wen's definition \cite{Wen2016ddy1610.03911} and References therein.
The gapped and gauged topological order in the colloquial sense can have fractionalized
excitations such as anyons \cite{Wilczek1990BookFractionalstatisticsanyonsuperconductivity}.
Some of the quantum vacua we look for in 4d and 5d may be regarded as a certain version of 
topological quantum computer \cite{Kitaev2003, Kitaev2006}.
}),\\
$\bullet$ Gapless, e.g., conformal field theory (CFT). 
There are also novel cases where the anomaly and symmetry together enforces the robustness of gapless ground states \cite{CWang1401.1142, Wan2018djlW2.1812.11955, Cordova2019bsd1910.04962}.\\
$\bullet$ 
Symmetry-preserving topological quantum field theory (TQFT): anomalous symmetry-enriched topological orders. 
\item Symmetry-extension \cite{Wang2017locWWW1705.06728}:
The symmetry-extension in any dimension is a helpful intermediate stepstone, 
to construct another earlier scenario: \emph{symmetry-preserving TQFT}, via gauging the extended-symmetry \cite{Wang2017locWWW1705.06728}.
\end{enumerate} 
We present the possible ways of matching of global anomalies of ``SM and GUT with extra symmetries,'' in \Sec{sec:HiddenTopologicalSectors}.
The condensed matter realization of 't Hooft matching by an anomalous symmetry-preserving gapped TQFT is especially exotic but truly important to us.
It is also known as the surface topological order in the condensed matter literature. {In a lower dimension, the 2+1d surface topological order 
is pointed out firstly by Senthil-Vishwanath \cite{VishwanathSenthil1209.3058}. The surface topological order  developments are nicely reviewed in \cite{Senthil1405.4015}.}
The particular symmetric gapped boundary spin-TQFT construction of fermionic SPTs given in \cite{GuoJW1812.11959} 
will be especially helpful for our construction.\footnote{Our result (3+1d boundary TQFT of 4+1d bulk) may be regarded 
as a \emph{one higher dimensional} fundamental physics realization of the condensed matter systems 
in \Refe{Fidkowski1305.5851, Metlitski20141406.3032} (2+1d boundary TQFT of 3+1d bulk).}

It turns out that a certain $\Z_{16}$ global anomaly for SM with an extra discrete $X=5({\bf B}- {\bf L})-4Y$ symmetry
may not be matched by the 15 Weyl fermions per generation.
The resolution, other than introducing a sterile right-handed neutrinos (the 16th Weyl fermions), 
can also be that introducing a new gapped topological sector matching some of the anomaly,
but preserves the discrete $\Z_{4,X}$ symmetry.\footnote{More precisely,
it turns out that as suggested by \cite{Hsieh2018ifc1808.02881,Cordova1912.13069}, 
we can only construct 
a $\Z_{4,X}$-symmetric gapped 4d TQFT
to saturate the anomaly of $\Z_{16}$ class from the 5d bordism group $\Omega_5^{\Spin \times_{\Z_2} \Z_4}=\Z_{16}$, 
when the anomaly index is an \emph{even} integer $\upnu \in \Z_{16}$. 
For example, the SM with three generations of 15 Weyl fermions has the anomaly index $\upnu = - 3 \in \Z_{16}$, 
we can introduce one right-handed neutrino to absorb part of the anomaly $\upnu =1 \in \Z_{16}$.
Then, we also introduce a $\Z_{4,X}$-symmetric gapped 4d TQFT to absorb the remained anomaly $\upnu =2 \in \Z_{16}$.
See more details on the anomaly in \Sec{Sec:discretesymmetries}.
See the construction of $\upnu =2 \in \Z_{16}$ TQFT in \Sec{sec:Z16-nu=2-TQFT}.
}

\item {\bf Ultra Unification: Grand Unification $+$ Topological Force/Matter} in \Sec{sec:UU}: 

In certain cases of anomaly matching, we require new hidden gapped topological sectors beyond SM and GUT. We may term the
unification including SM, Grand Unification plus additional topological sectors with Topological Force and Matter
as {Ultra Unification}.
We then suggest possible resolutions to sterile neutrinos, neutrino oscillations, 
and Dark Matter, in \Sec{sec:UU}.

\end{enumerate}


\noindent
{\bf Notations:} 
Throughout our work, 
{we follow the same notations as  \cite{WanWang2018bns1812.11967, WW2019fxh1910.14668}.
The imaginary number is $\ii\equiv \sqrt{-1}$.
The TQFT stands for a topological quantum field theory,
and the iTQFT  stands for an invertible topological quantum field theory.  
Follow the Freed-Hopkins notation \cite{Freed2016}, 
we denote the group $G_1 \times_N G_2 \equiv (G_1 \times G_2)/N$ where $N$ is a common normal subgroup of the groups $G_1$ and $G_2$.
We use the standard notation for characteristic classes \cite{milnor1974characteristic}: $c_i$ for the Chern class, $e_n$ for the Euler class, 
$p_i$ for the Pontryagin class, and $w_i$ for the Stiefel-Whitney class. 
We abbreviate $c_i(G)$, $e_n(G)$, $p_i(G)$, and  $w_i(G)$ for the characteristic classes of the associated vector bundle $V_G$ of the principal $G$ bundle 
(normally denoted as  $c_i(V_{G})$, $e_n(V_{G})$, $p_i(V_{G})$, and $w_i(V_{G})$). For simplicity, we denote
the Stiefel-Whitney class of the tangent bundle $TM$ of a manifold $M$ as $w_j \equiv w_j(TM)$; 
if we do not specify $w_j$ with which bundle, then it is for $TM$.
The PD is defined as taking the Poincar\'e dual.}
{All the product notations between cohomology classes are cup product ``$\smile$'', 
such as $w_2 w_3  \equiv   w_2(TM)w_3(TM)=  w_2(TM)\smile w_3(TM)$.
All the product notations between a cohomology class $\CA$ and any fermionic topological invariant ${\eta}$ (say, the 
4d eta invariant $\eta$ of Atiyah-Patodi-Singer  \cite{Atiyah1975jfAPS, Atiyah1976APS, Atiyah1980APS}), namely
 $\CA {\eta}$,
are defined as the value of ${\eta}$ on the submanifold of $M$ which represents the Poincar\'e dual of $\CA$.
Notice that here for $\CA {\eta}$, it is crucial to have $\CA$ as a cohomology class, so we can define its Poincar\'e dual of $\CA$ as $\text{PD}(\CA)$.
 In other words, the  $\CA {\eta} \equiv {\eta} (\text{PD}(\CA))$. 
 Similarly we define the product notations between a cohomology class and other fermionic invariants also via the Poincar\'e dual PD of cohomology class.
Here are some other examples of fermionic invariants appeared in this work:  \\
$\bullet$ The $\tilde{\eta}$ is a mod 2 index of 1d Dirac operator as a cobordism invariant of $\Omega_1^{\Spin}=\Z_2$.\\ 
$\bullet$ The ${\eta}'$ is a mod 4 index of 1d Dirac operator as a cobordism invariant of $\Omega_1^{\Spin\times \Z_4}=\Z_4$.\\ 
$\bullet$ The Arf invariant \cite{Arf1941} is a mod 2 cobordism invariant of $\Omega_2^{\Spin}=\Z_2$.
The Arf appears to be the low energy iTQFT of a 1+1d Kitaev fermionic chain \cite{Kitaev2001chain0010440},
whose boundary hosts a single 0+1d real Majorana zero mode on each of open ends.\\
$\bullet$ The Arf-Brown-Kervaire (ABK) invariant \cite{brown1972generalizations, KT1990} is a mod 8 cobordism invariant of $\Omega_2^{\Pin^-}=\Z_8$.
The ABK appears to be the low energy iTQFT 
of a 1+1d Fidkowski-Kitaev fermionic chain \cite{FidkowskifSPT1, FidkowskifSPT2} 
protected by time reversal and fermion parity symmetry $\Z_2^T \times \Z_2^F$,
whose boundary also hosts some number of 0+1d real Majorana zero modes.
\\
For cobordism or cohomology invariants, we may implicitly make a convention that cohomology classes are pulled back to the manifold $M$.
For example, $\CA_{{\Z_2}} \in \H^1(M,\Z_2)$ is the first cohomology class on $M$ with a $\Z_2$ coefficient,
as a generator from  $\CA_{{\Z_2}} \in\H^1(\B(\Z_4/\Z_2^F),\Z_2)$ of ${\Spin \times_{\Z_2^F} \Z_4}$ pullback to $M$ as the former expression.}

\quad


\newpage
\begin{table}[!t]
$\hspace{-19.8mm}
  \begin{tabular}{ccccccc c c c c c c c c}
    \hline
    $\begin{array}{c}
    \textbf{SM}\\ 
   \textbf{fermion}\\
   \textbf{spinor}\\ 
   \textbf{field}
       \end{array}$
   & ${\SU(3)}$& ${\SU(2)}$ & $\U(1)_Y$ & $\U(1)_{Y_W}$ 
   & $\U(1)_{\tilde Y}$
   & $\U(1)_{\rm{EM}}$ 
    & $\U(1)_{{ \mathbf{B}-  \mathbf{L}}}$  & $\U(1)_{X}$ & 
    $\Z_{4,X}$
    & $\Z_{2}^F$  & SU(5) & Spin(10) \\
        \hline\\[-4mm]
    $\bar{d}_R$& $\bar{\mathbf{3}}$& $\mathbf{1}$ & 1/3 &2/3  & 2 & 1/3 & $-1/3$ & $-3$ &1   &   1 &  \multirow{2}{*}{
     $\overline{\bf{5}}$}  &  \multirow{6}{*}{
     ${\bf{16}}$} \\
\cline{1-11} $l_L$& $\mathbf{1}$& $\mathbf{2}$& $-1/2$& $-1$  & $-3$ & 0 or $-1$ & $-1$ & $-3$  &1  &   1 \\
\cline{1-12}  $q_L$& ${{\mathbf{3}}}$& $\mathbf{2}$& 1/6 & 1/3 &1 & 2/3 or $-1/3$ &  1/3  &1 & 1 &    1 & \multirow{3}{*}{
     ${\bf{10}}$} \\
\cline{1-11} $\bar{u}_R$& $\bar{\mathbf{3}}$& $\mathbf{1}$& $-2/3$ & $-4/3$ & $-4$ & $-2/3$ & $-1/3$ & $1$  &1 &   1 \\
\cline{1-11} $\bar{e}_R= e_L^+$& $\mathbf{1}$& $\mathbf{1}$& 1 & 2 & 6 & 1 & 1  &1 &   1 &    1\\
\cline{1-12} $\bar{\nu}_R= {\nu}_L $& $\mathbf{1}$& $\mathbf{1}$& 0 & 0 & 0 & 0 & 1  & 5 &   1 &   1 & \;\;${\bf{1}}$ &\\
    \hline\end{tabular}
$
   \caption{We show the quantum numbers of $15 +1 = 16$ left-handed Weyl fermion spinors in each of three generations of matter fields in SM. 
The $15$ of 16 Weyl fermion are $ \overline{\bf 5} \oplus {\bf 10}$  of SU(5); namely,
$(\overline{\bf 3},{\bf 1},1/3)_L \oplus ({\bf 1},{\bf 2},-1/2)_L \sim \overline{\bf 5}$
and $({\bf 3},{\bf 2}, 1/6)_L \oplus (\overline{\bf 3},{\bf 1}, -2/3)_L \oplus ({\bf 1},{\bf 1},1)_L  \sim {\bf 10}$  of SU(5).
The $1$ of 16 is presented neither in the standard GSW SM nor in the SU(5) GUT, but it is within  ${\bf 16}$ of the SO(10) GUT.
The numbers in the Table entries indicate the quantum numbers associated to the representation of the groups given in the top row. 
We show the first generation of SM fermion matter fields in  Table \ref{table:SMfermion}. There are 3 generations, triplicating Table \ref{table:SMfermion}, in SM.
All fermions have the fermion parity  $\Z_{2}^F$ representation 1.}
 \label{table:SMfermion}
\end{table}
%
%
%
%
%
\begin{table}[!t]
$ \hspace{14mm}
  \begin{tabular}{c c c c c c c c c c c }
    \hline
    $\begin{array}{c}
    \textbf{SM}\\ 
   \textbf{boson}\\
      \textbf{scalar or}\\ 
      \textbf{vector }
   \textbf{field}
       \end{array}$
   & ${\SU(3)}$& ${\SU(2)}$ & $\U(1)_Y$   
    & $\U(1)_{{ \mathbf{B}-  \mathbf{L}}}$  & $\U(1)_{X}$ & 
    $\Z_{4,X}$   \\
        \hline\\[-4mm]
    \hline
    $\begin{array}{c}
   \textbf{SM Higgs scalar}\\
  \phi_H
       \end{array}$
     & $\mathbf{1}$ & $\mathbf{2}$& 1/2  & 0 & $-2$ &   2  \\
    \hline
    \hline
    \multicolumn{7}{l}{\bf{SM gauge vector bosons} ($1+3+8 =12$ generators)}
      \\
    \hline
     $\begin{array}{c}
   \textbf{photon} \;(\gamma)\\
 A_Y \text{ or } A
       \end{array}$
     &  ${}$ & \multicolumn{2}{c}{\multirow{3}{*}{
     ${3\;\;\;\;+\;\;\;1\;}$}}     & 0 & 0 &   0  \\
     \cline{1-2} \cline{5-7} 
      $\begin{array}{c}
   \textbf{Weak}\\
  W^{1,2,3} \text{ or } W^{\pm},  Z^{0}
       \end{array}$
     &    &  &    & 0 & 0 &   0  \\
     \hline     
     $\begin{array}{c}
      \textbf{Strong \rm{(g)}}\\
  A^\ra \tau^\ra 
  \end{array}$
     & ${8}$  &   &  & 0 & 0 & 0  \\
     \hline
     \hline
    \multicolumn{7}{l}{\bf{BSM Georgi-Glashow gauge vector bosons ($2 \times (3 \times 2 )=12$ generators)}}
      \\
    \hline
   \textbf{X}$^+$ ($+\frac{4}{3}|e|$) 
     &  \multirow{2}{*}{
     $\bar{\mathbf{3}}$}  & \multirow{2}{*}{
     $\mathbf{2}$}   &  \multirow{2}{*}{
     $5/6$}     & \multirow{2}{*}{
     $2/3$}  & 0 &   0 
     \\
       \cline{1-1}
        \textbf{Y}$^+$  ($+\frac{1}{3}|e|$)
     &   &   &  &   & 0 &   0 \\
       \cline{1-1}
         \textbf{Y}$^-$  ($-\frac{1}{3}|e|$)
     &  \multirow{2}{*}{
     ${\mathbf{3}}$}  &  \multirow{2}{*}{
     $\mathbf{2}$}   &   \multirow{2}{*}{
     $-5/6$}    & \multirow{2}{*}{
     $-2/3$}  & 0 &   0 \\
       \cline{1-1}
        \textbf{X}$^-$ ($-\frac{4}{3}|e|$)
     &   &   &  &  & 0 & 0 \\
     \hline
    \end{tabular}
$
   \caption{We show quantum numbers of the electroweak Higgs boson $\phi_H$, and
   24 gauge bosons corresponding to 24 Lie algebra $su(5)$ generators of  Lie group SU(5). 
   Hereby the representation numbers of gauge fields, we really mean the  
  gauge fields \emph{transform in the adjoint} of the gauge group.
  (But the gauge fields are \emph{not} in the adjoint or other $\mathbf{R}$ representations of the gauge group.) 
  Also the readers should not be confused with the symmetry charge $X$ (in the Italic form) and its gauge boson $X_g$, 
with the Georgi-Glashow (GG) model gauge boson \textbf{X} (in the boldface text form).
The \textbf{X}  and \textbf{Y} gauge bosons can carry certain SU(3), SU(2) and U(1) quantum numbers, 
because the \textbf{X}  and \textbf{Y} sit outside the SM gauge group.
All bosons have the fermion parity  $\Z_{2}^F$ representation 0.
   See also the caption in Table \ref{table:SMfermion}.}
    \label{table:SMboson}
\end{table}


\newpage
\section{Overview on Standard Models, Grand Unifications, and Anomalies}
\label{sec:2Overview}
\subsection{SM and GUT: local Lie algebras to global Lie groups, and representation theory }

\label{sec:2OverviewSMGUT}

We shall first overview the local Lie algebra the representation theory of matter field contents,
and the global Lie groups of of SM and GUT.
Then we will be able to be precise about 
the spacetime symmetry group $G_{\text{spacetime}}$ and internal symmetry group $G_{\text{internal}}$ relevant for SM and GUT physics, 
\begin{enumerate}[leftmargin=-4.mm, label=\textcolor{blue}{[\Roman*]}., ref={[\Roman*]}]
\item The local gauge structure of Standard Model is  the Lie algebra $u(1) \times su(2) \times su(3)$.
This means that the Lie algebra valued 1-form gauge fields take values in the Lie algebra generators of  $u(1) \times su(2) \times su(3)$.
There are $1 + 3 + 8 = 12$ Lie algebra generators.
The 1-form gauge fields are the 1-connections of the principals $G_{\text{internal}}$-bundles that we should determine.

\item  Fermions are the spinor fields,
as the \emph{sections of the spinor bundles} of the spacetime manifold. For the left-handed Weyl spinor $\Psi_L$,
it is a doublet spin-1/2 representation of spacetime symmetry group $G_{\text{spacetime}}$ (Minkowski $\Spin(3,1)$ or Euclidean $\Spin(4)$), denoted as
\bea
\Psi_L \sim {\bf 2}_L \text{ of } \Spin(3,1) = \SL(2,\C), \quad\text{ or }\quad \Psi_L \sim  {\bf 2}_L \text{ of } \Spin(4) = \SU(2)_L \times  \SU(2)_R.
\eea 
These Spin 
groups are the double-cover also the universal-cover version of the Lorentz group $\SO(3,1)_+$ or Euclidean rotation $\SO(4)$, extended by the fermion parity $\Z_2^F$. 
In the first generation of SM, the matter fields as Weyl spinors $\Psi_L$ contain:
\begin{itemize}
\item
The left-handed up and down quarks ($u$ and $d$) form a doublet $\begin{pmatrix}
u\\
d
\end{pmatrix}_L$ in ${\bf 2}$ for the SU(2)$_{\text{weak}}$, and they are in ${\bf 3}$ for the SU(3)$_{\text{strong}}$.
\item
The right-handed up and down quarks, each forms a singlet, $u_R$ and $d_R$, in ${\bf 1}$ for the SU(2)$_{\text{weak}}$. They are in ${\bf 3}$ for the SU(3)$_{\text{strong}}$.
\item
The left-handed electron and neutrino form a doublet $\begin{pmatrix}
\nu_e\\
e
\end{pmatrix}_L$ in ${\bf 2}$ for the SU(2)$_{\text{weak}}$, and they are in ${\bf 1}$ for the SU(3)$_{\text{strong}}$.
\item
The right-handed electron forms a singlet $e_R$ in ${\bf 1}$ for the SU(2)$_{\text{weak}}$, 
and it is in ${\bf 1}$ for the SU(3)$_{\text{strong}}$.
\end{itemize}
There are two more generations of quarks: 
charm and strange quarks ($c$ and $s$),
and top and bottom quarks ($t$ and $b$).
There are also two more generations of leptons:
muon and its neutrino ($\mu$ and $\nu_\mu$), 
and 
tauon and its neutrino ($\tau$ and $\nu_\tau$). 
So there are three generations (i.e., families) of quarks and leptons:
\bea
\Bigg(
\begin{pmatrix}
u\\
d
\end{pmatrix}_L \times {\bf 3}_{\text{color}}, \quad\quad \quad u_R \times {\bf 3}_{\text{color}}, \quad\quad\quad\quad d_R \times {\bf 3}_{\text{color}}, 
\quad\quad\quad
\begin{pmatrix}
\nu_e\\
e
\end{pmatrix}_L,\quad\quad  \quad e_R 
\quad  \Bigg),\\
\Bigg(
\begin{pmatrix}
c\\
s
\end{pmatrix}_L \times {\bf 3}_{\text{color}}, \quad\quad \quad c_R \times {\bf 3}_{\text{color}}, \quad\quad\quad\quad s_R \times {\bf 3}_{\text{color}}, 
\quad\quad\quad
\begin{pmatrix}
\nu_\mu\\
\mu
\end{pmatrix}_L,\quad\quad  \quad \mu_R 
\quad  \Bigg),\\
\Bigg(
\begin{pmatrix}
t\\
b
\end{pmatrix}_L \times {\bf 3}_{\text{color}}, \quad\quad \quad t_R \times {\bf 3}_{\text{color}}, \quad\quad\quad\quad b_R \times {\bf 3}_{\text{color}}, 
\quad\quad\quad
\begin{pmatrix}
\nu_\tau\\
\tau
\end{pmatrix}_L,\quad\quad  \quad \tau_R 
\quad  \Bigg).
\eea
In short, for all of them as three generations, we can denote them as:
\bea
\Bigg(
\begin{pmatrix}
u\\
d
\end{pmatrix}_L \times {\bf 3}_{\text{color}}, \quad\quad \quad u_R \times {\bf 3}_{\text{color}}, \quad\quad\quad\quad d_R \times {\bf 3}_{\text{color}}, 
\quad\quad\quad
\begin{pmatrix}
\nu_e\\
e
\end{pmatrix}_L,\quad\quad  \quad e_R 
\quad  \Bigg)\times \text{3 generations}. \quad
\eea
In fact, all the following \emph{four} kinds of
\bea
G_{\text{internal}}=\frac{\SU(3)\times \SU(2)\times \U(1)}{\Z_q}
\eea
with $q=1,2,3,6$ are compatible with the above representations of 
fermion fields.\footnote{See an excellent exposition on
$G_{\text{internal}}=\frac{\SU(3)\times \SU(2)\times \U(1)}{\Z_q}$
with $q=1,2,3,6$
in a recent work by Tong \cite{Tong2017oea1705.01853}.}
These $15  \times 3$ Weyl spinors can be written in the following more succinct forms of representations for any of the internal symmetry group $G_{\text{internal}}$ with $q=1,2,3,6$:
\bea \label{eq:rep-3generations}
\Bigg( ({\bf 3},{\bf 2}, 1/6)_L,({\bf 3},{\bf 1}, 2/3)_R,({\bf 3},{\bf 1},-1/3)_R,({\bf 1},{\bf 2},-1/2)_L,({\bf 1},{\bf 1},-1)_R  \Bigg)\times \text{3 generations} \nn\\
\Rightarrow
\Bigg( ({\bf 3},{\bf 2}, 1/6)_L,(\overline{\bf 3},{\bf 1}, -2/3)_L,(\overline{\bf 3},{\bf 1},1/3)_L, ({\bf 1},{\bf 2},-1/2)_L,({\bf 1},{\bf 1},1)_L  \Bigg)\times \text{3 generations}.
\eea
The triplet given above is listed by their  representations:
\bea \label{eq:rep}
\text{(SU(3) representation, SU(2) representation, hypercharge $Y$)}.
\eea

For example,
$({\bf 3},{\bf 2}, 1/6)$ means that ${\bf 3}$ in SU(3), ${\bf 2}$ in SU(2) and 1/6 for hypercharge.
In the second line of \Eq{eq:rep-3generations}, we transforms the right-handed Weyl spinor 
$\Psi_R \sim {\bf 2}_R \text{ of } \Spin(3,1)$
to its left-handed $\Psi_L \sim {\bf 2}_L \text{ of } \Spin(3,1)$, while we flip their representation \Eq{eq:rep}
to its complex conjugation representation.\footnote{Note that ${\bf 2}$ and $\overline{\bf 2}$ are the same representation in SU(2), see, e.g., in the context of Yang-Mills gauge theories with discrete symmetries and cobordism \cite{1711.11587GPW}.} 
If we include the right-handed neutrinos (say ${\nu_{e}}_R$, ${\nu_{\mu}}_R$, and ${\nu_{\tau}}_R$), they are all in the representation 
\bea \label{eq:nuR-rep}
({\bf 1},{\bf 1},0)_R
\eea with no hypercharge.
We can also represent a right-handed neutrino by the left-handed (complex) conjugation version 
\bea  \label{eq:nuL-rep}
({\bf 1},{\bf 1},0)_L.
\eea
Also the complex scalar Higgs field $\phi_H$ is in a representation 
\bea
({\bf 1},{\bf 2}, 1/2).
\eea

In the Higgs condensed phase of SM, the conventional Higgs vacuum expectation value (vev) is chosen to be $\langle \phi_H \rangle =\frac{1}{\sqrt 2}
\begin{pmatrix}
0\\
v
\end{pmatrix}$, which vev has a $Q_{\rm{EM}}=0$.

{Note that 
our hypercharge $Y$ is given conventionally by the relation: $Q_{\rm{EM}}=T_3 +Y$ where 
$Q_{\rm{EM}}$ is the unbroken (not Higgsed) electromagnetic gauge charge
and $T_3= \frac{1}{2}
\begin{pmatrix}
1 & 0\\
0 & -1
\end{pmatrix}$ is a generator of SU(2)$_{\text{weak}}$. However, some other conventions are common, we list down three conventions
\bea  \label{eq:QEM}
Q_{\rm{EM}}=T_3 +Y=T_3 +\frac{1}{2} Y_W=T_3 +\frac{1}{6} \tilde{Y}.
\eea  
In the $Q_{\rm{EM}}=T_3 +\frac{1}{6} \tilde{Y}$ version, we have the integer quantized $\tilde{Y} = 6 Y$. We can rewrite \eq{eq:rep-3generations} as:
\bea \label{eq:rep-3generations-Sec}
&&\hspace{-18mm}
\Bigg( ({\bf 3},{\bf 2}, Y= 1/6)_L,({\bf 3},{\bf 1}, Y= 2/3)_R,({\bf 3},{\bf 1}, Y= -1/3)_R,({\bf 1},{\bf 2},Y=-1/2)_L,({\bf 1},{\bf 1},Y= -1)_R  \Bigg)\times \text{3 generations} \nn\\
=
&&\hspace{-6mm}
\Bigg( ({\bf 3},{\bf 2}, Y= 1/6)_L,(\overline{\bf 3},{\bf 1}, Y= -2/3)_L,(\overline{\bf 3},{\bf 1},Y= 1/3)_L, ({\bf 1},{\bf 2},Y= -1/2)_L,({\bf 1},{\bf 1},Y= 1)_L  \Bigg)\times 
\text{3 generations}\nn\\
=
&&\hspace{-6mm}
\Bigg( ({\bf 3},{\bf 2}, Y_W=1/3)_L,({\bf 3},{\bf 1},Y_W=4/3)_R,({\bf 3},{\bf 1},Y_W=-2/3)_R,({\bf 1},{\bf 2},Y_W=-1)_L,({\bf 1},{\bf 1},Y_W=-2)_R  \Bigg)
\times \text{3 generations} \nn\\
=
&&\hspace{-6mm}
\Bigg( ({\bf 3},{\bf 2}, Y_W= 1/3)_L,(\overline{\bf 3},{\bf 1}, Y_W= -4/3)_L,(\overline{\bf 3},{\bf 1},Y_W= 2/3)_L, ({\bf 1},{\bf 2},Y_W= -1)_L,({\bf 1},{\bf 1},Y_W= 2)_L  \Bigg)\times 
\text{3 generations}\nn\\
 \label{eq:rep-3generations-tildeY-Sec}
=
&&\hspace{-6mm}
\Bigg( ({\bf 3},{\bf 2}, \tilde{Y}=1)_L,({\bf 3},{\bf 1}, \tilde{Y}=4)_R,({\bf 3},{\bf 1},\tilde{Y}=-2)_R,({\bf 1},{\bf 2},\tilde{Y}=-3)_L,({\bf 1},{\bf 1},\tilde{Y}=-6)_R  \Bigg)\times \text{3 generations} \nn\\
=
&&\hspace{-6mm}
\Bigg( ({\bf 3},{\bf 2}, \tilde{Y}=1)_L,(\overline{\bf 3},{\bf 1}, \tilde{Y}=-4)_L,(\overline{\bf 3},{\bf 1},\tilde{Y}=2)_L, ({\bf 1},{\bf 2},\tilde{Y}=-3)_L,({\bf 1},{\bf 1},\tilde{Y}=6)_L  \Bigg)\times \text{3 generations}.
\label{eq:Weyl-rep}
\eea
The right hand neutrino $\nu_R$ is in a representation 
\bea
({\bf 1},{\bf 1}, Y=0)
=({\bf 1},{\bf 1}, Y_W=0)
=({\bf 1},{\bf 1}, \tilde{Y} =0).
\eea
times some generation number.
Also the complex scalar Higgs field $\phi_H$ is in a representation 
\bea
({\bf 1},{\bf 2}, Y=1/2)
=({\bf 1},{\bf 2}, Y_W=1)
=({\bf 1},{\bf 2}, \tilde{Y} =3),
\eea
where $\tilde{Y} =3 Y_W= 6 Y$.
}
We organize the above data in Table \ref{table:SMfermion}.

\item If we include the $3 \times 2 + 3 + 3 + 2 +1 =15$ left-handed Weyl spinors from one single generation, 
we can combine them as a multiplet of $\overline{\bf 5}$ and {\bf 10} left-handed Weyl spinors of SU(5):
\bea
(\overline{\bf 3},{\bf 1},1/3)_L \oplus ({\bf 1},{\bf 2},-1/2)_L &\sim& \overline{\bf 5} \text{ of } \SU(5),\\
  ({\bf 3},{\bf 2}, 1/6)_L \oplus (\overline{\bf 3},{\bf 1}, -2/3)_L \oplus ({\bf 1},{\bf 1},1)_L  &\sim& {\bf 10} \text{ of } \SU(5).
\eea
Hence these are matter field representations of the SU(5) GUT with a SU(5) gauge group.
Other than the electroweak Higgs $\phi_H$ (which $\phi_H$ can be part of a scalar Higgs in the fundamental 
${\bf 5}$ of SU(5) GUT), we may also introduce a different GUT Higgs field $\phi_{GG}$ 
to break down SU(5) to $\frac{\SU(3) \times   \SU(2) \times \U(1)}{\Z_6}$.
The $\phi_{GG}$ is in the adjoint representation of SU(5) as 
\bea
 {\bf 24} =({\bf 8},{\bf 1}, {Y}=0) \oplus ({\bf 1},{\bf 3}, {Y}=0) \oplus  ({\bf 1},{\bf 1}, {Y}=0) \oplus({\bf 3},{\bf 2}, {Y}=-\frac{5}{6})
 \oplus(\bar{\bf 3},{\bf 2}, {Y}=\frac{5}{6}).
\eea
\item If we include an extra right-handed neutrino plus
the $3 \times 2 + 3 + 3 + 2 +1 =15$ left-handed Weyl spinors from one single generation, 
we can combine them as a multiplet of 16 left-handed Weyl spinors:
\bea
\quad  \Psi_L \sim {\bf 16}^+ \text{ of } \Spin(10),
\eea
which sits at the 16-dimensional irreducible spinor representation of Spin(10).
(In fact,  ${\bf 16}^+$ and ${\bf 16}^-$-dimensional irreducible spinor representations 
together form a ${\bf 32}$-dimensional reducible spinor representation of Spin(10).)
Namely, instead of an SO(10) gauge group,
we should study the SO(10) GUT with a Spin(10) gauge group.

\item  {\bf Lie algebra generators and gauge bosons}:
We can count the number of Lie algebra generators to represent the local 1-form gauge field of gauge bosons.
For example, there are $1+3+8 =12$ independent Lie algebra generators thus gauge bosons for the Lie algebra $u(1) \times su(2) \times su(3)$.
There are $24$ independent Lie algebra generators (thus gauge bosons) for the Lie algebra $su(5)$,
and $45$ Lie algebra generators (thus gauge bosons) for the Lie algebra $so(10)$.
In this work, we focus on SM and SU(5) GUT. We list down their quantum numbers in Table \ref{table:SMboson}.

\item {\bf $\U(1)_{{ \mathbf{B}-  \mathbf{L}}}$ symmetry and 
$\U(1)_{X}$ symmetry}: \label{RemarkU1}
For SM with ${{G}_{\text{internal}} }=G_{\text{SM}_q}$ of $q=1,2,3,6$,
we have a ${ \mathbf{B}-  \mathbf{L}}$ (baryon minus lepton numbers) $\U(1)_{{ \mathbf{B}-  \mathbf{L}}}$ global symmetry.
For SU(5) GUT with  ${{G}_{\text{internal}} }=\SU(5)$, we can have a $\U(1)_{X}$ symmetry \cite{Wilczek1979hcZee},
where the $X$ charge is related to ${ \mathbf{B}-  \mathbf{L}}$ via
\bea
X &\equiv&
5({ \mathbf{B}-  \mathbf{L}})-4Y=
5({ \mathbf{B}-  \mathbf{L}})-2Y_W 
 ,\\
\tilde X &\equiv& 3 X=
5 \cdot 3 ({ \mathbf{B}-  \mathbf{L}})-2 \cdot 3 Y_W
= 
5 \cdot 3 ({ \mathbf{B}-  \mathbf{L}})-2  \tilde Y  =  
5  ({ \tilde{\mathbf{B}}-  \tilde{\mathbf{L}}})-2  \tilde Y,\\
\tilde Y &=& 3 Y_W = 6 Y.
\eea
Here we have used \Eq{eq:QEM}. 

\item {{\bf  $\Z_{4, X}$ symmetry, $\Z_2^F$ fermion parity, and discrete symmetries}: \label{RemarkZ4X}\\
We will learn that it is natural to consider a $\Z_{4, X}$ subgroup of $\U(1)_{X}$, when we attempt to embed the SU(5) GUT to SO(10) GUT. In fact, the
 $\Z_{4, X}$ can be regarded as the $\Z_4$ center of the Spin(10) gauge group as $\text{Z}(\Spin(10))=\Z_4
 = \Z_{4,X} \supset \Z_{2}^F$ \cite{GarciaEtxebarriaMontero2018ajm1808.00009}.
Since Spin(10) is fully dynamically gauged in SO(10) GUT,  the $\Z_{4, X}$ is also dynamically gauged.
In summary, we have
\bea
&&\text{$\bullet$ $\U(1)_{{ \mathbf{B}-  \mathbf{L}}}$ can be a global symmetry (maybe anomalous) in the SMs for all $q=1,2,3,6$}.\cr
&&\text{$\bullet$ $\U(1)_{X}$ can be a global symmetry (maybe anomalous) in the SU(5) GUT}.\cr
&&\text{$\bullet$ $\Z_{4, X}$ can be chosen to be \emph{either} a global \emph{or} a gauge symmetry for the SM and SU(5) GUT.}\cr
&&\text{\quad \quad\quad But $\Z_{4, X}$ is a dynamical gauge symmetry in the SO(10) GUT.}\cr
&&\text{$\bullet$ $\Z_{2}^F$ 
 fermion parity has a generator $(-1)^F$ with the fermion number $F$, which
 can be a global symmetry} \cr
&&\text{\quad \quad\quad in the SMs (for all $q=1,2,3,6$) and in the SU(5) GUT.} \cr
 &&\text{\quad \quad\quad But the $\Z_{2}^F$ is naturally chosen to be a dynamical gauge symmetry in the SO(10) GUT.
 }\nn
\eea    
For the SO(10) GUT, we may call  $\Z_{4, X}$ a gauge symmetry colloquially, but we should remind readers that a 
gauge symmetry is not really a (global) symmetry, but only a gauge redundancy.
Note that:
\bea
&&\text{$\bullet$ All fermions from leptons and quarks have a $\Z_{4,X}$ charge 1, see \Table{table:SMfermion}.}\cr
&&\text{$\bullet$ All gauge bosons have a $\Z_{4,X}$ charge 0, but the Higgs have a $\Z_{4,X}$ charge 2, see \Table{table:SMboson}.}\cr
&&\text{$\bullet$ The $\Z_{2}^F$ gives $(-1)$ sign to all fermions (additive notation: charge 1 in  $\Z_{2}^F$ in \Table{table:SMfermion}).}\cr
&&\text{$\bullet$ The $\Z_{2}^F$ gives $(+1)$ sign to all bosons (additive notation: charge 0 in  $\Z_{2}^F$  in \Table{table:SMboson}).
} \nn
\eea
The above facts about $\Z_{4, X}$ charges for bosons or fermions (in \Table{table:SMfermion} and  \Table{table:SMboson})
can be understood from 
a group extension (as a short exact sequence):
$0 \to \Z_{2}^F  \to \Z_{4,X}  \to (\Z_{2}^\CA \equiv \frac{\Z_{4,X}}{\Z_{2}^F})  \to 0$.
The fermion parity $\Z_{2}^F$ is a normal subgroup of $\Z_{4,X}$, while the $\Z_{2}^\CA$ is only a quotient group.  
So the fermions, with an odd $\Z_{2}^F$ charge, must carry also an odd charge of $\Z_{4,X}$.
The bosons, with an even (i.e., no) $\Z_{2}^F$ charge, must carry also an even charge of $\Z_{4,X}$.\footnote{{Given
the $\Z_{4,X}$ charge state
$| X\rangle$ with $X=0,1,2,3$
as a representation $z^X$.
Note that $z \in \U(1)$ with $|z|=1$, where we embed the normal subgroup $\Z_2^F \subset  \Z_{4,X} \subset \U(1)$.\\
$\bullet$ The $\Z_{4,X}$ symmetry generator $\U_{\Z_{4,X}}$ acts on $| X\rangle$,
which becomes $\U_{\Z_{4,X}} | X\rangle = \ii^X | X\rangle$ with $z=\ii$. \\
$\bullet$ The subgroup  $\Z_{2}^F$ symmetry generator 
$\U_{\Z_{2}^F}= (\U_{\Z_{4,X}})^2$
can also act on $| X\rangle$,
which becomes $\U_{\Z_{2}^F}  | X\rangle  = (\U_{\Z_{4,X}})^2 | X\rangle = \ii^{2X} | X\rangle = (-1)^X | X\rangle $.
Thus, we read the fermion parity $(-1)^F$,
the $| 1\rangle$ and $| 3\rangle$ has $-1$ (thus odd 1 in $\Z_{2}^F$), while 
the $| 0\rangle$ and $| 2\rangle$ has $+1$ (thus even 0 in $\Z_{2}^F$).
This is also consistent with the fact that the group extension 
$0 \to \Z_{2}^F  \to \Z_{4,X}  \to \Z_{2}^\CA \equiv \frac{\Z_{4,X}}{\Z_{2}^F}  \to 0$
requires a nontrivial 2-cocycle $\H^2(\B \Z_{2}^\CA, \Z_{2}^F) = \Z_2$ to specify the extension to $\Z_{4,X}$.}
}
}

These extra symmetries are well-motivated in the earlier pioneer works
 \cite{Ibanez1991hvRossPLB, Banks1991xjDine9109045, Csaki1997awMurayama9710105, Dreiner2005rd0512163} and References therein 
the recent work \cite{GarciaEtxebarriaMontero2018ajm1808.00009, Hsieh2018ifc1808.02881, GuoJW1812.11959}.
Extra discrete symmetries can be powerful giving rise to new anomaly cancellation constraints. 
Part of the new ingredients we will survey are the new global anomalies for discrete symmetries of SMs, in \Sec{Sec:discretesymmetries}.

Let us discuss how these extra groups can be embedded into the total groups in Remarks
\ref{RemarkEmbed1},
\ref{RemarkEmbed2},
\ref{RemarkEmbed3},
and \ref{RemarkEmbed4}.
\item 
\label{RemarkEmbed1}
We find the Lie group embedding for the internal symmetry of GUTs and Standard Models 
 \cite{WW2019fxh1910.14668, WangWen2018cai1809.11171}:
\bea
\SO(10) 
\supset 
 \SU(5) 
\supset 
\frac{\U(1) \times \SU(2)\times\SU(3)}{\Z_6}.\\
\Spin(10) 
\supset 
 \SU(5) 
\supset 
\frac{\U(1) \times \SU(2)\times\SU(3)}{\Z_6}.
\eea
Only $q=6$, but not other $q=1,2,3$, for $G_{\text{SM}_q}$ can be embedded into Spin(10) nor SU(5).
So from the GUT perspective, it is natural to consider the Standard Model gauge group 
$\frac{\U(1) \times \SU(2)\times\SU(3)}{\Z_6}$.\footnote{In fact, the author
believe and declare that $G_{\text{SM}_6} \equiv \frac{\U(1) \times \SU(2)\times\SU(3)}{\Z_6}$
is the correct and natural gauge group of SM.}
\item 
\label{RemarkEmbed2}
We also find the following group embedding for the spacetime and internal symmetries of GUTs and Standard Models
(\Refe{WW2019fxh1910.14668, WangWen2018cai1809.11171}, see also \cite{WanWangv2} for the derivations):
\bea \label{eq:SMembed1}
{\frac{\Spin(d) \times
\Spin(10)}{{\Z_2^F}} 
\supset 
\Spin(d) \times \SU(5) 
\supset 
\Spin(d) \times \frac{\SU(3) \times   \SU(2) \times \U(1)}{\Z_6} }.\\
 \label{eq:SMembed2}
{{\Spin(d) \times
\Spin(10)} 
\supset 
\Spin(d) \times \SU(5) 
\supset 
\Spin(d) \times \frac{\SU(3) \times   \SU(2) \times \U(1)}{\Z_6}}.
\eea
\item
\label{RemarkEmbed3}
For an extra $\U(1)_{{ \mathbf{B}-  \mathbf{L}}}$ or $\U(1)_{X}$ symmetry, we need to consider Spin${}^c \equiv \Spin(d) \times_{\Z_2^F} \U(1)$ structure.
We find the embedding: 
{
\bea  
\Spin^c(d)\times \SU(5) 
\supset 
\Spin^c(d)  \times \frac{\SU(3) \times   \SU(2) \times \U(1)}{\Z_6}.
\eea
}
\item \label{RemarkEmbed4}
For an extra $\Z_{4,X}$ symmetry,
we need to consider $\Spin(d) \times_{\Z_2} \Z_4 =\Spin(d) \times_{\Z_2^F} \Z_{4,X}  $ structure.
In order to contain these groups embedded in $\frac{\Spin(d) \times
\Spin(10)}{{\Z_2^F}}$, it is more naturally to consider:
\bea 
 \label{eq:SMembed3}
{\frac{\Spin(d) \times
\Spin(10)}{{\Z_2^F}} 
\supset 
\Spin(d) \times_{\Z_2^F} \Z_{4,X} \times \SU(5) 
\supset 
\Spin(d) \times_{\Z_2^F} \Z_{4,X} \times \frac{\SU(3) \times   \SU(2) \times \U(1)}{\Z_6} }.
\eea

\end{enumerate}
\Refe{WW2019fxh1910.14668, WangWen2018cai1809.11171, GarciaEtxebarriaMontero2018ajm1808.00009, 2019arXiv191011277D}
 study the cobordism theory of some of these SM, BSM, and GUT groups.
 We shall particular pay attention to $d=5$ for
$\Omega^{d=5}_{G} \equiv
\Omega^{d=d}_{({\frac{{G_{\text{spacetime} }} \ltimes  {{G}_{\text{internal}} }}{{N_{\text{shared}}}}})} 
\equiv
\TP_{d=5}(G),$ since the 5d cobordism invariants of $\TP_{d=5}(G)$ can classify the 4d invertible anomalies of the total group $G$. 
We particularly apply the results in \cite{WW2019fxh1910.14668}, for anomaly constraints and anomaly matchings of BSM physics,
in the following \Sec{sec:DynamicalGaugeAnomalyCancellation}.

\subsection{Classifications of anomalies and their different uses}
\label{sec:2OverviewSMGUTdifferentanomalies}

\begin{figure}[h!] 
  \centering
      ({\it 1})  \includegraphics[width=3.in]{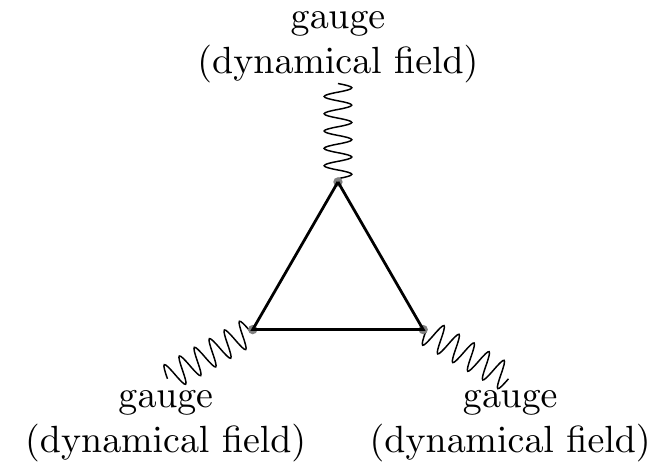}
  ({\it 2})  \includegraphics[width=3.in]{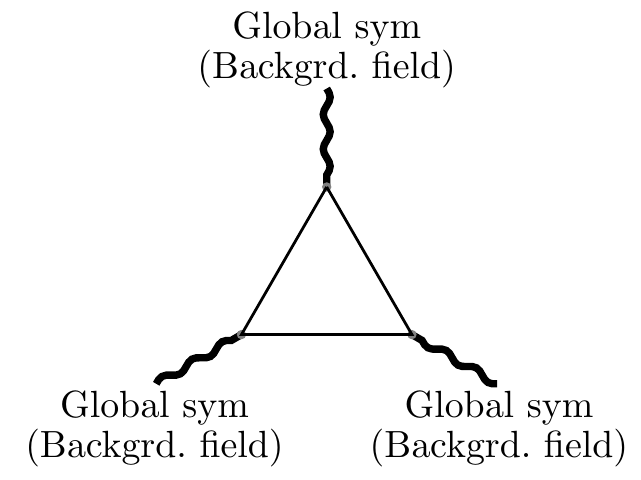} \\[2mm]
  ({\it 3}) \includegraphics[width=3.in]{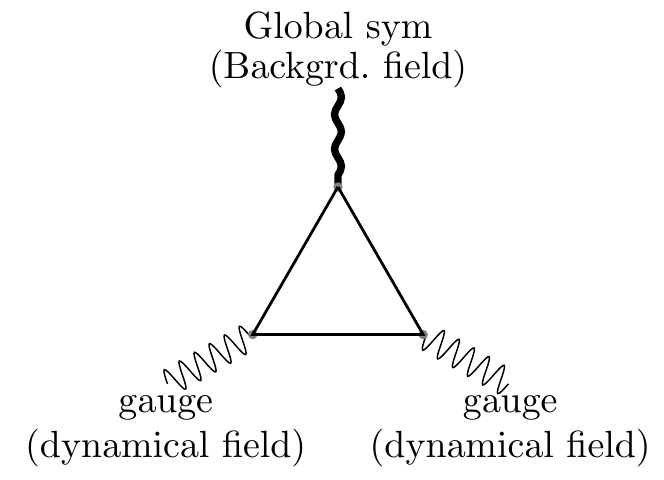} 
  ({\it 4}) \includegraphics[width=3.in]{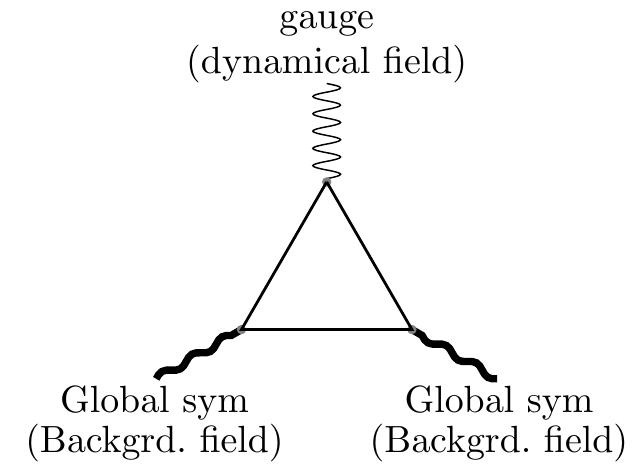}
  \caption{Examples of perturbative local anomalies of $\Z$ classes in 4d that can be captured by the free part of cobordism group 
  $\Omega^{d=5}_{G} 
\equiv
\TP_{d=5}(G)$ in \Eq{eq:TPG}, which is in fact descended from the free part of bordism group $\Omega_{d=6}^{G}$.
  ({\it 1}) Dynamical gauge anomaly. 
  ({\it 2}) 't Hooft anomaly of background (Backgrd.) fields. 
  ({\it 3}) Adler-Bell-Jackiw (ABJ \cite{Adler1969gkABJ, Bell1969tsABJ}) type of anomalies. 
  ({\it 4}) Anomaly that involves two background fields of global symmetries and one dynamical gauge field.
  }
  \label{fig:triangle}
\end{figure}

To start, let us mention different concepts of anomalies and their different uses.  See for example \Fig{fig:triangle} for
 perturbative local anomalies captured by Feynman-Dyson graphs.
Generally, we also have nonperturbative global anomalies not captured by Feynman graphs, but their uses are similar as follows:
\begin{enumerate}[leftmargin=-4.mm, label=\textcolor{blue}{({\it\arabic*})}., ref={({\it\arabic*})}]
\item  \label{remark:dynamicalanomaly}
Dynamical gauge anomaly of ${{G}_{\text{internal}} }$ (e.g., \Fig{fig:triangle} ({\it 1})): 
  Anomaly matching cancellation must be zero for a dynamical gauge theory of its group ${{G}_{\text{internal}} }$. 
\item  \label{remark:tHooftanomaly}
 't Hooft anomaly (e.g., \Fig{fig:triangle} ({\it 2})):
 Anomaly matching of background symmetry of $G$ including ${{G}_{\text{internal}}}$ for their background fields, surprisingly, need \emph{not} to be 
 zero for a QFT:\\
 \begin{enumerate}[leftmargin=6.mm, label=\textcolor{blue}{($\bullet$ \roman*)}., ref={($\bullet$ \roman*)}]
\item
 If the 't Hooft anomaly of a QFT of any $G$ background fields turn out to be exactly zero, this means that these global symmetries of QFT can be realized local and \emph{onsite} (or on an $n$-simplex for a higher generalized global $n$-symmetry \cite{Gaiotto2014kfa1412.5148}) on a lattice. 
This also means that we do not need to regularize the QFT living on the boundary of Symmetry-Protected/Enriched Topological states 
\cite{Chen2011pg1106.4772, Wen2013oza1303.1803,Wen2013ppa1305.1045, Wang2017locWWW1705.06728, WangWen2018cai1809.11171}.
In fact, the QFT may be well-defined as the low energy theory of a quantum local lattice model in its own dimensions, as a high-energy ultraviolet (UV) completion ---
if it is free from all anomalies, via a cobordism argument \cite{WangWen2018cai1809.11171}.\\
\item
If the 't Hooft anomaly of a QFT of any $G$ background fields turn out to be not zero nor cancelled:
Then this can be regarded as several meanings and implications on QFT dynamics:\\
$\bullet$ {\bf First}, the symmetries of QFT in fact are \emph{non-local} or \emph{non-onsite} (or \emph{non-on-$n$-simplex} for a higher generalized global $n$-symmetry) on a lattice. This means the obstruction of gauging such  \emph{non-local}  symmetries, thus this 
obstruction is equivalent with the definition of 't Hooft anomaly \cite{tHooft1979ratanomaly}.  

$\bullet$ {\bf Second}, the QFT can have a hidden sector of the same dimension. We can match the 't Hooft anomaly of QFT  with additional sector $S'$ of the same dimension,
but with the $S'$ sector with the opposite 't Hooft anomaly.
So, the combined system,  the QFT and $S'$, can have a no 't Hooft anomaly at all.

$\bullet$ {\bf Third}, the QFT can have a hidden sector of one higher dimension. In order to have the symmetries of QFT with 't Hooft anomaly to be \emph{local} or \emph{onsite} instead,
we need to append
and regularize this $d$d QFT living on the boundary of $(d+1)$d Symmetry-Protected/Enriched Topological states (SPTs/SETs) 
\cite{Chen2011pg1106.4772, Wen2013oza1303.1803, Wang2017locWWW1705.06728, 
WangWen2018cai1809.11171}.\footnote{Symmetry-Protected Topological states (SPTs) are the short-range entangled states as a generalization of the free non-interacting topological insulators and superconductors \cite{2010RMP_HasanKane,2011_RMP_Qi_Zhang, RyuSPT, Kitaevperiod, Wen1111.6341}
with interactions.
Symmetry-Enriched Topological states (SETs) are topological ordered states enriched by global symmetries.
 The enthusiastic readers can overview the condensed matter terminology bridging to QFT in \Refe{1711.11587GPW} for QFT theorists,
or the excellent condensed matter reviews \cite{Senthil1405.4015, Wen2016ddy1610.03911}.} 

This is related to the Callan-Harvey anomaly inflow \cite{1984saCallanHarvey} of $(d+1)$d bulk and $d$d boundary with their spacetime dimensions differed by one.

$\bullet$ {\bf Fourth}, 
(some part of) the global symmetry can also be broken spontaneously or explicitly, 
so that the 't Hooft anomaly is matched by \emph{symmetry breaking}.

$\bullet$ {\bf Fifth}, 
Of course, we can also have a certain combination of the First, Second, Third, and Fourth scenarios above, in order to make sense of QFT with 't Hooft anomaly.
\end{enumerate}

\item  \label{remark:ABJanomaly}
Adler-Bell-Jackiw (ABJ \cite{Adler1969gkABJ, Bell1969tsABJ}) type of anomalies (e.g., \Fig{fig:triangle} ({\it 3})):
The non-conservations of the global symmetry current $\CJ$ is coupled to the background non-dynamical field $\CA$
via the action term $\int \CA \wedge \star \CJ \equiv \int \dd^d x(\CA_{\mu}  \CJ^{\mu})$ with the Hodge dual star $\star$.
The non-conservation of current is proportional to the anomaly factor in $d$d spacetime
$$
\dd (\star \CJ ) \propto (F_a)^{d/2} \propto  F_a \wedge F_a \wedge \dots.
$$
Here $a$ are dynamical gauge fields; we can also modify the equation appropriate for several different dynamical 1-form or higher-form gauge fields. 
For $d=4$, the ABJ anomaly formula is precisely captured by \Fig{fig:triangle} ({\it 3}).

\item   \label{remark:4anomaly}
Anomaly that involves two background fields of global symmetries and one dynamical gauge field (e.g., \Fig{fig:triangle} ({\it 4})).

\end{enumerate}

Indeed it is obvious to observe that the anomalies \ref{remark:dynamicalanomaly} are tighten to anomalies \ref{remark:tHooftanomaly}.
\begin{enumerate}[label=\textcolor{blue}{(\greek*)}., ref={(\greek*)}]
\item \label{alp}
Anomalies from \ref{remark:dynamicalanomaly} can be related to anomalies from \ref{remark:tHooftanomaly} via the gauging principle.
\item \label{bet}
Anomalies from \ref{remark:tHooftanomaly} can be related to anomalies from \ref{remark:dynamicalanomaly} via the ungauging principle, or via gauging the higher 
symmetries  \cite{Gaiotto2014kfa1412.5148}.
\end{enumerate}
Thus if we learn the gauge group of a gauge theory (e.g., SM, GUT or BSM),
we may identify its ungauged global symmetry group as an internal symmetry group, say ${{G}_{\text{internal}} }$ via 
ungauging.\footnote{By gauging or ungauging, also depending on the representation of the matter fields that couple to the gauge theory,
we may gain or lose symmetries or higher symmetries \cite{Gaiotto2014kfa1412.5148}.
It will soon become clear, 
for our purpose, 
we need to primarily focus on 
the ordinary (0-form) internal global symmetries
and their anomalies.
}

\begin{figure}[h!] 
  \centering
  \includegraphics[width=4.in]{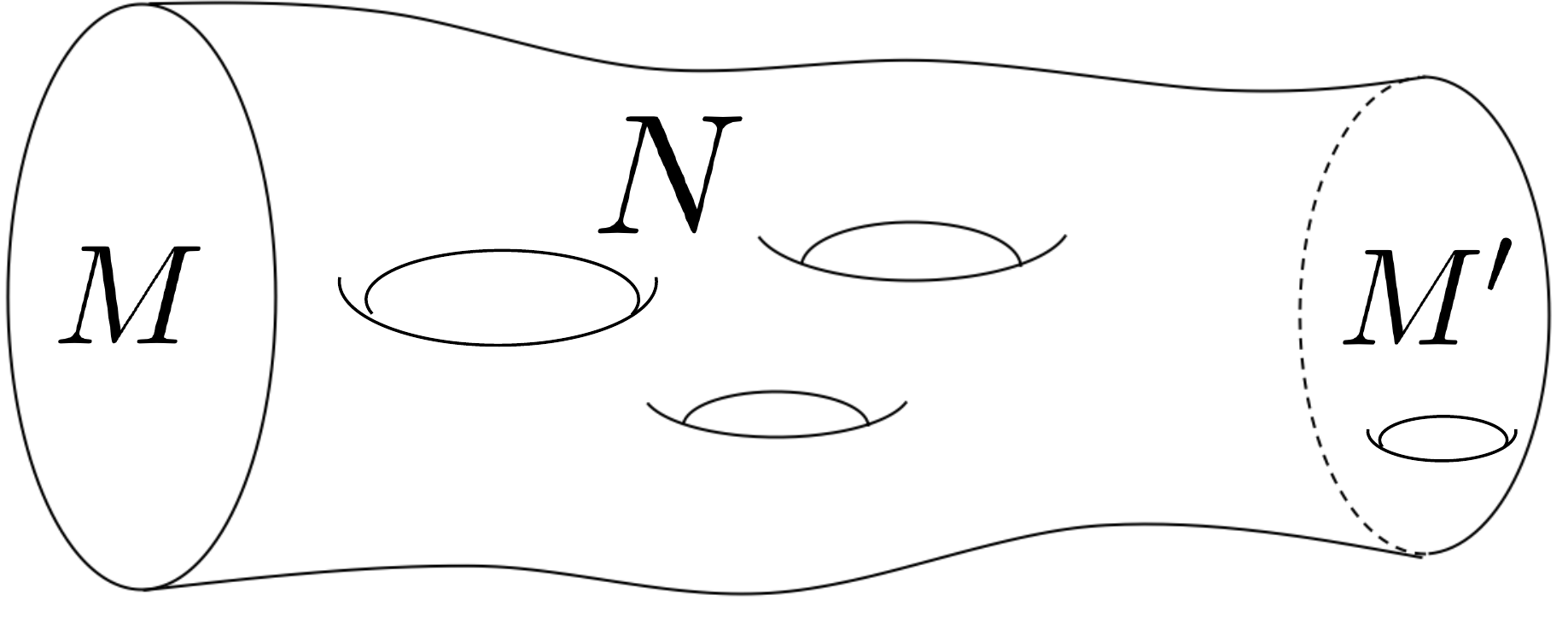}
  \caption{General nonperturbative global anomalies \emph{not} captured by Feynman graphs can still be characterized by
  generic curved manifolds with mixed gauge and gravitational background probes.
  The figure shows a bordism between manifolds. Here $M$ and $M'$ are two closed $d$-manifolds, $N$ is a compact $d+1$-manifold whose boundary is the disjoint union of $M$ and $M'$, {namely $\prt N= M \sqcup M'$.} If there are additional $G$-structures on these manifolds, then the $G$-structure on  $N$ is required to be compatible with the $G$-structures on $M$ and $M'$. 
  If there are additional maps from these manifolds to a fixed topological space, then the maps are also required to be compatible with each other. If these conditions are obeyed, then $M$ and $M'$ are called bordant equivalence 
  and $N$ is called a bordism between $M$ and $M'$. }\label{fig:Cobordism2006}
\end{figure}

Now let us discuss the classifications of anomalies.
By ``all invertible quantum anomalies'' obtain from a cobordism classification, we mean the inclusion of: 
\begin{enumerate}[leftmargin=2mm, label=\textcolor{blue}{(\roman*)}., ref={(\roman*)}]
\item
{\bf Perturbative local anomalies} captured by perturbative Feynman graph loop calculations, 
classified by the integer group $\mathbb{Z}$ classes, or the  free classes in mathematics. 
Some selective examples from QFT or gravity are:
\begin{enumerate} [label=\textcolor{blue}{(\arabic*)}:, ref={(\arabic*)}]
\item
Perturbative fermionic anomalies from chiral fermions with U(1) symmetry, originated from Adler-Bell-Jackiw (ABJ) anomalies \cite{Adler1969gkABJ,Bell1969tsABJ}
with $\mathbb{Z}$ classes.
\item Perturbative bosonic anomalies from bosonic systems with U(1) symmetry
 with $\mathbb{Z}$ classes \cite{WanWang2018bns1812.11967}.
 \item Perturbative gravitational anomalies \cite{AlvarezGaume1983igWitten1984}.
\end{enumerate}
\item {\bf  Non-perturbative global anomalies}, classified by a product of finite groups such as $\mathbb{Z}_n$, or the torsion classes  in mathematics.
Some selective examples from QFT or gravity are:
\begin{enumerate} [label=\textcolor{blue}{(\arabic*)}:, ref={(\arabic*)}]
\item An SU(2) anomaly of Witten in 4d or in 5d \cite{Witten1982fp} with a $\mathbb{Z}_2$ class, which is a gauge anomaly. 
\item A new SU(2) anomaly  in 4d or in 5d \cite{WangWenWitten2018qoy1810.00844} with a different $\mathbb{Z}_2$ class, which is a mixed gauge-gravity anomaly.
\item  Higher 't Hooft anomalies of $\Z_2$ class for a pure 4d SU(2) YM theory  
with a second-Chern-class topological term \cite{Gaiotto2017yupZoharTTT, Wan2018zqlWWZ1812.11968, Wan2019oyr1904.00994} (or the so-called SU(2)$_{\theta =\pi}$ YM):
The higher anomaly involves a discrete 0-form time-reversal symmetry and a 1-form center $\Z_2$-symmetry.  
The first anomaly is discovered in \cite{Gaiotto2017yupZoharTTT}; later the anomaly is refined via a mathematical well-defined 5d {bordism invariant as its topological term}, with additional new  $\Z_2$ class anomalies found for Lorentz symmetry-enriched 
four siblings of Yang-Mills gauge theories \cite{Wan2018zqlWWZ1812.11968, Wan2019oyr1904.00994}.
\item Global gravitational anomalies \cite{Witten1985xe} detected by exotic spheres.

\item Bosonic anomalies: Many types of 
bosonic anomalies in diverse dimensions are global anomalies \cite{1403.0617, WangSantosWen1403.5256, Kapustin1404.3230, 1405.7689, JWangthesis}.
These bosonic anomalies only require bosonic degrees of freedom,
but without the requirement of chiral fermions. Many such bosonic anomalies
are related to group cohomology or generalized group cohomology theory, living on the boundary of SPTs \cite{Chen2011pg1106.4772}, 
closely related to Dijkgraaf-Witten topological gauge theories \cite{Dijkgraaf1989pzCMP}.  
\end{enumerate} 
\end{enumerate} 
Our present work explore global anomalies (generically \emph{not} captured by Feynman graphs).
In order to help the readers to digest the physical meanings of global anomalies, 
we should imagine the computation of anomalies on generic curved spacetime manifolds with mixed gauge and gravitational background probes,
in the cobordism theory setting (as in \Fig{fig:Cobordism2006}).



\section{Dynamical Gauge Anomaly Cancellation}
\label{sec:DynamicalGaugeAnomalyCancellation}

In this section, we explicitly check the dynamical gauge anomaly cancellations of various SMs
 with four gauge group $G_{\text{SM}_q}\equiv{(\SU(3)\times \SU(2)\times \U(1))}/\Z_q$ with $q=1,2,3,6$.
The cobordism classifications of $G_{\text{SM}_q}$'s 4d anomalies are done in \cite{WW2019fxh1910.14668, 2019arXiv191011277D}.
The 4d anomalies can be written as 5d cobordism invariants, which are 5d invertible TQFT (iTQFT). 
These 5d cobordism invariants/iTQFT are derived in \cite{WW2019fxh1910.14668},
we summarized the classifications and invariants in Table \ref{table:SU3SU2U1}.

\begin{figure}[h!] 
  \centering
      (i)  \includegraphics[width=2.0in]{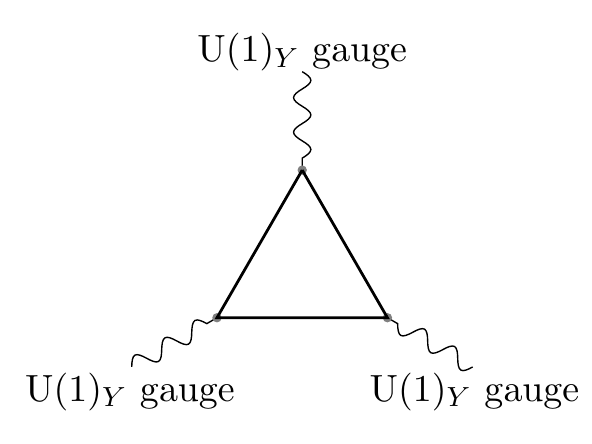}
  (ii)  \includegraphics[width=2.0in]{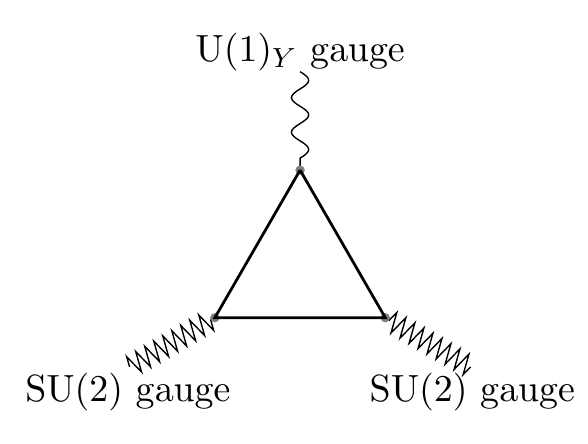} 
  (iii)  \includegraphics[width=2.0in]{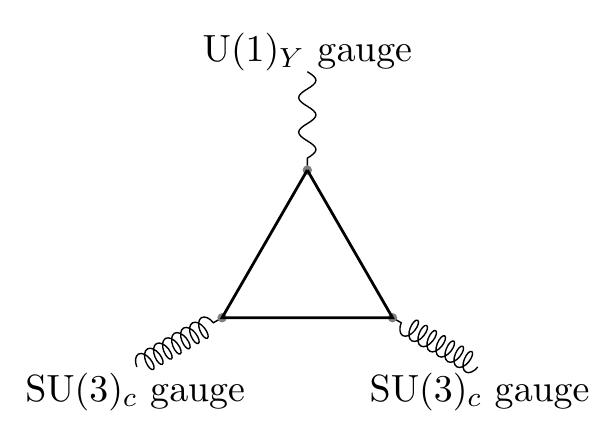}
  \\[2mm]
  (iv) \includegraphics[width=1.8in]{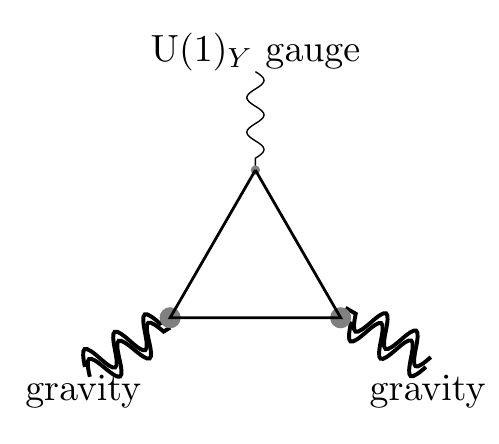} 
  (v) \includegraphics[width=2.2in]{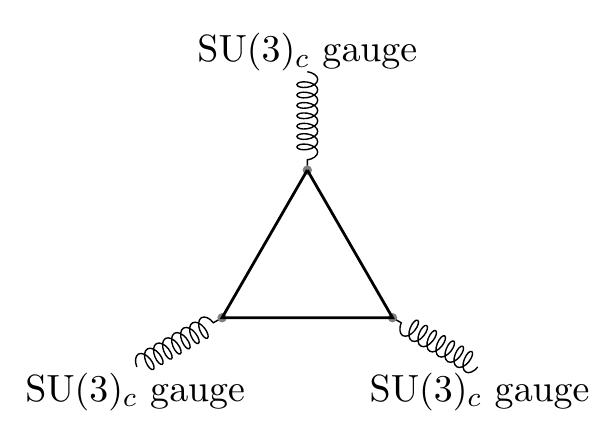}
  \caption{Examples of dynamical gauge anomaly cancellations in SM. In fact, the 5
   perturbative local anomalies from
    perturbative one-loop triangle Feynman diagrams
    precisely match
  anomaly classification of $\Z^5$ obtained from the cobordism group calculations in Table \ref{table:SU3SU2U1} and 
  \Refe{WW2019fxh1910.14668}.}
  \label{fig:triangle-dynamical}
\end{figure}

Moreover, as \Refe{2019arXiv191011277D} points out correctly already, we will show that dynamical gauge anomaly
cancellation indeed holds to be true for all  $G_{\text{SM}_q}$ with $q=1,2,3,6$.
\Refe{2019arXiv191011277D} has not written down the 5d cobordism invariants nor explicit anomaly polynomials.
However, by simply looking at the group classifications of anomalies, \Refe{2019arXiv191011277D} argues that there is only the famous Witten SU(2) nonperturbative global anomaly
of $\Z_2$ class \cite{Witten1982fp} ,
other anomalies all are perturbative local anomalies captured by Feynman graphs.
So why do we bother to do the calculations again, if \Refe{2019arXiv191011277D} has found all dynamical gauge anomalies cancel for $G_{\text{SM}_q}$?  
There are multiple reasons:\\[-10mm]
\begin{itemize}
\item First, we will show that the 5d cobordism invariants obtained in \cite{WW2019fxh1910.14668} indeed match with the anomaly polynomials.
For perturbative local anomalies of $\Z$ classes, see \Fig{fig:triangle-dynamical}, 
we indeed can show that the 5d cobordism invariants map to some one-loop Feynman graph calculations known in the 
standard QFT textbooks \cite{Weinberg1996Vol2, PeskinSchroeder1995book, Zee2003book, Srednicki2007book}. 
(In contrast, \Refe{2019arXiv191011277D} focus on global anomalies, and pays less attention on perturbative local anomalies of $\Z$ classes. Follow \Refe{WW2019fxh1910.14668}, 
we will fill in this
gap by considering all local and all global anomalies.)
We should work through all these correspondences carefully to gain a solid confidence for our understanding of cobordism classifications of anomalies.
\item Second, we can learn how to translate the cobordism data from math into the anomalies in physics.
For example, in Table \ref{table:SU3SU2U1}, we see that the 5d cobordism invariant $c_2(\SU(2))\tilde\eta$ in fact captures the 4d boundary theory has the 
Witten SU(2) nonperturbative global anomaly \cite{Witten1982fp}. 
\item Third, the most important thing, we will discover new phenomena later when we include the additional discrete symmetries into SM and GUT in \Sec{Sec:discretesymmetries}.
In fact, we will discover entirely new physics that previously have never been figured out in the past.
\end{itemize}

\noindent
{\bf Notations:} 
Throughout our work,  we write the three SU(2) Lie algebra generator ${\sigma^\ra}$ of the rank-2 matrix of fundamental representation
satisfying $\Tr[\frac{\sigma^\ra}{2} \frac{\sigma^\rb}{2}] =\frac{1}{2} \delta_{\ra \rb}$ with $\ra, \rb \in \{1,2,3\}$.
We write the eight SU(3) Lie algebra generator ${\tau^\ra}$ of the the rank-3 matrix of fundamental representation
satisfying $\Tr[\frac{\tau^\ra}{2} \frac{\tau^\rb}{2}] =\frac{1}{2} \delta_{\ra \rb}$ with $\ra, \rb \in \{1,2,\dots,8\}$.

\subsection{Summary of anomalies from a cobordism theory and Feynman diagrams} 
\label{sec:localanomaliesFeynman}

\begin{table}[H]
\centering
\hspace*{-7mm}
\begin{tabular}{c c c }
\hline
\multicolumn{3}{c}{Cobordism group $\TP_d(G)$ with $G_{\text{SM}_q}\equiv{(\SU(3)\times \SU(2)\times \U(1))}/\Z_q$ with $q=1,2,3,6$}\\
\hline
\hline
 & classes & cobordism invariants\\
\hline
\hline\\[-2mm]
\multicolumn{3}{c}{$G=\Spin\times {G_{\text{SM}_1}}$}\\[2mm]
\hline
5d & $\Z^5\times\Z_2$ & 
$\begin{matrix}
\mu(\text{PD}(c_1(\U(1)))), \quad \text{CS}_1^{\U(1)}c_1(\U(1))^2, \quad
\text{CS}_1^{\U(1)}c_2(\SU(2)) {\sim c_1(\U(1))\text{CS}_3^{\SU(2)}},\\
 \text{CS}_1^{\U(1)}c_2(\SU(3)) {\sim c_1(\U(1))\text{CS}_3^{\SU(3)}}, \quad\frac{\text{CS}_5^{\SU(3)}}{2}, \quad c_2(\SU(2))\tilde\eta 
 \end{matrix}$\\
\hline
\hline\\[-2mm]
\multicolumn{3}{c}{$G=\Spin\times {G_{\text{SM}_2}}$}\\[2mm]
\hline
5d & $\Z^5$ &  
$\begin{matrix}\mu(\text{PD}(c_1(\U(2)))), 
\quad 
\text{CS}_1^{\U(2)}c_1(\U(2))^2, 
\quad
\frac{\text{CS}_1^{\U(2)}c_2(\U(2))}{2} {\sim}  
\frac{c_1(\U(2)) \text{CS}_3^{\U(2)}}{2},\\ 
 \text{CS}_1^{\U(2)}c_2(\SU(3))
{\sim c_1(\U(2)) \text{CS}_3^{\SU(3)}}, 
\quad
 \frac{\text{CS}_5^{\SU(3)}}{2}
\end{matrix}$\\
\hline
\hline\\[-2mm]
\multicolumn{3}{c}{$G=\Spin\times {G_{\text{SM}_3}}$}\\[2mm]
\hline
5d & $\Z^5\times\Z_2$ & 
$\begin{matrix}\mu(\text{PD}(c_1(\U(3)))), 
\quad
\text{CS}_1^{\U(3)}c_1(\U(3))^2,
\quad
\text{CS}_1^{\U(3)}c_2(\SU(2)) {\sim c_1(\U(3)) \text{CS}_3^{\SU(2)}},
 \\
\frac{\text{CS}_1^{\U(3)}c_2(\U(3))+\text{CS}_5^{\U(3)}}{2}
{\sim 
\frac{
 c_1(\U(3)) \text{CS}_3^{\U(3)}
+\text{CS}_5^{\U(3)}}{2}
},
\quad
\text{CS}_5^{\U(3)}, 
\quad
c_2(\SU(2))\tilde\eta
 \end{matrix}$ \\
\hline
\hline\\[-2mm]
\multicolumn{3}{c}{$G=\Spin\times {G_{\text{SM}_6}}$}\\[2mm]
\hline
5d & $\Z^5$ & 
$\begin{matrix}\mu(\text{PD}(c_1(\U(2)))) \sim \mu(\text{PD}(c_1(\U(3)))),
\quad
\text{CS}_1^{\U(3)}c_1(\U(3))^2,
\quad
\frac{\text{CS}_1^{\U(3)}c_2(\U(2))}{2} {\sim \frac{c_1(\U(3)) \text{CS}_3^{\U(2)}}{2}},
 \\
\frac{\text{CS}_1^{\U(3)}c_2(\U(3))+\text{CS}_5^{\U(3)}}{2}
 {\sim 
\frac{ c_1(\U(3)) \text{CS}_3^{\U(3)}
+\text{CS}_5^{\U(3)}}{2}
 },  \quad
 \text{CS}_5^{\U(3)}
\end{matrix}$ \\
\hline
\hline
\end{tabular}
\caption{The 4d anomalies can be written as 5d cobordism invariants
of
$\Omega^{d=5}_{G} \equiv
\TP_{d=5}(G)$, 
which are 5d iTQFTs. 
These 5d cobordism invariants/iTQFTs are derived in \cite{WW2019fxh1910.14668}.
We summarized the group classifications of 4d anomalies and their 5d cobordism invariants.
The anomaly classification of $\Z^5$ means that there are 5 perturbative local anomalies (of $\Z$ classes descended from the 6d bordism group
$\Omega_{d=6}^{G}$),
precisely match 5 perturbative one-loop triangle Feynman diagrams in \Fig{fig:triangle-dynamical}.
The anomaly classification of $\Z_2$ means that there is a 1 nonperturbative global anomaly, which
turns out to be Witten SU(2) anomaly \cite{Witten1982fp}.
The $c_j(G)$ is the $j$th Chern class of the associated vector bundle of the principal $G$-bundle.
The $\mu$ is the 3d Rokhlin invariant.
If $\partial M^4=M^3$, then $\mu(M^3)=(\frac{\sigma-\rF\cdot\rF}{8} )(M^4)$, thus
$\mu(\text{PD}(c_1(\U(1))))$ is related to $\frac{c_1(\U(1))(\sigma-\rF \cdot \rF )}{8}$.
Here $\cdot$ is the intersection form of $M^4$.
The $\rF $ is the characteristic 2-surface \cite{Saveliev} in a 4-manifold $M^4$, it obeys the condition $\rF\cdot x=x\cdot x\mod2$ for all $x\in\H_2(M^4,\Z)$. 
By the Freedman-Kirby theorem, we have $(\frac{\sigma-\rF\cdot\rF}{8} )(M^4)=\text{Arf}(M^4,\rF)\mod2$.
The PD is defined as the Poincar\'e dual.
The Arf is a 2d Arf invariant, whose condensed matter realization is the 1+1d Kitaev fermionic chain \cite{Kitaev2001chain0010440}.
The $\tilde\eta$ is a mod 2 index of 1d Dirac operator.
In \Sec{sec:Wittenanomaly} and \Refe{WW2019fxh1910.14668},
we propose that the $\Z_2$ class 5d cobordism invariant $c_2(\SU(2))\tilde\eta$
corresponds to the 4d Witten  SU(2) anomaly \cite{Witten1982fp}.
See our notational conventions in Sec.~1 and Sec.~1.2.4 of \Refe{WW2019fxh1910.14668}.
The symbol ``$\sim$'' here means the equivalent rewriting of cobordism invariants on a closed 5-manifold.}
 \label{table:SU3SU2U1}
\end{table}

\newpage

\subsection{U(1)${}_Y^3$: 4d local anomaly from 5d $\text{CS}_1^{{\mathrm{U}}(1)}c_1({\mathrm{U}}(1))^2$ and 6d $c_1({\mathrm{U}}(1))^3$}
\label{sec:U(1)Y3}

We read from \Refe{WW2019fxh1910.14668} and Table \ref{table:SU3SU2U1} for the $\Z$ class of the 5d cobordism invariants of the following:\\
5d $\text{CS}_1^{\U(1)}c_1(\U(1))^2$ for $G_{\text{SM}_1}$,
5d $\text{CS}_1^{\U(2)}c_1(\U(2))^2$ for  $G_{\text{SM}_2}$ and  $G_{\text{SM}_6}$,
while 5d $\text{CS}_1^{\U(3)}c_1(\U(3))^2$ for  $G_{\text{SM}_3}$ and $G_{\text{SM}_6}$.
These 5d cobordism invariants correspond to the 4d perturbative local anomalies captured by the one-loop Feynman graph:
\bea \label{eq:anomaly-U1YU1YU1Y}
\includegraphics[width=2.2in]{anomaly-U1YU1YU1Y-.pdf}.
\eea
Without losing generality, we focus on the 4d cubic anomaly (U(1)$_Y$)$^3$ from 5d $\text{CS}_1^{\U(1)}c_1(\U(1))^2$, which also descends from 6d $c_1(\U(1))^3$
 of bordism group $\Omega_6$ in \Refe{WW2019fxh1910.14668}.
Plug in data in \Sec{sec:2OverviewSMGUT}, it is standard to check the anomaly \Eq{eq:anomaly-U1YU1YU1Y} vanishes,\footnote{When
 we switch between from the $L$-chiral fermion to the $R$-chiral fermion (anti-chiral fermion), there could be an additional minus 
sign.} 
\bea
\sum_\rq \Tr[(\hat Y_\rq)^3  ]
&=& \sum_{
{\rq_L},
{\rq_R}
} (Y_{\rq_L})^3 -
(Y_{\rq_R})^3\cr
&=& 
 \frac{1}{2} \delta_{ab} N_{\text{generation}} \cdot \bigg( N_c \cdot \Big( 2 \cdot (1/6)^3 + (-2/3)^3+ (1/3)^3  \Big) + 2 \cdot (-1/2)^3 +   (1)^3 +   (0)^3   \bigg)
\cr
&=&  N_{\text{generation}} \cdot ( - N_c +3) \cdot (1/4),  
\eea
which is 0 when $N_c=3$ for 3 colors as we have. The $N_{\text{generation}}$ (or  $N_{\text{family}}$) counts the number of generations (same as families).

\subsection{U(1)${}_Y$-SU(2)${}^2$: 4d local anomaly  from 5d $\text{CS}_1^{{\rm U}(1)}c_2({\rm SU}(2))$ and 6d $c_1({\rm U}(1))c_2({\rm SU}(2))$}

\label{sec:U(1)YSU(2)2}

We read from \Refe{WW2019fxh1910.14668} and Table \ref{table:SU3SU2U1} for a $\Z$ class of 5d cobordism invariants of the following:\\
5d $\text{CS}_1^{\U(1)}c_2(\SU(2))$ for $G_{\text{SM}_1}$,
5d $\frac{\text{CS}_1^{\U(2)}c_2(\U(2))}{2} {\sim}  
\frac{c_1(\U(2)) \text{CS}_3^{\U(2)}}{2}$ for  $G_{\text{SM}_2}$,\\
 5d $\text{CS}_1^{\U(3)}c_2(\SU(2)) {\sim c_1(\U(3)) \text{CS}_3^{\SU(2)}}$ for  $G_{\text{SM}_3}$, and 
5d $\frac{\text{CS}_1^{\U(3)}c_2(\U(2))}{2} {\sim \frac{c_1(\U(3)) \text{CS}_3^{\U(2)}}{2}}$ for $G_{\text{SM}_6}$.
These 5d cobordism invariants correspond to the 4d perturbative local anomalies captured by the one-loop Feynman graph:
\bea \label{eq:anomaly-U1YSU2SU2}
\includegraphics[width=2.2in]{anomaly-U1YSU2SU2-.pdf}.
\eea
Without losing generality, we focus on the 4d anomaly {U(1)$_Y$-SU(2)$^2$} from 
5d $\text{CS}_1^{\U(1)}c_2(\SU(2))$, which also descends from 6d $c_1(\U(1))c_2(\SU(2))$ 
 of bordism group $\Omega_6$ in \Refe{WW2019fxh1910.14668}.
Plug in data in \Sec{sec:2OverviewSMGUT}, we check the anomaly \Eq{eq:anomaly-U1YSU2SU2} vanishes, 
\bea 
\sum_\rq \Tr[\hat Y_\rq \sigma^a \sigma^b  ]
&=& \frac{1}{2} \delta_{ab}
(\sum_{
{\rq_L},
{\rq_R}
} (Y_{\rq_L}) -
(Y_{\rq_R}))
\cr
&=& 
 \frac{1}{2} \delta_{ab}N_{\text{generation}} \cdot \bigg( N_c \cdot \Big( 2 \cdot (1/6)   \Big) + 2 \cdot (-1/2)    \bigg)
\cr
&=& \frac{1}{2} \delta_{ab} N_{\text{generation}} \cdot ( N_c/3 -1),   
\eea
which is 0 when $N_c=3$.

\subsection{U(1)${}_Y$-SU(3)${}_c^2$: 4d local anomaly  from 5d $\text{CS}_1^{{\rm U}(1)}c_2({\rm SU}(3))$ and 6d $c_1({\rm U}(1))c_2({\rm SU}(3))$}

\label{sec:U(1)YSU(3)2}

We read from \Refe{WW2019fxh1910.14668} and Table \ref{table:SU3SU2U1} for a $\Z$ class of 5d cobordism invariants of the following:\\
5d $\text{CS}_1^{\U(1)}c_2(\SU(3)) {\sim c_1(\U(1))\text{CS}_3^{\SU(3)}}$ for $G_{\text{SM}_1}$,
5d $ \text{CS}_1^{\U(2)}c_2(\SU(3))
{\sim c_1(\U(2)) \text{CS}_3^{\SU(3)}}$ for $G_{\text{SM}_2}$,
 5d $\frac{\text{CS}_1^{\U(3)}c_2(\U(3))+\text{CS}_5^{\U(3)}}{2}
{\sim 
\frac{
 c_1(\U(3)) \text{CS}_3^{\U(3)}
+\text{CS}_5^{\U(3)}}{2}
}$ for  $G_{\text{SM}_3}$, and 
5d $\frac{\text{CS}_1^{\U(3)}c_2(\U(3))+\text{CS}_5^{\U(3)}}{2}
 {\sim 
\frac{ c_1(\U(3)) \text{CS}_3^{\U(3)}
+\text{CS}_5^{\U(3)}}{2}
 }$ for $G_{\text{SM}_6}$.
(Note that part of the additional contribution from $\text{CS}_5^{\U(3)}$ or $\text{CS}_5^{\SU(3)}$ will be separately discussed later in \Sec{sec:SU(3)3}
and \Eq{eq:anomaly-SU3SU3SU3}.)
These 5d cobordism invariants correspond to the 4d perturbative local anomalies captured by the one-loop Feynman graph:
\bea \label{eq:anomaly-U1YSU3SU3}
\includegraphics[width=2.2in]{anomaly-U1YSU3SU3-.pdf}.
\eea
Without losing generality, we focus on the 4d anomaly {U(1)$_Y$-SU(3)$_c^2$} from 
5d $\text{CS}_1^{\U(1)}c_2(\SU(3)) {\sim c_1(\U(1))\text{CS}_3^{\SU(3)}}$, which also descends from 6d $c_1(\U(1))c_2(\SU(3))$
 of bordism group $\Omega_6$ in \Refe{WW2019fxh1910.14668}.
Plug in data in \Sec{sec:2OverviewSMGUT}, we check the anomaly \Eq{eq:anomaly-U1YSU3SU3} vanishes, 
%
%
\bea  \label{eq:anomaly-U1YSU3SU3}
\sum_\rq \Tr[\hat Y_\rq \tau^a \tau^b  ]
&=& \sum_{
{\rq_L},
{\rq_R}
}  \Tr[\hat Y_{\rq_L} \tau^a \tau^b  ] -
\Tr[\hat Y_{\rq_R} \tau^a \tau^b  ]
= \frac{1}{2} \delta_{ab}
(\sum_{
{\rq_L},
{\rq_R}
} (Y_{\rq_L}) -
(Y_{\rq_R}))
\cr
&=&  \frac{1}{2} \delta_{ab} N_{\text{generation}} \cdot \bigg( N_c \cdot \Big( 2 \cdot (1/6) + (-2/3)+ (1/3)  \Big)  \bigg)
\cr
&=&  \frac{1}{2} \delta_{ab} N_{\text{generation}} \cdot N_c \cdot 0=0 . 
\eea

\subsection{U(1)$_Y$-(gravity)$^2$: 4d local anomaly from 5d $\mu(\text{PD}(c_1({\rm U}(1))))$ and 6d $\frac{c_1({\rm U}(1))(\sigma-\rF \cdot \rF )}{8}$}

\label{sec:U(1)Ygrav2}
We read from \Refe{WW2019fxh1910.14668} and Table \ref{table:SU3SU2U1} for a $\Z$ class of 5d cobordism invariants of the following:\\
5d $\mu(\text{PD}(c_1(\U(1))))$ for $G_{\text{SM}_1}$,
5d $\mu(\text{PD}(c_1(\U(2))))$ for $G_{\text{SM}_2}$ and $G_{\text{SM}_6}$,
while
 5d $\mu(\text{PD}(c_1(\U(3))))$ for  $G_{\text{SM}_3}$  and $G_{\text{SM}_6}$. 
 These 5d cobordism invariants correspond to the 4d perturbative local anomalies captured by the one-loop Feynman graph:
\bea \label{eq:anomaly-U1Ygravgrav}
\includegraphics[width=2.2in]{anomaly-U1Ygravgrav-.pdf}.
\eea
Without losing generality, we focus on the 4d anomaly U(1)$_Y$-gravity$^2$ of U(1)$_Y$-gravitational anomaly 
from 
5d $\mu(\text{PD}(c_1(\U(1))))$, 
which also descends from 6d $\frac{c_1(\U(1))(\sigma-\rF \cdot \rF )}{8}$
 of bordism group $\Omega_6$ in \Refe{WW2019fxh1910.14668}.
Plug in data in \Sec{sec:2OverviewSMGUT}, we check the anomaly \Eq{eq:anomaly-U1Ygravgrav} vanishes, 
%
%
%
\bea
\sum_\rq \Tr[\hat Y_\rq ]
&=& \sum_{
{\rq_L},
{\rq_R}
} (Y_{\rq_L})-
(Y_{\rq_R})\cr
&=& N_{\text{generation}} \cdot \bigg( N_c \cdot \Big( 2 \cdot (1/6) + (-2/3)+ (1/3)  \Big) + 2 \cdot (-1/2) +   (1) +   (0)   \bigg)
\cr
&=&  N_{\text{generation}} \cdot ( 0 \cdot N_c +0) =0.  
\eea
We remark that if we view the gravity as dynamical fields, then \Eq{eq:anomaly-U1Ygravgrav} checks the dynamical gauge anomaly cancellation of the
\Fig{fig:triangle} ({\it 1}) and Remark  \ref{remark:dynamicalanomaly};
if we view the gravity as background probe fields, then \Eq{eq:anomaly-U1Ygravgrav} checks the anomaly cancellation of the type of
\Fig{fig:triangle} ({\it 4})  and Remark  \ref{remark:4anomaly}.

\subsection{SU(3)$_c^3$: 4d local anomaly from 5d $\frac{1}{2}{\text{CS}_5^{\SU(3)}}$ and 6d $\frac{1}{2}{c_3(\SU(3))}$}

\label{sec:SU(3)3}

We read from \Refe{WW2019fxh1910.14668} and Table \ref{table:SU3SU2U1} for a $\Z$ class of 5d cobordism invariants of the following:\\
5d $\frac{1}{2}{\text{CS}_5^{\SU(3)}}$ for $G_{\text{SM}_1}$ and $G_{\text{SM}_2}$,
and 5d ${\text{CS}_5^{\U(3)}}$ for  $G_{\text{SM}_3}$ and $G_{\text{SM}_6}$.
(Note that part of the contributions from  $\text{CS}_5^{\U(3)}$ also occur in \Sec{sec:U(1)YSU(3)2}.)
These 5d cobordism invariants correspond to the 4d perturbative local anomalies captured by the one-loop Feynman graph:
\bea \label{eq:anomaly-SU3SU3SU3}
\includegraphics[width=2.2in]{anomaly-SU3SU3SU3-.pdf}.
\eea
Without losing generality, we focus on the 4d anomaly {SU(3)$_c^3$} 
from 
5d $\frac{1}{2}{\text{CS}_5^{\SU(3)}}$, 
which also descends from 6d $\frac{1}{2}{c_3(\SU(3))}$
 of bordism group $\Omega_6$ in \Refe{WW2019fxh1910.14668}.
Plug in data in \Sec{sec:2OverviewSMGUT}, we check the anomaly \Eq{eq:anomaly-SU3SU3SU3} vanishes.
%
%
In fact, in the context of SM physics,  even without checking explicitly, 
it is clear that this {SU(3)$_c^3$} anomaly \Eq{eq:anomaly-SU3SU3SU3} must vanish,
since the color {SU(3)$_c$} is vector gauge theory not chiral gauge theory respect to the color charge. 
We recall that only U(1)$_Y$ and SU(2)$_{\text{weak}}$ are chiral gauge theories in SM.

Readers may ask what happen to the {SU(2)$^3$} anomaly by replacing the gauge fields in \Eq{eq:anomaly-SU3SU3SU3} to SU(2), since
SU(2)$_{\text{weak}}$ is chiral? The answer is that {SU(2)$^3$} anomaly does not exist thus must vanish,
because there is no such corresponding 5d cobordism invariant found in \Refe{WW2019fxh1910.14668} and Table \ref{table:SU3SU2U1}. 
In fact, for SU(2) and SO($N$) group, all representations have zero 4d perturbative local anomalies, thus
they must have none of $\Z$ classes of 5d cobordism invariants, agreed with \Refe{WW2019fxh1910.14668}.

\subsection{Witten SU(2) anomaly: 4d $\Z_2$ global anomaly from 5d $c_2(\SU(2))\tilde\eta$ and 6d $c_2(\SU(2))\text{Arf}$}
\label{sec:Wittenanomaly}

{The old SU(2) anomaly of Witten in 4d \cite{Witten1982fp}}
is summarized in \cite{WangWenWitten2018qoy1810.00844}  for the context we need.
We read from \Refe{WW2019fxh1910.14668} and Table \ref{table:SU3SU2U1} for a $\Z_2$ class of 5d cobordism invariant
and suggest the 4d SU(2) anomaly corresponds to 5d $c_2(\SU(2))\tilde\eta $, and descends from 6d $c_2(\SU(2))\text{Arf}$
 of bordism group $\Omega_6$ in \Refe{WW2019fxh1910.14668}.\footnote{As explained in the Notation in the end of \Sec{sec:Intro}:
 The $\tilde{\eta}$ is a mod 2 index of 1d Dirac operator as a cobordism invariant of $\Omega_1^{\Spin}=\Z_2$.
 The Arf invariant \cite{Arf1941} is a mod 2 cobordism invariant of $\Omega_2^{\Spin}=\Z_2$,
 whose realization is the 1+1d Kitaev fermionic chain \cite{Kitaev2001chain0010440}.}
 
\begin{enumerate} 
\item
5d $c_2(\SU(2))\tilde\eta$ and 4d Witten SU(2) anomaly: This 4d anomaly is a mod 2 index of $\Z_2$ class
counts the spin-$2r +1/2$ (or ${\bf 4r} + {\bf 2}$ in the dimension of representation) 
Weyl spinor as fermion doublet under SU(2) \cite{WangWenWitten2018qoy1810.00844}.
From \Sec{sec:2OverviewSMGUT}, there are four of spin-$2r +1/2$ fermions from 
$({\bf 3},{\bf 2}, 1/6)_L$ and $({\bf 1},{\bf 2},-1/2)_L$, multiplied by $N_{\text{generation}}$. So we check overall the Witten anomaly vanishes in SM:
$$
(\text{even number})\vert_{\text{of ${\bf 2}$}} \mod 2 = 0.
$$

\item N\"aively, 5d $c_2(\SU(2))\tilde\eta$ only presents for $G_{\text{SM}_1}$ and $G_{\text{SM}_3}$,
but not for $G_{\text{SM}_2}$ and $G_{\text{SM}_6}$. Readers may wonder how does
Witten anomaly vanish for SM of $q=2,6$? \\
$\bullet$ \Refe{Davighi2020bvi2001.07731} explains nicely and accurately  that
the Witten anomaly mutates from a $\Z_2$ class  global anomaly 
into a subclass of perturbative local $\Z$ class when we changes the SU(2) group to the U(2) group, namely 
for $q=2,6$.\\
$\bullet$ \Refe{WW2019fxh1910.14668} gives a formal explanation as follows:
The difference between $q=1$ and $q=2$ case is parallel to the difference between $q=3$ and $q=6$ case.
So without losing generality, we focus on the difference between $q=1$ and $q=2$ cases.
The Witten anomaly from $c_2(\SU(2))\tilde\eta$ of $\Z_2$ for $q=1$ vanishes in $q=2$, 
while in contrast the {$\text{CS}_1^{\U(1)}c_2(\SU(2))
\sim
c_1(\U(1))\text{CS}_3^{\SU(2)}
$} of $\Z$ for $q=1$ becomes {
$\frac{1}{2}\text{CS}_1^{\U(2)}c_2(\U(2))
\sim
\frac{1}{2}c_1(\U(2))\text{CS}_3^{\U(2)}
$} 
in $q=2$. The $\text{CS}_1^{\U(1)}c_2(\SU(2))$ in 5d comes from $c_1(\U(1))c_2(\U(2))$ in 6d.
Here the $c_1(\U(2))c_2(\U(2))\mod2$ is the quotient $\Z_2$, the $c_1(\U(2))c_2(\U(2))$ is the total $\Z$, and the $\frac{1}{2}c_1(\U(2))c_2(\U(2))$ is the normal $\Z$ in the short exact sequence $0 \to \Z \overset{2}{\to} \Z \to \Z_2 \to 0$. 
Let us 
express the levels of those cobordism invariants as
$k_{q=2}$, $k_{q=1}$, and ${k}_{q=1}'$
respectively. Indeed we can also understand
the short exact sequence as the classes of the levels:
$0 \to 
k_{q=1} \in \Z \overset{2}{\to} 
k_{q=2} \in  \Z 
\to k_{q=1}'  \in  \Z_2 \to 0$.
There is yet another way to explain why $c_1(\U(2))c_2(\U(2))\mod2$ vanishes for $q=2$:
Via Wu formula, the $c_1(\U(2))c_2(\U(2))=\Sq^2c_2(\U(2))=(w_2+w_1^2)c_2(\U(2))=0\mod2$ on the Spin 6-manifolds.
In short, the $k_{q=2} \in  \Z$ of local anomalies now also carry information of the 
$k_{q=1}'  \in  \Z_2$ of the Witten SU(2) global anomaly.

\end{enumerate}

In summary of \Sec{sec:DynamicalGaugeAnomalyCancellation}, by checking
five $\Z$ classes of local anomalies and one
$\Z_2$ class of Witten SU(2) global anomaly, we have shown that the for SM with $G_{\text{SM}_q}$ of $q=1,2,3,6$
are indeed free from all {dynamical gauge anomalies, thus dynamical gauge anomaly cancellation holds.}

As we have checked, the $G_{\text{SM}_q}$ is a healthy chiral gauge theory by its own with a dynamical gauge group $G_{\text{SM}_q}$.\footnote{It has been long
sought that $G_{\text{SM}_q}$ is a healthy chiral gauge theory by its own with a dynamical gauge group $G_{\text{SM}_q}$ 
free from all dynamical gauge anomalies. However, it is only until very recently by the systematic computations of cobordism groups
in \cite{GarciaEtxebarriaMontero2018ajm1808.00009} (which checks $q=1$), \cite{Davighi2020bvi2001.07731} and \cite{WW2019fxh1910.14668} 
(which the two papers checks $q=1,2,3,6$)
completing the full checks on 
$G_{\text{SM}_q}$.
Without a cobordism classification of anomalies,
previous literature either only check perturbative anomalies,
or may still miss additional global anomaly constraints (as we shall see new anomaly constraints in \Sec{Sec:discretesymmetries}).}
However, what if we include additional global symmetries or gauge sectors?
Such as the ${ \mathbf{B}-  \mathbf{L}}$
or 
$X \equiv
5({ \mathbf{B}-  \mathbf{L}})-4Y$? This motivates us to explore further  in the next section \Sec{Sec:discretesymmetries}.

\section{Anomaly Matching for SM and GUT with Extra $({\mathbf{B}-  \mathbf{L}})$ Symmetries} 
\label{Sec:discretesymmetries}

\begin{figure}[h!] 
  \centering
      (i)  \includegraphics[width=2.0in]{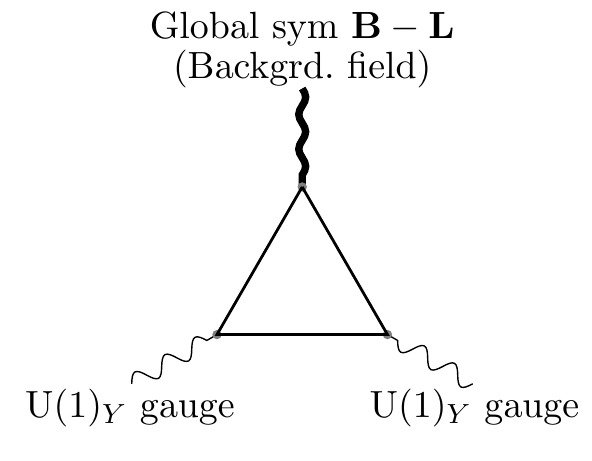}
  (ii)  \includegraphics[width=2.0in]{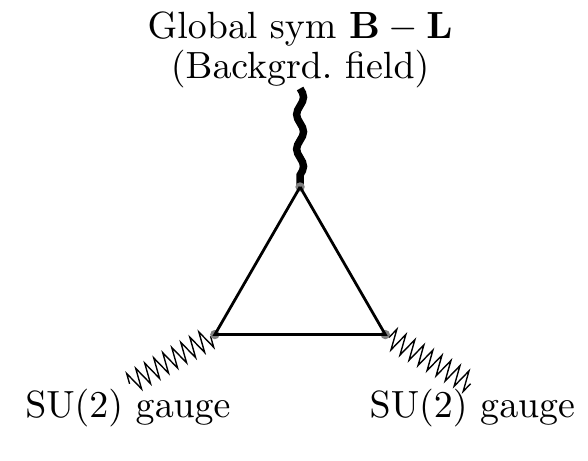}
  (iii)  \includegraphics[width=2.0in]{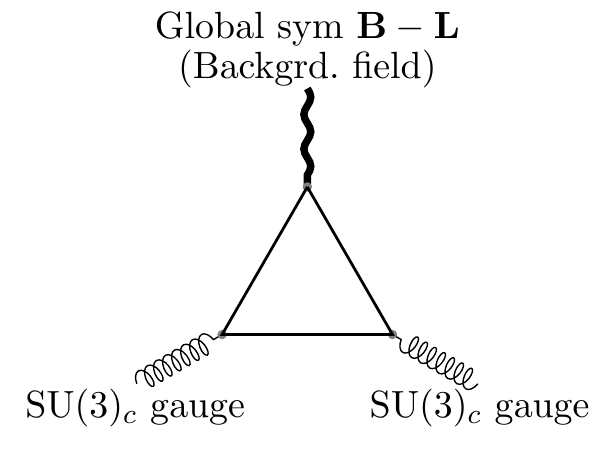}
  \\[2mm]
  (iv) \includegraphics[width=1.8in]{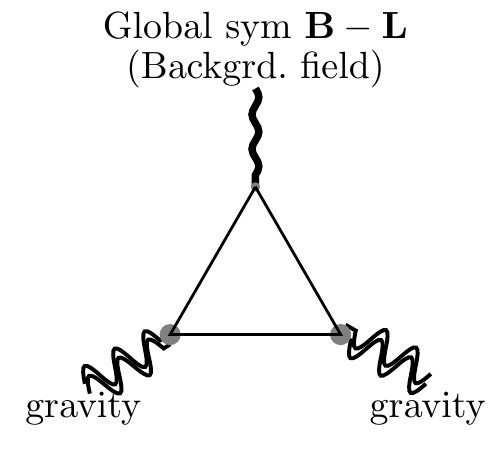} 
  (v) \includegraphics[width=1.8in]{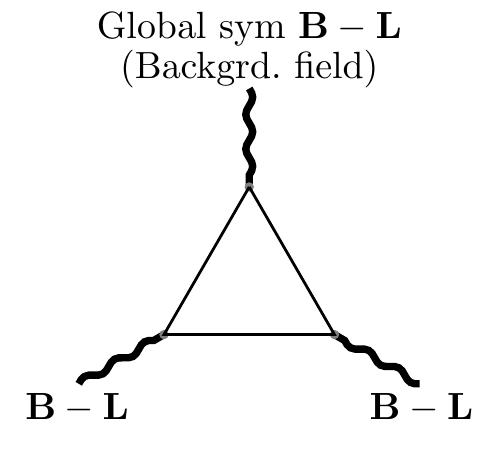}
  (vi)\includegraphics[width=2.2in]{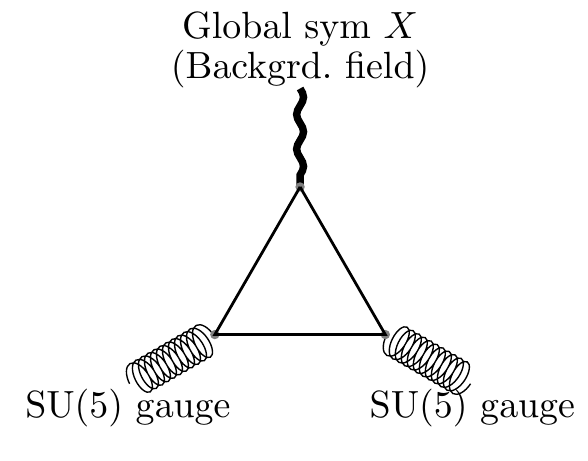} 
  \caption{Examples of anomaly constraint for SM (or GUT) with extra symmetries such as ${{ \mathbf{B}-  \mathbf{L}}}$ or $X \equiv
5({ \mathbf{B}-  \mathbf{L}})-4Y$. 
We only show
   perturbative local anomalies from
    perturbative one-loop triangle Feynman diagrams discussed in \Eq{eq:Spinccobordism}.
    We will explore nonperturbative global anomalies (not captured by  Feynman diagrams) in later sections. 
    Assume the gravity contributes as background field: \newline
   $\bullet$ If ${{ \mathbf{B}-  \mathbf{L}}}$ or $X$ is not gauged, (i), (ii), (iii), and (vi) are ABJ anomalies of \Fig{fig:triangle} ({\it 3}) and  Remark \ref{remark:ABJanomaly};\newline
   (iv) and (v) are 't Hooft anomalies of \Fig{fig:triangle} ({\it 2}) and  Remark \ref{remark:tHooftanomaly}.\newline
      $\bullet$ If ${{ \mathbf{B}-  \mathbf{L}}}$ or $X$ is gauged, (i)-(iv), (vi) are dynamical gauge anomalies of \Fig{fig:triangle} ({\it 1}) and  Remark \ref{remark:dynamicalanomaly};\newline
           (v) is an anomaly of \Fig{fig:triangle} ({\it 4}) and  Remark \ref{remark:4anomaly}.\newline
    If these anomalies are not matched, we can still saturate the anomalies by proposing new sectors appending to the QFT; 
    we will explore those new physics in \Sec{sec:HiddenTopologicalSectors}.}
    \label{fig:triangle-B-L}
\end{figure}

\subsection{SM and GUT with extra continuous symmetries and a cobordism theory}

In \Sec{sec:2OverviewSMGUT} Remark \ref{RemarkEmbed3},
for SM and SU(5) GUT with an extra $\U(1)_{{ \mathbf{B}-  \mathbf{L}}}$ or $\U(1)_{X}$ symmetry, we need to consider Spin${}^c \equiv \Spin(d) \times_{\Z_2} \U(1)$ structure.
We find the embedding: 
$$  
\Spin^c(d)\times \SU(5) 
\supset 
\Spin^c(d)  \times \frac{\SU(3) \times   \SU(2) \times \U(1)}{\Z_6}.
$$
\Refe{2019arXiv191011277D} checks that the 5d bordism group:
\bea
\Omega_5^{\Spin^c}( \frac{{\SU(3)\times \SU(2)\times \U(1)}}{\Z_q})=0,
\eea
which means no global anomalies.
\Refe{WanWangv2} computes the following cobordism groups $\TP_5$ and bordism groups $\Omega_6$:
\bea \label{eq:Spinccobordism}
\begin{array}{ll}
\TP_5({\Spin^c}\times \SU(5))=\Z^4, &  \Omega_6^{\Spin^c}(\SU(5))=\Z^4.\\[3mm]
\TP_5({\Spin^c}\times \frac{{\SU(3)\times \SU(2)\times \U(1)}}{\Z_q})=\Z^{11},& \Omega_6^{\Spin^c}( \frac{{\SU(3)\times \SU(2)\times \U(1)}}{\Z_q})=\Z^{11}.
\end{array}
\eea
\Refe{WanWangv2} finds that these $\TP_5$ and $\Omega_6$ only contain $\Z$ classes, thus they only correspond to 4d local anomalies
 captured by the one-loop Feynman graph shown in \Fig{fig:triangle-B-L}. 
 
 We emphasize that if $\U(1)_{{ \mathbf{B}-  \mathbf{L}}}$ or $\U(1)_{X}$ is free of all anomalies,
then we can dynamically gauge this symmetry. In that case, we should regard the corresponding gauge field as a Spin$^c$ connection
instead of the familiar U(1) gauge fields,
since the original theory requires to be defined on Spin$^c$ manifolds.
(When we mention gauge fields for gauging $\U(1)_{{ \mathbf{B}-  \mathbf{L}}}$ or $\U(1)_{X}$, what we really have in mind is the 
Spin$^c$ connection.)
The $\U(1)_{{ \mathbf{B}-  \mathbf{L}}}$ or $\U(1)_{X}$ is free of all anomalies if they are free from perturbative local anomalies given by \Eq{eq:Spinccobordism},
since they do not have global anomalies.

 Are all the perturbative local anomalies canceled for these {SM and GUT with extra continuous symmetries}?
 The baryon and lepton local currents densities are
\bea
j_{\bfB}^\mu= \frac{1}{3} (\bar q_L \gamma^\mu q_L + \bar u_R \gamma^\mu u_R + \bar d_R \gamma^\mu d_R), \quad  
j_{\bfL}^\mu= (\bar l_L  \gamma^\mu l_L +\bar e_R  \gamma^\mu e_R +N_{\nu_R} \bar \nu_R  \gamma^\mu \nu_R).
\eea
Here $N_{\nu_R}$ is the number of right-handed neutrinos in one generation whose representation is given in \Eq{eq:nuR-rep}:
\\
$\bullet$ In the standard GSW SM, we have $N_{\nu_R}=0$.\\
$\bullet$ In the SU(5) GUT, it is common to have $N_{\nu_R}=0$.\\
$\bullet$ In the SO(10) GUT, we have $N_{\nu_R}=1$.\\[2mm]
 We used to believe that there are no perturbative local anomalies for an additional U(1) if this U(1)
 is the $\U(1)_{{ \mathbf{B}-  \mathbf{L}}}$ or $\U(1)_{X}$. Let us check explicitly in the next subsections.
 We should pay attention on the anomaly cancellation and its dependence on $N_{\nu_R}=0$ or $1$. 
 What we will check is the conservation of the $\U(1)_{{ \mathbf{B}-  \mathbf{L}}}$ current,
 \bea
\dd \star (j_{\bfB}- j_{\bfL})= \prt_\mu (j_{\bfB}^\mu - j_{\bfL}^\mu) \dd^d x,
 \eea
 by taking into account all possible anomaly contributions from cobordism considerations.

\subsubsection{($\bf{B}-\bf{L}$)-U(1)$_Y^2$: 4d local anomaly}
Plug in data from Table \ref{table:SMfermion} to check the 4d local anomaly of $(\bfB-\bfL)$-U(1)$_Y^2$:
\bea  \label{eq:anomaly-BLU1YU1Y}
\includegraphics[width=2.2in]{anomaly-BLU1YU1Y-.pdf},
\eea
\noindent
we find the anomaly factor contributes to \footnote{The minus sign can be interpreted either from the anti-quark of the right-handed $R$-chiral fermion (anti-chiral fermion).}
\bea
j_{\bfB} &:& N_{\text{generation}} \cdot  ({N_c}/{3}) \cdot \bigg( 
 2 \cdot (1/6)^2 - (-2/3)^2 - (1/3)^2\bigg) 
= N_{\text{generation}} \cdot  ({N_c}/{3}) \cdot (1/2),\cr
j_{\bfL}&:& N_{\text{generation}} \cdot \bigg( 
 2 \cdot (-1/2)^2 + 1^2\bigg)
 = N_{\text{generation}}   \cdot (1/2), 
 \eea
 such that  $\dd \star j_{\bfB} \neq 0 $ and  $\dd \star j_{\bfL} \neq 0 $,
but $\dd \star (j_{\bfB}- j_{\bfL}) =0$ if $N_c=3$ as the color number it is.

\subsubsection{($\bf{B}-\bf{L}$)-SU(2)$^2$: 4d local anomaly}

Plug in data from Table \ref{table:SMfermion} to check the 4d local anomaly of $(\bfB-\bfL)$-SU(2)$^2$:
\bea  \label{eq:anomaly-BLSU2SU2}
\includegraphics[width=2.2in]{anomaly-BLSU2SU2-.pdf}
\eea
\noindent
we find the anomaly factor contributes to
\bea
j_{\bfB} &:& N_{\text{generation}} \cdot  ({N_c}/{3}) \cdot \bigg( 
\Tr[\frac{\sigma^a}{2} \frac{\sigma^b}{2}] \bigg) 
= N_{\text{generation}} \cdot  ({N_c}/{3}) \cdot (\delta_{ab} /2),  \cr
j_{\bfL}&:&  N_{\text{generation}} \cdot \bigg( 
\Tr[\frac{\sigma^a}{2} \frac{\sigma^b}{2}] \bigg)
 = N_{\text{generation}}   \cdot (\delta_{ab} /2) ,
 \eea
such that  $\dd \star j_{\bfB} \neq 0 $ and  $\dd \star j_{\bfL} \neq 0 $,
but $\dd \star (j_{\bfB}- j_{\bfL}) =0$ if $N_c=3$ as the color number it is.

 \subsubsection{($\bf{B}-\bf{L}$)-SU(3)$^2_c$: 4d local anomaly}

Plug in data from Table \ref{table:SMfermion} to check the 4d local anomaly of $(\bfB-\bfL)$-SU(3)$^2_c$:
\bea  \label{eq:anomaly-BLSU3SU3}
\includegraphics[width=2.2in]{anomaly-BLSU3SU3-.pdf}
\eea
\noindent
we find the anomaly factor contributes to
\bea
j_{\bfB} &:& N_{\text{generation}}   \cdot \bigg( 
(2 -1 -1) \Tr[\frac{\tau^a}{2} \frac{\tau^b}{2}] \bigg) 
= N_{\text{generation}} \cdot 0 \cdot   (\delta_{ab} /2) =  0, \cr
j_{\bfL}&:&  N_{\text{generation}} \cdot 0
 = 0, 
\eea
such that  $\dd \star j_{\bfB} =\dd \star j_{\bfL}=\dd \star (j_{\bfB}- j_{\bfL}) =0$.

  \subsubsection{($\bf{B}-\bf{L}$)-(gravity)$^2$: 4d local anomaly}
\label{sec:BLgravity2}

 Plug in data from Table \ref{table:SMfermion} to check the 4d local anomaly of $(\bfB-\bfL)$-(gravity)$^2$:
\bea  \label{eq:anomaly-BLgravgrav}
\includegraphics[width=2.2in]{anomaly-BLgravgrav-.pdf}
\eea
\noindent
we find the anomaly factor contributes to
\bea
j_{\bfB} &:& N_{\text{generation}} \cdot  ({N_c}/{3}) \cdot \bigg( 
 2  - 1 - 1\bigg) 
= 0.  \\
j_{\bfL}&:&  N_{\text{generation}} \cdot \bigg( 
 2 - 1 - N_{\nu_R}\bigg)
 = N_{\text{generation}}   \cdot (1- N_{\nu_R}) .
 \eea
 It turns out that  $\dd \star j_{\bfB} =0$ but $\dd \star j_{\bfL}\neq 0$ unless  $N_{\nu_R}=1$. Same for $\dd \star (j_{\bfB}- j_{\bfL}) =0$ only if
 $N_{\nu_R}=1$. Perturbative anomaly seems to suggest one right-handed neutrino $N_{\nu_R}=1$ to 
 saturate the ($\bf{B}-\bf{L}$) current non-conservation.
 Are there other ways to saturate this ABJ type anomaly?
When the ($\bf{B}-\bf{L}$) becomes a \emph{discrete} symmetry, 
we will be able to resolve the anomaly matching issue with other novel possibilities in \Sec{sec:HiddenTopologicalSectors}.

  \subsubsection{($\bf{B}-\bf{L}$)$^3$: 4d local anomaly}
\label{sec:BL3}

 Plug in data from Table \ref{table:SMfermion} to check the 4d local anomaly of $(\bfB-\bfL)^3$:
\bea  \label{eq:anomaly-BLBLBL}
\includegraphics[width=2.in]{anomaly-BLBLBL-.pdf}
\eea
\noindent
we find the anomaly factor contributes to
\bea
j_{\bfB} &:&  N_{\text{generation}} \cdot  {N_c} \cdot(1/{3})^3 \cdot \bigg( 
 2  - 1 - 1\bigg) 
= 0.  \\
j_{\bfL}&:&   N_{\text{generation}}  \cdot(1)^3 \cdot \bigg( 
 2 - 1- N_{\nu_R} \bigg)
 = N_{\text{generation}}   \cdot (1- N_{\nu_R}) 
 \eea
 It turns out that  $\dd \star j_{\bfB} =0$ but $\dd \star j_{\bfL}\neq 0$ unless  $N_{\nu_R}=1$. Same for $\dd \star (j_{\bfB}- j_{\bfL}) =0$ only if
 $N_{\nu_R}=1$. Perturbative anomaly seems to suggest one right-handed neutrino $N_{\nu_R}=1$ to 
 saturate the ($\bf{B}-\bf{L}$) current non-conservation.
 Are there other ways to saturate this ABJ type anomaly?
 When the ($\bf{B}-\bf{L}$) becomes a \emph{discrete} symmetry, 
 we will resolve the anomaly matching with other novel possibilities in \Sec{sec:HiddenTopologicalSectors}.

  \subsubsection{$X$-SU(5)$^2$: 4d local anomaly}

Recall in \Sec{sec:2OverviewSMGUT}, the $\U(1)_{{ \mathbf{B}-  \mathbf{L}}}$ is not a proper symmetry of SU(5) GUT.
The ``baryon minus lepton number symmetry'' of SU(5) GUT is $\U(1)_X$.
 Plug in data from Table \ref{table:SMfermion} to check 4d local anomaly of $X$-SU(5)$^2$:
\bea  \label{eq:anomaly-XSU5SU5}
\includegraphics[width=2.2in]{anomaly-XSU5SU5-.pdf}
\eea
\noindent
we find the anomaly factor contributed from the representation ${\bf{R}}$ of fermions in SU(5)
as the anti-fundamental ${\bf{R}} = \bar{\bf 5}$
and anti-symmetric ${\bf{R}} = {10}$, from the 15 Weyl fermions 
$\bar{\bf 5}\oplus {\bf  10}$ in one generation. Let us check the
$X$ current conservation or violation by ABJ type anomaly:
\bea \label{eq:XSU(5)2}
\dd \star (j_{X}) 
\propto
\sum_{\bf{R}} X_{\bf{R}} \cdot  \Tr_{\bf{R}}[ F_{\SU(5)} \wedge F_{\SU(5)}]
\propto 
\sum_{\bf{R}} X_{\bf{R}} \cdot c_2({\SU(5)}).
\eea
Here $c_2({\SU(5)})$ is the second Chern class of SU(5),
which is also related to the 4d instanton number of SU(5) gauge bundle.
For $\bar{\bf 5}\oplus {\bf  10}$ with $N_{\text{generation}}$,  from Table \ref{table:SMfermion},
we get the $\U(1)_X$ charges for 
$$X_{\bar{\bf 5}}=-3, \quad X_{\bf  10}=1,$$ 
so\footnote{To evaluate the $c_2$ or the instanton number
in different representations,  ${\bf{R}}_1$ and  ${\bf{R}}_2$, 
we use the fact that
\bea
\Tr_{{\bf{R}}_1}[F \wedge F] / \Tr_{{\bf{R}}_2}[F \wedge F]  = (C_2({{\bf{R}}_1})d({{\bf{R}}_1})) / (C_2({{\bf{R}}_2})d({{\bf{R}}_2})),
\eea
here 
 $C_2({\bf{R}})$ and $d({\bf{R}})$ are respectively the quadratic
Casimir and the dimension  of an irreducible representation ${\bf{R}}$.
For the representation ${\bf{R}}$ of SU($N$), we have 
\bea
\begin{array}{ll l}
\hline
{\bf{R}}      &      d       &       C_2 \\
\hline
\text{Fundamental}     &   N         &   N^2-1\\
\text{Antisymmetric}      &   N(N-1)/2   & 2(N+1)(N-2)\\
\hline
\end{array}.
\eea
For SU(5) with $N=5$, we get
$\Tr_{{\bf{10}}}[F \wedge F]=(N-2)\Tr_{\bar{\bf{5}}}[F \wedge F]=3\Tr_{\bar{\bf{5}}}[F \wedge F]$.
\label{ft:SUNrep}
}
\bea
\dd \star (j_{X}) 
\propto
N_{\text{generation}} \Big(
X_{\bar{\bf 5}}\Tr_{\bar{\bf{5}}}[F \wedge F]
+ X_{\bf  10}\Tr_{{\bf{10}}}[F \wedge F]
\Big)
=N_{\text{generation}} \cdot 0 =0
\eea
vanishes. We confirm that the $\U(1)_X$ symmetry is ABJ anomaly free at least perturbatively in SU(5) GUT.

\subsection{SM and GUT with extra discrete symmetries and a cobordism theory}
\label{sec:SMandGUTdiscrete}

In the subsection, we aim to digest better how robust are the anomalies from
\Sec{sec:BLgravity2} and \Sec{sec:BL3} that seems only to be matched with a right-handed neutrino (the 16th Weyl spinor) per generation. 
These anomalies are not dynamical gauge anomalies if ($\bf{B}-\bf{L}$)
and $X$ 
are only global symmetries --- namely the theory is only suffered from 't Hooft anomaly which only results in nonlocal or non-onsite ($\bf{B}-\bf{L}$) symmetry.
However, the ($\bf{B}-\bf{L}$) and $X$ have to be gauged in the SO(10) GUT. If fact, the discrete 
$\Z_{4,{X}}  \subset \U(1)_{X}$ acts as the 
$\Z_4$ center of Spin(10),
\bea
\Z_{4,{X}} = Z(\Spin(10))  \subset \Spin(10)
\eea
needs to be dynamically gauged in the SO(10) GUT. 
This fact motivates \Refe{GarciaEtxebarriaMontero2018ajm1808.00009} to use the
$\Omega_5^{\Spin \times_{\Z_2} \Z_4}=\Z_{16}$ to argue the 16 chiral fermions for the 4d GUT in one generation.
This fact also motivates \Refe{WW2019fxh1910.14668, WanWangv2} to check the following cobordism groups $\TP_5(G)$ 
with $G \supset {\Spin \times_{\Z_2^F} \Z_{4,X}}$ summarized in Table \ref{table:SU3SU2U1-discrete}.
\begin{table}[H]
\centering
\hspace*{-18mm}
\begin{tabular}{c c c }
\hline
\multicolumn{3}{c}{
 $\begin{array}{c}
\text{ Cobordism group $\TP_d(G)$ with} 
\text{ 
$G_{\text{SM}_q}\equiv{(\SU(3)\times \SU(2)\times \U(1))}/\Z_q$ 
and $q=1,2,3,6$} 
\end{array}$
}\\
\hline
\hline
$d$d & classes & cobordism invariants\\
\hline
\hline\\[-2mm]
\multicolumn{3}{c}{$G=\Spin\times_{\Z_2} \Z_4  \times {G_{\text{SM}_1}}$}\\[2mm]
\hline
5d & $\Z^5\times\Z_2\times\Z_4^2\times\Z_{16}$ & 
$\begin{matrix}\mu(\text{PD}(c_1(\U(1)))), \quad
\text{CS}_1^{\U(1)} c_1(\U(1))^2, \quad
{\text{CS}_1^{\U(1)}c_2(\SU(3)) \sim c_1(\U(1))\text{CS}_3^{\SU(2)}},\\ 
 {\text{CS}_1^{\U(1)}c_2(\SU(3)) \sim c_1(\U(1))\text{CS}_3^{\SU(3)}}, \quad
 \frac{(\CA_{{\Z_2}})^2\text{CS}_3^{\SU(3)}+\text{CS}_5^{\SU(3)}}{2},
\\
(\CA_{{\Z_2}}) c_2(\SU(3)),\quad 
c_2(\SU(2))\eta', \quad 
c_1(\U(1))^2\eta', \quad\eta(\text{PD}(\CA_{{\Z_2}}))
\end{matrix}$\\
\hline
\hline\\[-2mm]
\multicolumn{3}{c}{$G=\Spin\times_{\Z_2} \Z_4 \times {G_{\text{SM}_2}}$}\\[2mm]
\hline
5d & $\Z^5\times\Z_2^2\times\Z_4\times\Z_{16}$  &
$\begin{matrix}
\mu(\text{PD}(c_1(\U(2)))),
\quad
{\text{CS}_1^{\U(2)}c_1(\U(2))^2},
\quad
\frac{(\CA_{{\Z_2}})^2\text{CS}_3^{\U(2)}+\text{CS}_1^{\U(2)}c_2(\U(2)) }{2}\sim
\frac{(\CA_{{\Z_2}})^2\text{CS}_3^{\U(2)}+c_1(\U(2))\text{CS}_3^{\U(2)}}{2},\\
\text{CS}_1^{\U(2)}c_2(\SU(3)) \sim c_1(\U(2))\text{CS}_3^{\SU(3)},
\quad \frac{(\CA_{{\Z_2}})^2\text{CS}_3^{\SU(3)}+\text{CS}_5^{\SU(3)}}{2},
\\
(\CA_{{\Z_2}})c_2(\SU(3)), \quad (\CA_{{\Z_2}})c_2(\U(2)), \quad c_1(\U(2))^2\eta', \quad \eta(\text{PD}(\CA_{{\Z_2}}))
\end{matrix}$ \\
\hline
\hline\\[-2mm]
\multicolumn{3}{c}{$G=\Spin\times_{\Z_2} \Z_4 \times {G_{\text{SM}_3}}$}\\[2mm]
\hline
5d & $\Z^5\times\Z_2\times\Z_4^2\times\Z_{16}$ & 
$\begin{matrix}\mu(\text{PD}(c_1(\U(3)))),
\quad
{c_1(\U(3))^2\text{CS}_1^{\U(3)}, 
\quad
\text{CS}_1^{\U(3)}c_2(\SU(2)) \sim c_1(\U(3))\text{CS}_3^{\SU(2)}},
\\
\frac{(\CA_{{\Z_2}})^2\text{CS}_3^{\U(3)}+\text{CS}_1^{\U(3)} c_2(\U(3))+\text{CS}_5^{\U(3)}}{2}
\sim
\frac{(\CA_{{\Z_2}})^2\text{CS}_3^{\U(3)}+c_1(\U(3))\text{CS}_3^{\U(3)}+\text{CS}_5^{\U(3)}}{2},
\quad
\text{CS}_5^{\U(3)},
\\
(\CA_{{\Z_2}})c_2(\U(3)), 
\quad
c_2(\SU(2))\eta',
\quad
c_1(\U(3))^2\eta',
\quad
\eta(\text{PD}(\CA_{{\Z_2}}))
\end{matrix}$\\
\hline
\hline\\[-2mm]
\multicolumn{3}{c}{$G=\Spin\times_{\Z_2} \Z_4 \times {G_{\text{SM}_6}}$}\\[2mm]
\hline
5d  & $\Z^5\times\Z_2^2\times\Z_4\times\Z_{16}$  &
$\begin{matrix}\mu(\text{PD}(c_1(\U(3)))),
\quad {c_1(\U(3))^2\text{CS}_1^{\U(3)},
\quad
\frac{(\CA_{{\Z_2}})^2\text{CS}_3^{\U(2)}+\text{CS}_1^{\U(3)}c_2(\U(2))}{2}
\sim 
\frac{(\CA_{{\Z_2}})^2\text{CS}_3^{\U(2)}+c_1(\U(3))\text{CS}_3^{\U(2)}}{2}},
\\
\frac{(\CA_{{\Z_2}})^2\text{CS}_3^{\U(3)}+\text{CS}_1^{\U(3)} c_2(\U(3))+\text{CS}_5^{\U(3)}}{2}
{\sim 
\frac{(\CA_{{\Z_2}})^2\text{CS}_3^{\U(3)}+c_1(\U(3))\text{CS}_3^{\U(3)}+\text{CS}_5^{\U(3)}}{2}},
\quad
\text{CS}_5^{\U(3)},\\
(\CA_{{\Z_2}})c_2(\U(3)),
\quad
(\CA_{{\Z_2}})c_2(\U(2)),
\quad
c_1(\U(3))^2\eta',
\quad
\eta(\text{PD}(\CA_{{\Z_2}}))
\end{matrix}$ \\
\hline
\hline\\[-2mm]
\multicolumn{3}{c}{$G=\Spin\times_{\Z_2} \Z_4 \times \SU(5)$}\\[2mm]
\hline
5d & $\Z\times\Z_2\times\Z_{16}$ & 
$\frac{(\CA_{{\Z_2}})^2\text{CS}_3^{\SU(3)}+\text{CS}_5^{\SU(3)}}{2}$, \quad
$(\CA_{{\Z_2}}) c_2(\SU(5))$,\quad
$\eta(\text{PD}(\CA_{{\Z_2}}))$\\
\hline
\hline
\end{tabular}
\caption{Our setup follows Table \ref{table:SU3SU2U1} and \Refe{WW2019fxh1910.14668, WanWangv2}.
The  $\CA_{{\Z_2}} \in \H^1(M,\Z_2)$ is the generator from  $\H^1(\B(\Z_4/\Z_2^F),\Z_2)$ of ${\Spin \times_{\Z_2^F} \Z_4}$.
The 
$\eta'$ is a $\Z_4$ valued 1d eta invariant which is the extension of a quotient $\CA_{{\Z_2}}$ by the normal 1d $\tilde\eta$.
So $(\CA_{{\Z_2}})\equiv (\CA_{{\Z_4}}) \mod 2$ is the quotient, while $\Z_{4,{X}}  \subset \U(1)_{X}$.
The $\eta(\text{PD}(\CA_{{\Z_2}}))$ is
the value of ${\eta} \in \Z_{16}$ on the Poincar\'e dual (PD) submanifold of $\CA_{{\Z_2}}$.}
 \label{table:SU3SU2U1-discrete}
\end{table}

{We aim to initiate a new approach on matching the nonperturbative global $\Z_{16}$ anomaly
 (descended from the perturbative local $\Z$ anomalies of
\Sec{sec:BLgravity2} and \Sec{sec:BL3}) for the missing neutrinos.}

The $\Z^5$ classes perturbative local anomalies are the same results parallel to \Sec{sec:localanomaliesFeynman} and Table \ref{table:SU3SU2U1}.
So in the following subsections, we only focus on checking global anomaly cancellations of $\Z_n$ classes.
Generically Feynman diagrams cannot characterize global anomalies (so we do not present Feynman diagrams below). 
But we can characterize global anomalies by generic curved manifolds as in \Fig{fig:Cobordism2006}
with gauge, gravity, or mixed gauge-gravity background fields.

\subsubsection{Witten anomaly $c_2(\SU(2))\tilde\eta$ vs $c_2(\SU(2))\eta'$: 4d $\Z_{2}$ vs $\Z_{4}$ global anomalies}

Follow \Sec{sec:Wittenanomaly}, {the old SU(2) anomaly of Witten in 4d} is a mod 2 class. It is a $\Z_{2}$ global anomaly given 
 by 5d $c_2(\SU(2))\tilde\eta $ and 6d $c_2(\SU(2))\text{Arf}$  in \cite{WW2019fxh1910.14668} and  Table \ref{table:SU3SU2U1-discrete}.
The $c_2(\SU(2))\tilde\eta$ as Witten SU(2) anomaly 
can be contributed by the 4d fermion doublet {\bf 2} under the SU(2) representation (or 
the (iso-)spin-$2r +1/2$ of SU(2) Weyl spinor for some non-negative integer $r \in \Z_{\geq 0}$).
So to check the anomaly cancellation,
we count the fermion doublet {\bf 2} under SU(2).
There are four of {\bf 2} of SU(2) fermions from 
$({\bf 3},{\bf 2}, \tilde{Y}=1)_L$ and $({\bf 1},{\bf 1},\tilde{Y}=6)_L$.
The anomaly from $c_2(\SU(2))\tilde \eta$ counts the number of 4d Weyl spinors 
of SU(2) fundamental ${\bf 2}$ mod 2.
From \eq{eq:Weyl-rep}
for $N_{\text{generation}}$, we have:
\bea
N_{\text{generation}} \cdot (3 + 1)  = 0 \mod 2.
\eea
There is also an extended $\Z_{4}$ global anomaly from the 5d cobordism invariant 
$c_2(\SU(2))\eta'$  in \cite{WW2019fxh1910.14668} and  Table \ref{table:SU3SU2U1-discrete} counting the number of 4d Weyl spinors 
of SU(2) fundamental ${\bf 2}$ mod 4.\footnote{As explained in the Notation in the end of \Sec{sec:Intro}:
The ${\eta}'$ is a mod 4 index of 1d Dirac operator as a cobordism invariant of $\Omega_1^{\Spin\times \Z_4}=\Z_4$.}
From \eq{eq:Weyl-rep}
for $N_{\text{generation}}$, we have:
\bea
N_{\text{generation}} \cdot (3 + 1)  = 0 \mod 4.
\eea
Therefore, we have checked no global SU(2) anomaly of $\Z_{2}$ or $\Z_{4}$ classes for SM with a Spin or ${\Spin \times_{\Z_2^F} \Z_4}$ structure.

\subsubsection{$(\CA_{{\Z_2}}) c_2(\SU(2))$: 4d $\Z_{2}$ global anomaly}

\noindent
The 4d $\Z_{2}$ global  anomaly from the 5d cobordism invariant  $(\CA_{{\Z_2}}) c_2(\SU(2))$, 
in  \cite{WW2019fxh1910.14668} and Table \ref{table:SU3SU2U1-discrete},
counts the number of 4d left-handed Weyl spinors of SU(2) fundamental ${\bf 2}$ mod 2.
Here  $(\CA_{{\Z_2}})\equiv (\CA_{{\Z_4}}) \mod 2$, where
 $\CA_{{\Z_2}} \in \H^1(M,\Z_2)$ is the generator from  $\H^1(\B(\Z_4/\Z_2^F),\Z_2)$ of ${\Spin \times_{\Z_2^F} \Z_4}$.
So $(\CA_{{\Z_2}})\equiv (\CA_{{\Z_4}}) \mod 2$ is the quotient, while $\Z_{4,{X}}  \subset \U(1)_{X}$.
The $(\CA_{{\Z_2}}) c_2(\SU(2))$ counts the number of 4d Weyl spinors 
of SU(2) fundamental ${\bf 2}$ mod 2.
From \eq{eq:Weyl-rep}
for $N_{\text{generation}}$, we have:
\bea
N_{\text{generation}} \cdot (3 + 1)  = 0 \mod 2.
\eea
Thus the anomaly vanishes. 
We have no obstruction to gauge the $\Z_4$ by making
$\CA_{{\Z_4}}$ dynamical at least from this anomaly cancellation.

\subsubsection{$(\CA_{{\Z_2}}) c_2(\SU(3))$: 4d $\Z_{2}$ global anomaly}

\noindent
The 4d $\Z_{2}$ global anomaly from the 5d cobordism invariant $(\CA_{{\Z_2}}) c_2(\SU(3))$, 
 in \cite{WW2019fxh1910.14668} and Table \ref{table:SU3SU2U1-discrete},
 counts the number of 4d left-handed Weyl spinors of SU(3) fundamental ${\bf 3}$ mod 2.
%
%
From \eq{eq:Weyl-rep},
we count $({\bf 3},{\bf 2}, \tilde{Y}=1)_L$,
$(\overline{\bf 3},{\bf 1}, \tilde{Y}=-4)_L$, and 
$(\overline{\bf 3},{\bf 1},\tilde{Y}=2)_L$ 
with {3 generations}.
For $N_{\text{generation}}$, we have:
\bea
N_{\text{generation}} \cdot (2 - 1 -1) = 0 \mod 4.
\eea
Thus there is no anomaly. We have no obstruction to gauge the $\Z_4$ by making
$\CA_{{\Z_4}}$ dynamical at least from this anomaly cancellation.

\subsubsection{$c_1(\U(1))^2\eta'$: 4d $\Z_{4}$ global anomaly}

The 4d $\Z_{4}$ global anomaly from a 5d cobordism invariant $c_1(\U(1))^2\eta'$  (in \Refe{WW2019fxh1910.14668} and  Table \ref{table:SU3SU2U1-discrete})
counts the sum of 4d Weyl spinors' (U(1) charge)$^2$ mod 4.
Let us apply $\tilde{Y}$ for U(1)$_{\tilde{Y}}$ charge
from \eq{eq:Weyl-rep}
with 
$N_{\text{generation}}$.
For each generation, we not only have the sum of U(1) charge:
\bea
3 \cdot 2  \cdot 1+ 3 \cdot 1  \cdot (-4) + 3 \cdot 1 \cdot 2 + 1 \cdot 2 \cdot  (-3)  + 1 \cdot 1 \cdot 6 =0  \mod 4.
\eea
but also have the sum of
(U(1) charge)$^2$:
\bea
3 \cdot 2 \cdot 1^2+ 3 \cdot 1  \cdot (-4)^2 + 3 \cdot 1 \cdot 2^2 + 1 \cdot 2 \cdot  (-3)^2  + 1 \cdot 2 \cdot (6)^2
=  3 \cdot 4 \cdot 13 =0 \mod 4.
\eea
The $c_1(\U(1))^2$ also counts the U(1) instanton number up to a proportional factor.
Thus there is no anomaly. We have no obstruction to gauge the $\Z_4$ by making the
$\CA_{{\Z_4}}$ dynamical at least from this anomaly cancellation.

\subsubsection{$(\CA_{{\Z_2}}) c_2(\SU(5))$: 4d $\Z_{2}$ global anomaly}

Similar to \Eq{eq:XSU(5)2}, we consider the discrete 4d $\Z_2$ global anomaly from the 
5d cobordism invariant $(\CA_{{\Z_2}}) c_2(\SU(5))$ in \cite{WW2019fxh1910.14668, WanWangv2} and Table \ref{table:SU3SU2U1-discrete}. Here $c_2({\SU(5)})$ is the second Chern class of SU(5),
which is also related to the 4d instanton number of SU(5) gauge bundle.
For $\bar{\bf 5}\oplus {\bf  10}$ with $N_{\text{generation}}$,  from Table \ref{table:SMfermion},
we get the $\Z_{4,X}$ charges for 
$$X_{\bar{\bf 5}}=-3 = 1 \mod 4, \quad X_{\bf  10}=1  \mod 4.$$ 
By footnote \ref{ft:SUNrep}, we compute the anomaly factor
\bea
&&N_{\text{generation}} \Big(
(X_{\bar{\bf 5}} \mod 4) \Tr_{\bar{\bf{5}}}[F \wedge F]
+ (X_{ {\bf 10}} \mod 4)\Tr_{{\bf{10}}}[F \wedge F]
\Big) \cr
&&=N_{\text{generation}} \cdot \Big( 1 \cdot 1 + 1 \cdot 3  \Big) =N_{\text{generation}} \cdot 4 =0 \mod 4.
\eea
This certainly vanishes for the mod 2 anomaly for SU(5) GUT.

There is also another $\Z$ class local anomalies for SU(5) GUT captured by 5d cobordism invariants
$\frac{(\CA_{{\Z_2}})^2\text{CS}_3^{\SU(3)}+\text{CS}_5^{\SU(3)}}{2}$, we can easily check that SU(5) GUT is free from 
any local anomaly given by another 5d cobordism invariant \cite{JW2008.06499}.

\subsubsection{$\eta(\text{PD}(\CA_{{\Z_2}}))$: 4d $\Z_{16}$ global anomaly}
 \label{sec:etaPDAZ16}
 
The 4d $\Z_{16}$ global anomaly given by a 5d cobordism invariant $\eta(\text{PD}(\CA_{{\Z_2}}))$,  in  \cite{WW2019fxh1910.14668} and Table \ref{table:SU3SU2U1-discrete},
counts the number mod 16 of 4d left-handed Weyl spinors 
($\Psi_L \sim {\bf 2}_L \text{ of } \Spin(3,1)$ {or}  $\Psi_L \sim  {\bf 2}_L \text{ of } \Spin(4) = \SU(2)_L \times  \SU(2)_R$).
From \eq{eq:Weyl-rep} with $N_{\text{generation}}$ (e.g., {3 generations}), 
for each generation, we have:
\bea
3 \cdot 2 + 3 \cdot 1 + 3 \cdot 1 + 1 \cdot 2   + 1 \cdot 1  = 15 = -1 \mod 16.
\eea 
For 1 generation, we need to saturates the anomaly: 
\bea \label{eq:nu-1}
\upnu =-1 \mod 16.
\eea
For 3 generations, we need
\bea
3 \Bigg( 3 \cdot 2 + 3 \cdot 1 + 3 \cdot 1 + 1 \cdot 2   + 1 \cdot 1  \Bigg)= 45 = -3 \mod 16.
\eea 
Therefore we need to saturates the anomaly: 
\bea
\upnu =-3 \mod 16.
\eea
For $N_{\text{generation}}$ generations, we need to saturates the anomaly: 
\bea \label{eq:Ngeneration}
\upnu =-N_{\text{generation}} \mod 16.
\eea
This anomaly can be canceled by adding new degrees of freedom
\bea \label{eq:NgenerationNnuR}
\upnu =N_{\text{generation}}\cdot (N_{\nu_R}=1) \mod 16.
\eea
The anomaly matching in this \Sec{sec:etaPDAZ16} seems to be matched with a right-handed neutrino (the 16th Weyl spinor) per generation, similar to
\Sec{sec:BLgravity2} and \Sec{sec:BL3}.
This also shows the robustness of \Sec{sec:BLgravity2} and \Sec{sec:BL3} even if we break down $\U(1)_{\bf{B}-\bf{L}}$ or $\U(1)_X$
down to $\Z_{4,\bf{B}-\bf{L}}$ or to $\Z_{4,X}$. Again this $\Z_4$ as the center  $Z(\Spin(10))$ of Spin(10) is important for the SO(10) GUT.

Are there other ways to match the anomaly other than introducing the right-handed neutrino (the 16th Weyl spinor) per generation? 
Let us explore the alternatives in the next subsection.

\subsection{How to match the anomaly? Preserving or breaking the $\Z_{4,X}$ symmetry?}
\label{sec:matchtheanomaly}

Let us summarize what we learn from the anomaly computation and matching in the previous sections.
We have shown that all anomalies presented in Table \ref{table:SU3SU2U1} and Table \ref{table:SU3SU2U1-discrete} can be cancelled,
except the additional anomaly constraint from \Sec{sec:BLgravity2}, \Sec{sec:BL3}, and \Sec{sec:etaPDAZ16} may not be matched unless we obey
\bea\label{eq:localglobal16}
\begin{array}{lll}
N_{\text{generation}}   \cdot (1- N_{\nu_R} + \text{ hidden sector }) &=& 0, \quad\quad\quad\quad\;\text{ from local anomalies of \Sec{sec:BLgravity2} and \ref{sec:BL3}}.\quad\\
N_{\text{generation}}   \cdot (1- N_{\nu_R} + \text{ hidden sector }) &=& 0 \mod 16, \text{ from a global anomaly of \Sec{sec:etaPDAZ16}.}
\end{array}
\eea
So the above n\"aively suggests we need the right-handed neutrino (the 16th Weyl spinor) $N_{\nu_R}=1$  per generation. 
However, by demanding only a discrete $\Z_{4,X}$ instead of a continuous $\U(1)_{X}$ symmetry,
the local $\Z$ class anomaly becomes a global $\Z_{16}$ class anomaly. If so,
we do have different ways to match the anomaly, other than introducing the right-handed neutrino (the 16th Weyl spinor) $N_{\nu_R}=1$.
Can we match the anomaly by {\bf additional new
hidden sectors} not yet discovered in SM or in SU(5) Georgi-Glashow GUT?
Let us focus on the $\Z_{4,X}$ symmetry 
for the sake of thinking $\Z_{4,X}=Z(\Spin(10))$ is gauged in the SO(10) GUT eventually.
Let us enumerate the possibilities to match the anomaly \Eq{eq:localglobal16}:
\begin{enumerate} [label=\textcolor{blue}{\arabic*}., ref={\arabic*}]
\item  {\bf Anomaly matched by a right-handed neutrino (the 16th Weyl spinor)} $N_{\nu_R}=1$:\\
For the right-handed neutrino $\nu_R$ to be massless (or gapless) while preserving the 
$\Z_{4,X}$ symmetry, we need to have $\nu_R$ to be a complex Weyl spinor (with a $\Z_{4,X}$ charge $-1 \mod 4$), instead of a real Majorana spinor, in order to have the 
$\Z_{4,X}$ symmetry transformation manifest:
\bea  \label{eq:nuRsym}
\nu_R \to \exp(-2 \pi \ii/4) \; \nu_R = (-\ii) \; \nu_R.
\eea
Since this is a sterile neutrino with a trivial representation of SM gauge group \Eq{eq:nuR-rep}, $({\bf 1},{\bf 1},0)_R$, we can rotate it to left-handed Weyl spinor
$\bar{\nu}_R= {\nu}_L$ with $({\bf 1},{\bf 1},0)_L$ and flips the $\Z_{4,X}$ representation to its complex conjugation:
\bea \label{eq:nuLsym}
\nu_L \to \exp(2 \pi \ii/4)\;\nu_L = (\ii) \;\nu_L.
\eea
What can the low energy dynamics of the ${\nu}_R$ be?
\begin{enumerate} [label=\textcolor{blue}{(\arabic*)}., ref={(\arabic*)}]
\item  {\bf  The $\Z_{4,X}$ preserving massless neutrino}: If the sterile neutrino ${\nu}_R$ remains massless, we can match the anomaly \Eq{eq:localglobal16}
by a symmetric gapless low energy theory with an action on a 4d spacetime $M^4$:
\bea
 \int_{M^4} \bar\nu_R (\ii \sigma^\mu \prt_\mu)\nu_R, \quad \text{ or equivalently }
 \quad
  \int_{M^4} \bar\nu_L (\ii \bar \sigma^\mu \prt_\mu)\nu_L,
\eea
with $\bar\sigma \equiv (1, \vec{\sigma})$. Note that $\nu_L$ has a $\Z_{4,X}$ charge 1 and 
$\bar\nu_L$ has a $\Z_{4,X}$ charge $-1$, so the action preserves the $\Z_{4,X}$ symmetry.
\item  {\bf  The $\Z_{4,X}$ preserving Dirac mass term, but the $\Z_{4,X}$ is spontaneously broken by the electroweak Higgs}: The Yukawa-Higgs-Dirac term preserves the  $\Z_{4,X}$ symmetry:
\bea
  \int_{M^4} \bar\nu_R  \phi_H  \nu_L + \bar\nu_L  \phi_H  \nu_R,
\eea
because $\nu_L$ (or $\bar\nu_R$) has a $\Z_{4,X}$ charge 1,  
$\nu_R$ (or $\bar\nu_L$) has a $\Z_{4,X}$ charge $-1$, and $\phi_H$ has a $\Z_{4,X}$ charge 2.
However, when the electroweak symmetry breaking sets in (at a lower energy around 246 G$e$V), 
the Higgs condenses, $\langle \phi_H \rangle \neq 0$, and spontaneously breaks the $\Z_{4,X}$ symmetry.

\item  {\bf  The $\Z_{4,X}$ explicit breaking Majorana mass term}:
If the sterile neutrino ${\nu}_R$ becomes gapped by {Majorana mass} $m_{\text{Maj}}$, the spacetime spinor is in a real Majorana representation.
Then, the  $\Z_{4,X}$ symmetry in \Eq{eq:nuRsym} and \Eq{eq:nuLsym} is explicitly broken.
We can still match the anomaly \Eq{eq:localglobal16}
by a $\Z_{4,X}$-symmetry-breaking gapped theory with an action on a 4d spacetime $M^4$:
\bea
 \int_{M^4=\prt M^5} \chi^{\rm T} (\ii \bar\sigma^\mu \nabla_\mu)\chi 
+\frac{\ii m_{\text{Maj}} }{2}( \chi^{\rm T} \sigma^2 \chi+ \chi^\dagger  \sigma^2  \chi^*  ),
\eea
where we have written the 4-component Majorana spinor as $\Psi_{\text{Maj}}=
\begin{pmatrix}
\chi\\
\ii \sigma^2 \chi^*
\end{pmatrix}$ with the transpose T,  complex conjugate $*$,  and complex conjugate transpose $\dagger$.
\end{enumerate} 
See more discussions on Dirac or Majorana masses for three generations of SM fermions in \Sec{sec:MajoranaMassorDiracMass}.

\item {\bf Anomaly matched by  new  additional or 
hidden sectors beyond SM}: Let us hypothesize many scenarios with different low energy dynamics following
\Refe{WW2019fxh1910.14668}'s Sec.~8.2:
\begin{enumerate} [label=\textcolor{blue}{(\arabic*)}., ref={(\arabic*)}]
\item \label{Z4XCFT}
$\Z_{4,X}$-symmetry-preserving anomalous gapless, free or interacting, 4d CFT.
\item \label{Z4XbreakCFT}
$\Z_{4,X}$-symmetry-breaking gapless, free or interacting, 4d CFT.
\item \label{Z4XTQFT}
$\Z_{4,X}$-symmetry-preserving anomalous \emph{long-range entangled} gapped 4d TQFT.
\item \label{Z4XbreakTQFT}
$\Z_{4,X}$-symmetry-breaking gapped 4d TQFT.
\item \label{Z45dSPT}
$\Z_{4,X}$-Symmetry-Protected Topological state (SPTs) in 5d
as a \emph{short-range entangled} gapped phase
 captured by a 5d cobordism invariant 
\bea
\exp\bigg(\frac{2\pi \ii}{16} \cdot(-N_{\text{generation}}) \cdot  \eta\big(\text{PD}(\CA_{{\Z_2}})  \big) \bigg).
\eea
\item \label{Z45dSET}
$\Z_{4,X}$-gauged-(Symmetry)-Enriched Topological state (SETs) in 5d coupled to gravity.
\end{enumerate} 
\end{enumerate} 
All the above theories are unitary by their own.
We propose that all the above scenarios and their proper \emph{linear combinations}, 
conventional or exotic, if existing, can saturate the anomaly \Eq{eq:localglobal16}.
Based on the contemporary knowledge of SM physics and experimental hints,\\[2mm]
$\bullet$ Scenario \ref{Z4XCFT} and \ref{Z4XbreakCFT} seem less practical, because
it is less likely to have any new gapless or interacting CFT that we do not observe below the T$e$V energy scale.\footnote{However,
if the hidden new gapless sectors are fully decoupled from SM forces, 
they may be weakly interacting massless or gapless matter 
account for the Dark Matter sector.} 
Also if there is spontaneous symmetry-breaking (SSB), 
for $\U(1)_{X}$ SSB, we expect to observe new Goldstone boson modes; for $\Z_{4,X}$ SSB, we may observe different vacua or domain walls between different vacua.
This shall be falsifiable in the experiments. \\[2mm]
$\bullet$
Scenario \ref{Z4XTQFT} is exotic but very interesting, which we discover new insights into the neutrino physics and Dark Matter.
Since the $\Z_{4,X}$-symmetry is a discrete chiral symmetry,
and the gapped 4d TQFT can be regarded as a gapped phase.
The anomalous $\Z_{4,X}$-symmetry preserving gapped 4d TQFT can be regarded 
as \emph{a gapped phase without $\Z_{4,X}$-chiral symmetry breaking}.
However, the 4d TQFT has additional generalized global symmetries \cite{Gaiotto2014kfa1412.5148}, 
known as the higher-symmetries, whose charged objects are extended objects (1d lines, or 2d surfaces, etc.).
\begin{itemize}[leftmargin=1.8mm]
\item[-] The higher-symmetries of the 4d TQFT can be spontaneously broken. In this sense, 
the spontaneously symmetry breaking (SSB) of higher-symmetries imply that the 4d TQFT is \emph{deconfined}. 
Below the topological order energy gap, we have the low energy 4d TQFT.
Above the topological order energy gap, 
the 4d TQFT hosts \emph{deconfined} excitations of fractionalized particles (from breaking 1d worldline with open 0d ends) or
fractionalized strings (from breaking 2d worldsheet with open 1d boundaries).

\item[-] If the higher-symmetries of the 4d TQFT have no SSB, then this leads to a \emph{gapped confined phase without $\Z_{4,X}$-chiral symmetry breaking}.
However, eventually in a quantum gravity theory,
all the global symmetries must be either \emph{gauged} or \emph{broken}.
Thus we should not have the option of SSB or symmetry-preserving.
Preferably,  in a quantum gravity theory, 
we can have either of the following options:\\ 
(1) the higher-symmetries of TQFT are dynamically \emph{gauged} due to a higher energy theory (above GUT),\\
(2) the higher-symmetries of TQFT are \emph{explicitly} broken.
\end{itemize} 
Moreover,  Scenario \ref{Z4XTQFT} can give rise to  Scenario \ref{Z4XbreakTQFT}, if we construct such an anomalous $\Z_{4,X}$-symmetry-preserving TQFT first,
we can break some of the symmetry to obtain the symmetry-breaking gapped TQFT.
We can also construct a mixed symmetry-extension and symmetry-breaking TQFT following \Refe{Wang2017locWWW1705.06728}.
So we should focus on Scenario \ref{Z4XTQFT} explored in \Sec{sec:Hidden4dnon-invertiblenon-abelianTQFT}.  \\[2mm]
$\bullet$
Scenario \ref{Z45dSPT} implies that
our 4d SM lives on the boundary of 
5d $\Z_{4,X}$-SPTs given by 5d $\eta(\text{PD}(\CA_{{\Z_2}}))$. 
If Scenario \ref{Z45dSPT}  describes our universe, we discover at least an extra dimension from the 5d theory.
We explore this 5d SPT or invertible TQFT theory in \Sec{sec:5dSPT}.  \\[2mm] 
$\bullet$
Scenario \ref{Z45dSET}
Above certain higher energy scale, the $\Z_{4,X}$ may be dynamically gauged such as in $\Z_{4,X}=Z(\Spin(10))$ of the SO(10) GUT, then
the 4d and 5d bulk are fully gauged and entangled together. The 5d bulk is in fact 
a 5d Symmetry-Enriched Topologically ordered state (SETs) that can be coupled to dynamical gravity.
We explore this 5d SETs coupled to dynamical gravity theory in \Sec{sec:gravity}.


\section{Beyond Known ``Fundamental'' Forces: Hidden Topological Force}
\label{sec:HiddenTopologicalSectors}

Follow \Sec{sec:matchtheanomaly}, now we propose new Scenarios, beyond SM and beyond Georgi-Glashow (GG) SU(5) GUT,
to match the $\Z_{16}$ global anomalies
 from \Sec{sec:etaPDAZ16}. 
We focus on the 
Scenario \ref{Z4XTQFT} (thus also Scenario \ref{Z4XbreakTQFT})  
for a hidden 4d non-invertible TQFT
in \Sec{sec:Hidden4dnon-invertiblenon-abelianTQFT},
and 
for a 5d SPTs or 5d SETs coupled to gravity in Scenario \ref{Z45dSPT} and \ref{Z45dSET} in \Sec{sec:gravity}.

\subsection{Hidden 5d invertible TQFT or 5d Symmetry-Protected Topological state (SPTs)}
\label{sec:5dSPT}

We can saturate the missing anomaly of $\upnu =- N_{\text{generation}}\mod 16$
in
\Eq{eq:Ngeneration} by a 5d invertible TQFT (or a 5d SPTs protected by 
$G=\Spin\times_{\Z_2} \Z_4 \times {G_{\text{internal}}}$-symmetry
in a condensed matter language) with a 5d partition function:
\bea
{\bf Z}_{\text{5d-iTQFT}}=\exp\bigg(\frac{2\pi \ii}{16} \cdot(-N_{\text{generation}}) \cdot  \eta\big(\text{PD}(\CA_{{\Z_2}})  \big) \bigg\rvert_{M^5}\bigg).
\eea
Here we define the eta invariant 
$$\eta \in \Z_{16}, \quad \text{ and }\quad   \eta\big(\text{PD}(\CA_{{\Z_2}})  \big) \in \Z_{16}$$
slightly different by a proportional factor in the math literature.
Let us overview quickly what we need about the 
{APS eta invariant $\upeta$} \cite{Atiyah1975jfAPS, Atiyah1976APS, Atiyah1980APS, gilkey1985eta, Witten:2015aba}.
Let us consider the 4d $\upeta$ invariant and then relate to our 4d $\eta$\footnote{The 
4d $\eta$ invariant has condensed matter realizations. It is known as the  
class DIII topological superconductor [TSC] in the free fermion limit with a $\Z$ classification  \cite{RyuSPT, Kitaevperiod, Wen1111.6341}. 
The $\upnu = 1$ of this TSC is the Balian-Werthamer (BW)  state of the B phase of
 ${}^3$He liquid (${}^3$He B) \cite{Volovik2003Book}. A  2+1d single Majorana fermion can live on the
surface protected by time reversal symmetry in the non-interacting free system.
Including the time-reversal $\Z_4^T$ symmetry preserving interactions (preserving the $\Pin^+$ structure), 
the 4d $\eta$ invariant is reduced from a free $\Z$ class to a torsion $\Z_{16}$ class.} 
invariant and 5d $\eta\big(\text{PD}(\CA_{{\Z_2}})  \big)$ invariant.

On an Euclidean signature curved spacetime manifold $M^d$, the path integral of a massive 
5d Dirac fermion spinor $\psi$ with a mass $m$
coupled to a background gauge field $A$ or a background gravity of a metric $g_{\mu \nu}$, the Euclidean path integral is
\begin{equation}
\int [{\cal D} {\psi}] [{\cal D}\bar{\psi}] \exp\left(-S_E \right)=\int [{\cal D} {\psi}] [{\cal D}\bar{\psi}] \exp\bigg(
-\int_{M^d} \dd^4x_E\, \sqrt{ \det g}
\big( \bar\psi (\slashed{D}_A+ m)\psi \big)
\bigg)=\det(\slashed{D}_A+ m),
\end{equation} 
where \textit{locally} $\slashed{D}_A \equiv e^\mu_{\mu'} \gamma^{\mu'} (\partial_\mu+\im \omega_\mu-\im A_\mu)$, $e^\mu_{\mu'}$ is a vielbein, 
with a spin connection $\omega_\mu$, and a gauge connection $A_\mu$ for a gauge bundle of a group $\tilde{G}$. More explicitly, the components are 
$\omega_\mu=\im \omega_\mu^{\lambda\nu}[\gamma^\lambda,\gamma^\nu]/8$ and $A_\mu=A^{\ra}_\mu \rm{T}^{\ra}$ with $\rm{T}^{\ra}$ generators of the Lie algebra $\text{Lie}(\tilde{G})$.  We also specify the transition functions that relate fields 
$\psi$ and $\bar{\psi}$ (two independent Grassmann fields)
 and $A$ on different patches in order to \textit{globally} define the Dirac operator $\slashed{D}_A$ on $M^d$. 
 The transition function also leaves the local expression of the partition function invariant. 
If we define $m \to \infty$ to be a trivial gapped vacua without any topological feature, then the
$m \to -\infty$ gapped vacua can host a nontrivial iTQFT or SPTs in $d$d. 
Follow the notations in \cite{1711.11587GPW}, the 
 $d$d partition function of nontrivial iTQFT or SPTs
can be defined via a ratio\footnote{The $\det(-\ii\slashed{D}_A)=\prod_\lambda \lambda$.
One can regularize it by $\det(-\ii\slashed{D}_A)= \prod_\lambda \frac{\lambda}{\lambda + \ii {\rm M}}$ via a Pauli-Villars regulator of mass ${\rm M}$.
The APS $\eta$-invariant \cite{Atiyah1975jfAPS, Atiyah1976APS, Atiyah1980APS} 
is a regularization of
$\sum_{\lambda} (\text{sign}\,\lambda)$.
}
\begin{equation}
{\bf Z}_{{\text{$d$d-iTQFT}}}[A]= \lim_{|m|\rightarrow \infty}\frac{\det(\slashed{D}_A-|m|)}{\det(\slashed{D}_A+|m|)}\equiv
\lim_{|m|\rightarrow \infty}\prod_\lambda \frac{\ii \lambda -|m|}{\ii \lambda +|m|}
=\frac{1}{2}(N_0+\lim_{s\rightarrow {0}+}\sum_{\lambda\neq 0} (\text{sign}\,\lambda)\cdot |\lambda|^{-s})
\end{equation} 
as a regularized sum of eigenvalues of Dirac operator.
Here $\slashed{D}_A$ is anti-Hermitian, so $-\ii\slashed{D}_A$ is Hermitian. 
The $\lambda$ are eigenvalues of $-\ii\slashed{D}_A$ so $\lambda \in \R$ are real. \\
$N_0$ are the number of the operator $-\ii\slashed{D}_A$'s zero modes.
Depending on the underlying $G$-structure of manifolds and $-\ii\slashed{D}_A$
(such as $G={\Pin^+}$ \cite{Witten:2015aba}, $G={\Pin^c}$ \cite{gilkey1985eta, Metlitski1510.05663}, or $G={\Pin^+\times_{\Z_2}\SU(2)}$ \cite{1711.11587GPW}),
the 4d iTQFT/SPT partition functions are respectively:
\bea
\Big({\bf Z}_{{\text{4d-iTQFT}}} \Big)^\upnu
\xrightarrow{|m|\rightarrow \infty}
\left\{\begin{array}{lll} 
\exp(\frac{2\pi \ii}{2} \cdot\upnu \cdot \upeta_{\Pin^+}), & \text{ with } \upeta_{\text{Pin}^+} \in \frac{1}{8}\Z, &\quad \upnu\in \Z_{16}.\\
\exp({2\pi \ii} \cdot\upnu \cdot \upeta_{\Pin^c}), & \text{ with }\upeta_{\text{Pin}^c} \in \frac{1}{8}\Z,  &\quad \upnu\in \Z_{8}.\\
\exp({2\pi \ii} \cdot\upnu \cdot \upeta_{\Pin^+\times_{\Z_2}\SU(2)}), & \text{ with }\upeta_{\Pin^+\times_{\Z_2}\SU(2)} \in \frac{1}{4}\Z, &\quad \upnu\in \Z_{4}.
\end{array}
\right.
\eea
However, in our work, we adjust above definitions to make $\eta_{\text{Pin}^+} \in \Z_{16}$ via $\eta_{\text{Pin}^+} = 8 \upeta_{\text{Pin}^+}$ mod 16, so:
\bea \label{eq:4dZ16iTQFT}
\Big({\bf Z}_{{\text{$4$d-iTQFT}}} \Big)^\upnu 
\xrightarrow{|m|\rightarrow \infty}
\exp(\frac{2\pi \ii}{16} \cdot\upnu \cdot \eta_{\Pin^+} \bigg\rvert_{M^4}), & \text{ with } \eta_{\text{Pin}^+} \in \Z_{16}, &\quad \upnu\in \Z_{16}.
\eea
Similarly, based on the Smith homomorphism \cite{2018arXiv180502772T, GarciaEtxebarriaMontero2018ajm1808.00009, Hsieh2018ifc1808.02881, GuoJW1812.11959}
\footnote{In fact, there are more Smith homomorphisms in any dimension. Related discussions along these maps are abundant, see 
\cite{2018arXiv180502772T} and also \cite{Kapustin1406.7329, GuoJW1812.11959, WanWangv2}:
\bea \label{eq:Smithmap}
\begin{array}{rrrl}
\Omega_8^{\Pin^+} =\Z_{32} \times \Z_2&\to& \Omega_7^{\Spin \times {\Z_2}}=\Z_{16}&\quad \text{generated by $\RP^7$},\\
\Omega_7^{\Spin \times {\Z_2}}=\Z_{16}&\to&
\Omega_6^{\Pin^-} =\Z_{16}& \quad \text{generated by $\RP^6$},\\
\Omega_6^{\Pin^-} =\Z_{16} &\to&
\Omega_5^{\Spin \times_{\Z_2} \Z_4}=\Z_{16}&\quad \text{generated by $\RP^5$},\\
\Omega_5^{\Spin \times_{\Z_2} \Z_4}=\Z_{16} &\to&
\Omega_4^{\Pin^+} =\Z_{16}& \quad \text{generated by $\RP^4$},\\
\Omega_4^{\Pin^+} =\Z_{16}
&\to& \Omega_3^{\Spin \times {\Z_2}}=\Z_{8}&\quad  \text{generated by $\RP^3$},\\
\Omega_3^{\Spin \times {\Z_2}}=\Z_{8}&\to&
\Omega_2^{\Pin^-} =\Z_{8}&\quad \text{generated by $\RP^2$},\\
\Omega_2^{\Pin^-} =\Z_{8}&\to&
 \Omega_1^{\Spin \times_{\Z_2} \Z_4}=\Z_{4}&\quad \text{generated by $\RP^1=S^1/\Z_2=S^1$},\\
 \Omega_1^{\Spin \times_{\Z_2} \Z_4}=\Z_{4} 
 &\to&
  \Omega_0^{\Pin^+} =\Z_{2} &\quad \text{generated by $\RP^0 = $ point}. 
  \end{array}
\eea
Notice that $\H^1(\RP^n,\Z_2)=\Z_2$, for example by using $n=5$, if we choose
${\CA_{{\Z_2}}}$ for $\H^1(\RP^n,\Z_2)=\Z_2$, then $\RP^4$ and  $\RP^5$ detect all $\upnu \in \Z_{16}$: 
\bea
\exp(\frac{2\pi \ii}{16} \upnu \left.   \eta\right\vert_{{M^4=\RP^4}}) &=&\exp(\frac{2\pi \ii}{16} \upnu ).\cr
\exp(\frac{2\pi \ii}{16}\upnu \left.\eta\big(\text{PD}(\CA_{{\Z_2}})\big) \right\vert_{M^5=\RP^5}) &=&\exp(\frac{2\pi \ii}{16} \upnu).
\eea
}
\bea \label{eq:5to4Smith}
\Omega_5^{\Spin \times_{\Z_2} \Z_4}=\Z_{16} \quad \text{(generated by $\RP^5$)}
\stackrel{ \cap  \CA_{{\Z_2}}}{\xrightarrow{\hspace*{1cm}}}
\Omega_4^{\Pin^+} =\Z_{16} \quad \text{(generated by $\RP^4$)},
\eea
we characterize the 5d iTQFT whose manifold generator for bordism group is $\RP^5$ at $\upnu=1$ below.
The cobordism invariant for any $\upnu$ corresponds to a 5d iTQFT partition function  
\bea \label{eq:5dSPT}
\Big({\bf Z}_{{\text{$5$d-iTQFT}}} \Big)^\upnu
\xrightarrow{|m|\rightarrow \infty}
\exp(\frac{2\pi \ii}{16} \cdot\upnu \cdot \eta(\text{PD}(\CA_{{\Z_2}})) \bigg\rvert_{M^5}), & \text{ with } \eta = \eta_{\text{Pin}^+} \in \Z_{16}, &\quad \upnu\in \Z_{16}.
\eea
Given a $G \supseteq {\Spin \times_{\Z_2^F} \Z_4}$ structure, the cohomology class 
$\CA_{{\Z_2}} \in \H^1(M,\Z_2)$ is the generator from  $\H^1(\B(\Z_4/\Z_2^F),\Z_2)$.
So $(\CA_{{\Z_2}})\equiv (\CA_{{\Z_4}}) \mod 2$ is the quotient for the $\Z_4$ gauge field $\CA_{{\Z_4}}$, which is the background field for the symmetry $\Z_{4,{X}}  \subset \U(1)_{X}$.
The $\eta(\text{PD}(\CA_{{\Z_2}}))$ is
the value of ${\eta} \in \Z_{16}$ on the Poincar\'e dual (PD) submanifold of the cohomology class
$\CA_{{\Z_2}}$. This PD is the precise meaning of \Eq{eq:5to4Smith} taking
${ \cap  \CA_{{\Z_2}}}$ from 5d to 4d. 

Thus, to summarize, the full 5d iTQFT and the 4d SM or SU(5) GG GUT model action $S_{\text{SM or GUT}}$
with $N_{\text{generation}}$
together can make the anomaly \Eq{eq:Ngeneration} matched via
the full 5d-4d coupled partition function 
\bea \label{eq:UU-GUT}
{\bf Z}_{\text{5d-4d}}
=
\exp(\frac{2\pi \ii}{16} \cdot\upnu \cdot \eta(\text{PD}(\CA_{{\Z_2}})) \rvert_{M^5})   \bigg\rvert_{\upnu=-N_{\text{generation}}}\cdot
\int [{\cal D} {\psi}] [{\cal D}\bar{\psi}][{\cal D} A]\cdots \exp\left( \ii S_{\text{SM or GUT}} \right  \rvert_{M^4}). \quad\quad
\eea
with a bulk-boundary correspondence $\prt {M^5}={M^4}$.
The full ${\bf Z}_{\text{5d-4d}}$ is gauge invariant, in particular also invariant under the background
$\Z_{4,X}$ transformation (at least at the higher-energy of GUT scale). 
This concludes the 
5d SPTs coupled to 4d SM or GUT in the Scenario \ref{Z45dSPT}.


\subsection{Hidden 4d non-invertible TQFT or 4d symmetry-enriched topological order}
\label{sec:Hidden4dnon-invertiblenon-abelianTQFT}

Now we explore the Scenario \ref{Z4XTQFT} (thus also Scenario \ref{Z4XbreakTQFT})  
for a hidden 4d non-invertible TQFT in \Sec{sec:Hidden4dnon-invertiblenon-abelianTQFT} to match the missing anomaly.
In fact to construct such an {\bf anomalous symmetry-preserving 4d TQFT} (here we preserve $\Z_{4,X}$ in ${\Spin \times_{\Z_2} \Z_4}$ structure),
we take the inspirations from the quantum condensed matter phenomenon in one lower dimension, 
known as the {\bf anomalous symmetry-enriched boundary topological order} in 3d (2+1d spacetime)
living on the boundary of 4d SPTs (3+1d spacetime).
In condensed matter thinking, it can be understood as saturating the 't Hooft anomaly by a symmetric gapped sector,
without breaking any symmetry breaking and without gapless modes,
via smearing the 't Hooft anomaly and anomalous symmetry into the long-range entanglement.
In the context of 3d boundary and 4d bulk, 
a novel surface topological order was firstly pointed out by
Vishwanath-Senthil in an insightful work \cite{VishwanathSenthil1209.3058}. Later on many people follow up on developing the 
surface topological order constructions (see overviews in \cite{Senthil1405.4015, Wen2016ddy1610.03911} and \cite{Wang2017locWWW1705.06728}).
We need to generalize this condensed matter idea to find an anomalous symmetric 4d TQFT living on the boundary of 5d fermionic SPTs \Eq{eq:5dSPT}.
In particular, we require the symmetry-extension method \cite{Wang2017locWWW1705.06728} and 
the spin-TQFT construction from a previous work \cite{GuoJW1812.11959}.

\subsubsection{Symmetry extension $[\Z_2] \to \Spin \times {\Z_{4,X}} \to \Spin \times_{\Z_2^F} {\Z_{4,X}}$ and a 4d $[\Z_2]$ gauge theory}
\label{sec:sym-ext}

First, we can rewrite the 5d iTQFT partition function \eq{eq:5dSPT} on a 5d manifold $M^5$ in terms of 
\begin{multline} \label{eq:5dSPT-4dTQFT}
\Big({\bf Z}_{{\text{$5$d-iTQFT}}} \Big)^\upnu
=
\exp(\frac{2\pi \ii}{16} \cdot\upnu \cdot \eta(\text{PD}(\CA_{{\Z_2}})) \bigg\rvert_{M^5})\\
= 
\exp(\frac{2\pi \ii}{16} \cdot\upnu \cdot  \Big( 8 \cdot\frac{p_1(TM)}{48}(\text{PD}(\CA_{{\Z_2}})) + 4 \cdot\text{Arf} (\text{PD}((\CA_{{\Z_2}})^3) ) 
+ 2 \cdot{\tilde{\eta}} (\text{PD}((\CA_{{\Z_2}})^4) )   + (\CA_{{\Z_2}})^5 \Big) \bigg\rvert_{M^5})\\
= 
\exp(\frac{2\pi \ii}{16} \cdot\upnu \cdot  \Big( 8 \cdot\frac{\sigma}{16}(\text{PD}(\CA_{{\Z_2}})) + 4 \cdot\text{Arf} (\text{PD}((\CA_{{\Z_2}})^3) ) 
+ 2 \cdot{\tilde{\eta}} (\text{PD}((\CA_{{\Z_2}})^4) )   + (\CA_{{\Z_2}})^5 \Big) \bigg\rvert_{M^5}),
\end{multline}
for a generic ${\upnu=-N_{\text{generation}}} \in \Z_{16}$.
Here are some explanations on the mathematical inputs of \eq{eq:5dSPT-4dTQFT} 
(also partly explained in the Notation in the end of \Sec{sec:Intro}):\\
$\bullet$ The $p_1(TM)$ is the first Pontryagin class of spacetime tangent bundle $TM$ of the manifold $M$.
Via the Hirzebruch signature theorem, we have 
$$\frac{1}{3}\int_{\Sigma^4} p_1(TM) =\sigma(\Sigma^4) =\sigma$$ 
on a 4-manifold $\Sigma^4$, where $\sigma$ is the signature of $\Sigma^4$.
So in \eq{eq:5dSPT-4dTQFT}, we evaluate the $\frac{1}{3}\int_{\Sigma^4} p_1(TM) =\sigma$ on the Poincar\'e dual  $\Sigma^4$ manifold of
the $(\CA_{{\Z_2}})$ cohomology class within the $M^5$.\\
{$\bullet$ The $\tilde{\eta}$ is a mod 2 index of 1d Dirac operator as a cobordism invariant of $\Omega_1^{\Spin}=\Z_2$.\\
$\bullet$  The Arf invariant \cite{Arf1941} is a mod 2 cobordism invariant of $\Omega_2^{\Spin}=\Z_2$,
 whose realization is the 1+1d Kitaev fermionic chain \cite{Kitaev2001chain0010440}.\\
$\bullet$ The $(\CA_{{\Z_2}})^5$ is a mod 2 class purely bosonic topological invariant, which corresponds to a 5d bosonic SPT phase given by
 the group cohomology class data $\H^5(\B\Z_2,\U(1))=\Z_2$, which is also one of the $\Z_2$ generators in $\Omega_5^{\SO}(\B\Z_2)$.
 }
 
Below we ask whether we can construct a fully gauge-invariant 5d-4d coupled partition function preserving the
${\Spin \times_{\Z_2} \Z_4}$ structure:
\bea \label{eq:5dSPT-4dTQFT-0}
\Big({\bf Z}_{{\text{$5$d-iTQFT}}} \Big)^\upnu
\Big({\bf Z}_{{\text{$4$d-TQFT}}} \Big).
\eea
Preserving the
${\Spin \times_{\Z_2} \Z_4}$ structure means 
that under the spacetime background transformation and the $\CA_{{\Z_2}}$ background gauge transformation,
the 5d-4d coupled partition function is still fully gauge invariant. 

\begin{enumerate}[leftmargin=-1.mm]
\item
When ${\upnu}$ is odd, such as ${\upnu}=1,3,5,7,\dots  \in \Z_{16}$, 
\Refe{Hsieh2018ifc1808.02881} suggested that the symmetry-extension method \cite{Wang2017locWWW1705.06728} cannot help to construct a symmetry-gapped TQFT.  
Furthermore, Cordova-Ohmori \cite{Cordova1912.13069} proves that a symmetry-preserving gapped TQFT phase is impossible for
this odd ${\upnu} \in \Z_{16}$ anomaly from $\Omega_5^{\Spin \times_{\Z_2} \Z_4}=\Z_{16}$.
The general statement in \cite{Cordova1912.13069} is that 
given an anomaly index $\upnu \in \Omega_5^{\Spin \times_{\Z_2} \Z_{4}}=\Z_{16}$,
we can at most construct a fully symmetric gapped TQFT 
\emph{if and only if} 
$$ 4 \mid 2 \upnu.$$
Namely, 4 has to be a divisor of $2 \upnu$.
Apparently, the $4 \mid 2 \upnu$ is true only when $\upnu$ is even,
while the $4 \mid 2 \upnu$ is false when $\upnu$ is odd.

Since ${\upnu=-N_{\text{generation}}}$, the case of ${\upnu}=1$ (for a single generation) and ${\upnu}=3$ (for three generations) are particularly important 
for the high energy physics phenomenology.
{This means that we are not able to
directly construct any 4d symmetric gapped TQFT that explicitly matches the same $\Z_{16}$ anomaly 
for one right-handed neutrino (${\upnu}=1$) or three right-handed neutrinos (${\upnu}=3$).}

\item When ${\upnu}$ is even, such as ${\upnu}=2,4,6,8,\dots  \in \Z_{16}$, 
\Refe{Hsieh2018ifc1808.02881} suggested that the symmetry-extension method \cite{Wang2017locWWW1705.06728} can trivialize the 't Hooft anomaly. 
Furthermore, Cordova-Ohmori \cite{Cordova1912.13069} shows that there is no obstruction to construct 
a symmetry-preserving gapped TQFT phase for any even ${\upnu_{\text{even}} } \in \Z_{16}$.
We can verify the claim by rewriting \eq{eq:5dSPT-4dTQFT} in terms of the $(\frac{\upnu_{\text{even}}}{2}) \in \Z_{8}$ index
with a 2d Arf-Brown-Kervaire (ABK) invariant:
\begin{multline} \label{eq:5dSPT-4dTQFT-Z8}
\Big({\bf Z}_{{\text{$5$d-iTQFT}}} \Big)^{\upnu_{\text{even}}}
=
\exp(\frac{2\pi \ii}{16} \cdot{\upnu_{\text{even}}} \cdot \eta(\text{PD}(\CA_{{\Z_2}})) \bigg\rvert_{M^5})
= 
\exp(\frac{2\pi \ii}{8} \cdot (\frac{{\upnu_{\text{even}}}}{2}) \cdot  \Big( 
\text{ABK} (\text{PD}((\CA_{{\Z_2}})^3) ) 
\Big) \bigg\rvert_{M^5}) \\
= 
\exp(\frac{2\pi \ii}{8} \cdot (\frac{{\upnu_{\text{even}}}}{2}) \cdot  \Big(  4 \cdot \text{Arf} (\text{PD}((\CA_{{\Z_2}})^3) ) 
+ 2 \cdot{\tilde{\eta}} (\text{PD}((\CA_{{\Z_2}})^4) )   + (\CA_{{\Z_2}})^5 \Big) \bigg\rvert_{M^5}) .
\end{multline}
Notice that \eq{eq:5dSPT-4dTQFT-Z8} can become trivialized if we can trivialize the $(\CA_{{\Z_2}})^2$ factor.
In fact, the topological term $(\CA_{{\Z_2}})^2$ can be trivialized by the symmetry extension \cite{Wang2017locWWW1705.06728} 
$$
0  \to \Z_2 \to {\Z_{4,X}} \to \frac{\Z_{4,X}}{\Z_2^F} \to 0.
$$
Namely, the cocycle topological term $(\CA_{{\Z_2}})^2$ in $\H^2(\B(\frac{\Z_{4,X}}{\Z_2^F}),\U(1))$ becomes a coboundary once we
lifting the $\frac{\Z_{4,X}}{\Z_2^F}$-gauge field $\CA_{{\Z_2}}$ 
to a ${\Z_{4,X}}$ gauge field in
$\H^2(\B{\Z_{4,X}},\U(1))$. So this suggests that the following symmetry extension for the spacetime-internal symmetry\footnote{In fact,
the recent works \cite{Wan2019sooWWZHAHSII1912.13504, PrakashJW2011.13921, PTWtoappear} suggest that many cobordism invariants given in \eq{eq:Smithmap} can be trivialized via a $\Z_2$ extension. For example, the following group extension to $\EPin(d)$ via 
\bea
1 \to \Z_2 \to \EPin(d) \to \Pin^{-}(d) \to 1
\eea
can trivialize the cobordism invariants when $\upnu = 4 \in  \Z_{8} = \Omega_2^{\Pin^-}$ at $d=2$,
and for any $\upnu \in\Z_{16} = \Omega_6^{\Pin^-}$ at $d=6$.
The $\EPin(d)$ is firstly introduced in \cite{Wan2019sooWWZHAHSII1912.13504}.
The following extension to $\Spin(d) \times\Z_4$ via
\bea
1 \to \Z_2 \to  \Spin(d) \times\Z_4 \to \Spin(d) \times\Z_2 \to 1
\eea
can trivialize the cobordism invariants  for any even $\upnu_{\text{even}} $ $\in  \Z_{8} = \Omega_3^{\Spin \times\Z_2}$ at $d=3$
and for any $\upnu \in\Z_{16} = \Omega_7^{\Spin \times\Z_2}$ at $d=7$.
The following group extension to $\EPin(d)$ via
\bea
1 \to \Z_2 \to \EPin(d) \to \Pin^{+}(d) \to 1
\eea
can trivialize the cobordism invariants  for any even $\upnu_{\text{even}} $ $\in  \Z_{16} = \Omega_4^{\Pin^+}$ at $d=4$.
The following extension to $\Spin(d) \times\Z_4$ via
\bea
1 \to \Z_2 \to  \Spin(d) \times\Z_4 \to \Spin(d) \times_{\Z_2} \Z_4 \to 1
\eea
can trivialize the cobordism invariants for any even $\upnu_{\text{even}} $ $\in  \Z_{16} = \Omega_5^{\Spin \times_{\Z_2} {\Z_4}}$ at $d=5$.
This is similar to what \Refe{RyanThorngren2001.11938} suggested that if the anomaly can be matched by a symmetric gapped boundary TQFT, then a
$\Z_2$ gauge theory almost always work.
}
\bea \label{eq:Z2-SpinxZ4}
1  \to [\Z_2] \to \Spin \times {\Z_{4,X}} \to \Spin \times_{\Z_2^F} {\Z_{4,X}} \to 1.
\eea
can fully trivialize any even ${\upnu_{\text{even}}} \in \Z_{16}$ cobordism invariant given in \eq{eq:5dSPT-4dTQFT-Z8}.
The $[\Z_2]$ means that we can gauge the normal subgroup $[\Z_2]$ in the total group $\Spin \times {\Z_{4,X}}$.
This symmetry extension \eq{eq:Z2-SpinxZ4} also means that a 4d $[\Z_2]$ gauge theory can preserve the 
$\Spin \times_{\Z_2^F} {\Z_{4,X}}$ symmetry while 
can also saturate the even ${\upnu_{\text{even}}} \in \Z_{16}$ anomaly.
This 4d $[\Z_2]$ gauge theory is the anomalous symmetric gapped non-invertible TQFT 
that we aim for in the Scenario \ref{Z4XTQFT}.

\end{enumerate}

 \subsubsection{4d anomalous symmetric $\Z_2$ gauge theory with $\upnu = 2 \in \Z_{16}$ preserving $\Spin \times_{\Z_2^F} {\Z_{4,X}}$}
\label{sec:Z16-nu=2-TQFT}

For example, when
${\upnu_{\text{even}}} =2 \in \Z_{16}$,
we have \eq{eq:5dSPT-4dTQFT-0} with the input of 5d bulk iTQFT \eq{eq:5dSPT-4dTQFT-Z8},
then we can explicitly construct the partition function on a 5d manifold $M^5$ with a 4d boundary 
$M^4 \equiv \partial M^5$ as,\footnote{Follow the Notation in the end of \Sec{sec:Intro},
we use the $\smile$ notation for the cup product between cohomology classes, or between a cohomology class and a fermionic topological invariant.
We use the $\cup$ notation for the surgery gluing the boundaries of two manifolds within relative homology classes.
So the $M_1 \cup M_2$ means gluing the boundary $\partial M_1 = \overline{\partial M_2}$ such that the common orientation of
$\overline{\partial M_2}$ is the reverse of ${\partial M_2}$.
}
\begin{multline} \label{eq:5dSPT-4dTQFT-explicit}
\Big({\bf Z}_{{\text{$5$d-iTQFT}}} \Big)^{{\upnu_{\text{even}}} =2 } \cdot
\Big({\bf Z}_{{\text{$4$d-$\Z_2$-gauge theory}}} \Big)  \\ 
=
 \sum_{c\in {\partial'}^{-1}(\partial [\text{PD}(\CA^3)])}\e^{\frac{2\pi\ii}{8}\text{ABK}(c\cup \text{PD}(\CA^3))} \quad\quad\quad\quad\quad\quad\\
\cdot \frac{1}{2^{|\pi_0(M^4)|}} \sum_{\substack{a\in C^1(M^4,\Z_2) ,\\b\in C^2(M^4,\Z_2)} }(-1)^{\int_{ M^4} b (\delta a+\CA^2)} \cdot \e^{\frac{2\pi\ii}{8}\text{ABK}(c\cup \text{PD}'(\CA a))}
\cdot
\end{multline} 
Here we write the mod 2 cohomology class $\Z_2$ gauge field as $\CA \equiv \CA_{{\Z_2}} \in \H^1(M^5, \Z_2)$.
We generalize the boundary TQFT construction in the Section 8 of \cite{GuoJW1812.11959}.
Here are some remarks about this 5d bulk iTQFT-4d boundary TQFT partition function \eq{eq:5dSPT-4dTQFT-explicit}:
\begin{enumerate} [leftmargin=.mm, label=\textcolor{blue}{\arabic*)}., ref={\arabic*)}]
\item 
The $(\text{PD}(\CA^3))$ is a 2d manifold taking the Poincar\'e dual (PD) of 3-cocycle $\CA^3$ in the $M^5$; 
but the 2d manifold $(\text{PD}(\CA^3))$ may touch the 
the 4d boundary $\partial M^5 =M^4$.
The 1d boundary $\partial (\text{PD}(\CA^3))$ can be regarded as the 1d intersection 
between the 2d $(\text{PD}(\CA^3))$ and the $M^4$.

\item \noindent 
More precisely, for a $\Spin \times_{\Z_2} \Z_4$ manifold $M^5$ with a boundary, 
we have used the Poincar\'e-Lefschetz duality for a manifold with boundaries:
\bea
\CA^2\in \H^3(M^5,\Z_2)\xrightarrow{\cong} \H_2(M^5, M^4,\Z_2)\ni \text{PD}(\CA^3).
\eea

\item \noindent
For any pair $(\rS,\rS')$, where $\rS'$ is a subspace of $S$, the short exact sequence of chain complexes 
\bea
0\to C_*(\rS')\to C_*(\rS)\to C_*(\rS,\rS')\to 0
\eea
induces a long exact sequence of homology groups
\bea
\cdots\to \H_n(\rS')\to \H_n(\rS)\to \H_n(\rS,\rS')\xrightarrow{\partial} \H_{n-1}(\rS')\to\cdots.
\eea
Here $C_n(\rS,\rS'):=C_n(\rS)/C_n(\rS')$, 
the $\H_n(\rS,\rS')$ is the relative homology group, and $\partial$ is the boundary map.\\
$\bullet$ Take $(\rS,\rS')=(\text{PD}(\CA^3),\partial \text{PD}(\CA^3))$, we denote the boundary map by $\partial$: 
\bea \label{eq:map-partial}
\H_2( \text{PD}(\CA^3), \partial \text{PD}(\CA^3) ) \xrightarrow{\partial} \H_{1}( \partial \text{PD}(\CA^3)).
\eea
Here $\text{PD}(\CA^3)$ is not a closed 2-manifold, but which has a boundary closed 1-manifold $\partial \text{PD}(\CA^3)$. \\
$\bullet$  Take $(\rS,\rS')=(M^4= \partial M^5, \partial \text{PD}(\CA^3))$, we denote another boundary map by $\partial_1$:
\bea \label{eq:map-partial1}
\H_2(M^4, \partial \text{PD}(\CA^3) ) \xrightarrow{\partial_1} \H_{1}( \partial \text{PD}(\CA^3)).
\eea 
Both $M^4= \partial M^5$ and $\partial \text{PD}(\CA^3)$ are closed manifolds, of 4d and 1d, respectively.

\item \noindent
Now, the $c$ is defined as a 2d surface living on the boundary $M^4$.
The ${\partial_1}c$ uses the boundary map \eq{eq:map-partial1}'s $\partial_1$ of $c$ on the $M^4$.
We can compensate the 2-surface $\text{PD}(\CA^3)$ potentially with a 1-boundary, 
by gluing it with $c$  to make a closed 2-surface.
To do so, we require both $\text{PD}(\CA^3)$ and $c$ share the same 1d boundary.

\item 
\noindent
 The ${c\in {\partial_1}^{-1}(\partial [\text{PD}(\CA^3)])}$ also means
\bea \label{eq:partial'c=partialPDA3}
{{\partial_1} c = \partial [\text{PD}(\CA^3)]  = [\partial \text{PD}(\CA^3)]}.
\eea
This is exactly the requirement that the bulk 2-surface $\text{PD}(\CA^3)$ (living in $M^5$) and
the boundary 2-surface $c$ (living on $M^4$) share the same 1d boundary. \\
\noindent
$\bullet$ The $[\text{PD}(\CA^3)]$ means the fundamental class and the relative homology class of the 2d manifold $\text{PD}(\CA^3)$.\\
 \noindent
$\bullet$ The {$\partial [\text{PD}(\CA^3)] = [\partial \text{PD}(\CA^3)]$} means the boundary of the fundamental class
(via the boundary $\partial$ map in \eq{eq:map-partial})
is equivalent to the fundamental class of the boundary $\partial$ of $\text{PD}(\CA^3)$.
Beware that the two $\partial$ operations in {$\partial [\text{PD}(\CA^3)] = [\partial \text{PD}(\CA^3)]$} have different meanings.

\item 
\noindent Note that when $M^5$ is a closed manifold with no boundary $M^4= \partial M^5= \emptyset$ thus $c = \emptyset$,
then the term 
$$\sum_{c\in {\partial_1}^{-1}(\partial [\text{PD}(\CA^3)])}\e^{\frac{2\pi\ii}{8}\text{ABK}(c\cup \text{PD}(\CA^3))}
\text{ is equivalently reduced to }
\e^{\frac{2\pi\ii}{8}\text{ABK}( \text{PD}(\CA^3))}
=\Big({\bf Z}_{{\text{$5$d-iTQFT}}} \Big)^{\upnu_{\text{even}}=2}.$$
This is a satisfactory consistent check, consistent with the 5d bulk-only iTQFT at ${\upnu_{\text{even}}}=2$ in \eq{eq:5dSPT-4dTQFT-Z8}.

\item \noindent
The 
$a\in C^1(M^4,\Z_2)$ means that $a$ is a 1-cochain,
and the $b\in C^2(M^4,\Z_2)$ means that {$b$ is a 2-cochain}.
The factor $(-1)^{\int_{ M^4} b (\delta a+\CA^2)}= 
\exp(\ii \pi {\int_{ M^4} b (\delta a+\CA^2)})$ gives the weight of the 
4d $\Z_2$ gauge theory. 
The $b \delta a$ term is the BF theory written in the mod 2 class.\footnote{The continuum QFT version of this $\Z_2$ gauge theory is
$\exp(\ii  {\int_{ M^4}\frac{2}{2 \pi} b \dd a+ \frac{1}{\pi^2} b \CA^2)})$, where
the $b$ integration over a closed 2-cycle, $\oiint b$, can be $n \pi$ with some integer $n \in \Z$.
The $a$ and $\CA$ integration over a closed 1-cycle, $\oint a$ and $\oint \CA$, can be $n \pi$ with some integer $n \in \Z$.}
The path integral sums over these distinct cochain classes.

The variation of $b$ gives the equation of motion $(\delta a+\CA^2) = 0 \mod 2.$
In the path integral, we can integrate out $b$ to give the same constraint $(\delta a+\CA^2) = 0 \mod 2.$
This is precisely the trivialization of the second cohomology class,
\bea \label{eq:trivializationA2}
\CA^2= (\CA_{{\Z_2}})^2 =\delta a \mod 2 
\eea
so the 2-cocycle becomes a 2-coboundary which splits to a 1-cochain $a$. This is exactly the condition imposed by the symmetry extension \eq{eq:Z2-SpinxZ4}:
$1  \to [\Z_2] \to \Spin \times {\Z_{4,X}} \to \Spin \times_{\Z_2^F} {\Z_{4,X}} \to 1$,
where the $\CA^2$ term in $ \Spin \times_{\Z_2^F} {\Z_{4,X}}$ becomes trivialized as a coboundary 
$\CA^2 =\delta a$ 
(so $\CA^2= 0$ in terms of a cohomology or cocycle class)
in $\Spin \times {\Z_{4,X}}$.

\item \noindent
The $\frac{1}{2^{|\pi_0(M^4)|}}$ factor mod out the gauge redundancy for the boundary 4d $\Z_2$ gauge theory. 
The $\pi_0(M^4)$ is the zeroth homotopy group of $M^4$, namely
the set of all path components of $M^4$.
Thus, we only sum over the gauge inequivalent classes in the path integral summation.

\item \noindent
The ${\ABK}(c\cup \text{PD}(\CA^3))$ is defined on a 2d manifold with $\Pin^-$ structure.
Recall the $M^5$ has the $\Spin \times_{\Z_2} {\Z_{4}}$ structure.
If $M^5$ is closed, then there is a natural Smith map \eq{eq:Smithmap} to induce the 2d $\Pin^-$ structure on the closed surface via $\text{PD}(\CA^3)$.
However, the $M^5$ has a boundary $M^4$, so $ \text{PD}(\CA^3)$ may not be closed  --- the previously constructed closed 2-surface $(c\cup \text{PD}(\CA^3))$
is meant to induce a 2d $\Pin^-$ structure.\footnote{We do not yet know whether it is always possible to induce a unique 2d $\Pin^-$ structure on
$(c\cup \text{PD}(\CA^3))$ for any possible pair of data $(M^5,M^4= \partial M^5)$ given any $\Pin^-$ $M^5$. However, we claim that it is possible
to find some suitable $M^4$ so that the 2d $(c\cup \text{PD}(\CA^3))$ has $\Pin^-$ induced, thus in this sense the ${\ABK}(c\cup \text{PD}(\CA^3))$ is defined.
\label{ft:inducePin-1}} 
Then we compute the ABK on this closed 2-surface $(c\cup \text{PD}(\CA^3))$.

\item \noindent
Let us explain 
the other term ${\text{ABK}(c\cup \text{PD}'(\CA a))}$ of the 4d boundary TQFT in \eq{eq:5dSPT-4dTQFT-explicit}.
The $\text{PD}'$ is the Poincar\'e dual on $M^4= \partial M^5$.
Since 
the fundamental classes of $M^5$ and $M^4= \partial M^5$ are related by
\bea
[M^5]\in \H_5(M^5, M^4,\Z_2)\xrightarrow{\partial}\H_4( M^4,\Z_2)\ni[ M^4]  = [\partial M^5].
\eea
Here we have the following relations:\footnote{{Let us clarify
the notations: $\partial$, $\partial'$, and $\partial_1$. 
The boundary notation $\partial$ may mean as (1) taking the boundary, or (2) in the boundary map of relative homology class in \eq{eq:map-partial}. 
It should be also clear to the readers 
that\\ 
$\bullet$ the $\partial$ is associated with the operations on objects living in the bulk $M^5$ or ending on the boundary $M^4$,\\
$\bullet$ while the $\partial'$ is associated with the operations on objects living on the boundary $M^4$ alone.\\
The $\partial_1$ is defined as another boundary map in \eq{eq:map-partial1}.}}
\bea \label{eq:PD-cap-relation}
\text{PD}&=&[M^5]\cap, \cr
\text{PD}' &=&[ M^4]\cap=[\partial M^5]\cap =\partial[ M^5]\cap,\\
{\partial'} \text{PD}'(\CA a) &=&
{\text{PD}'(\delta (\CA a))=\text{PD}'(\CA^3)=\partial [\text{PD}(\CA^3)]=\partial_1 c.}
\eea
$\bullet$ The cap product $\cap$ here is to define PD homology class, so $\text{PD}(\CA) = [M^5] \cap \CA$.\\
$\bullet$  Here we use $[M^4]=[\partial M^5]=\partial [M^5] $:
the fundamental class of boundary of $M^5$ gives
the boundary of fundamental class.\\
$\bullet$ {The \eq{eq:PD-cap-relation}'s first equality ${\partial'} \text{PD}'(\CA a)  =\text{PD}'(\delta (\CA a))$ uses
the coboundary operator $\delta$ on the cohomology class $\CA a$.}\\
$\bullet$ The \eq{eq:PD-cap-relation}'s second equality $\text{PD}'(\delta (\CA a))=\text{PD}'(\CA^3)$ 
uses the condition $\delta \CA =0$ and the trivialization condition \eq{eq:trivializationA2}: $\delta a = \cA^2$. \\
$\bullet$ The \eq{eq:PD-cap-relation}'s third equality $\text{PD}'(\CA^3)=\partial [\text{PD}(\CA^3)]$,
we use ``\emph{the naturality of the cap product}''. 
There are natural pushforward
and pullback maps on homology and cohomology, 
related by the projection formula, also known as ``the naturality of the cap product.''
\\
$\bullet$ The \eq{eq:PD-cap-relation}'s last equality $\partial [\text{PD}(\CA^3)]=\partial_1 c$ is based on \eq{eq:partial'c=partialPDA3}. 
Importantly, as a satisfactory consistency check, this also shows that the 2-surface $c$ obeys: 
\bea
\text{the $c$
in  ${\ABK}(c\cup \text{PD}(\CA^3))$ is the same $c$ in ${\text{ABK}(c\cup \text{PD}'(\CA a))}$.}
\eea
$\bullet$ Similarly as before, the union $(c\cup\text{PD}'(\CA a))$ is a closed 2-surface and we induce a $\Pin^-$ structure on this 2-surface.
So we can compute the ABK on this closed 2-surface $(c\cup\text{PD}'(\CA a))$ on the $M^4$.\footnote{Similar to Footnote \ref{ft:inducePin-1},
we do not yet know whether it is always possible to induce a unique 2d $\Pin^-$ structure on
$(c\cup\text{PD}'(Aa))$ on any $M^4$ given any $\Pin^-$ $M^5$. However, we claim that it is possible
to find suitable some $M^4$ so that the 2d $\Pin^-$ is induced, thus in this sense the ${\ABK}(c\cup\text{PD}'(Aa))$ is defined.
\label{ft:inducePin-2} }

\end{enumerate}

In summary, we have constructed the 4d $\Z_2$-gauge theory in \eq{eq:5dSPT-4dTQFT-explicit}
preserving the $(\Spin \times_{\Z_2^F} {\Z_{4,X}})$-structure, namely it is a $(\Spin \times_{\Z_2^F} {\Z_{4,X}})$-{symmetric} TQFT
but with ${\upnu_{\text{even}}}=2 \in \Z_{16}$ anomaly.
This can be used to compensate the anomaly 
$\upnu =- N_{\text{generation}}\mod 16$
with $N_{\text{generation}}=2$, two generations of missing right-handed neutrinos.
We could not however directly construct the symmetric gapped TQFT for 
$\upnu$ is odd (thus symmetric TQFTs not possible for $N_{\text{generation}}=1$ or $3$), due to the obstruction found in 
\cite{Hsieh2018ifc1808.02881, Cordova1912.13069}.

However, we can still compensate the 
$\upnu =- N_{\text{generation}} = -3\mod 16$ anomaly by:
\begin{enumerate}
\item
Appending the
symmetric gapped 4d TQFT with the anomaly ${\upnu}=2 \in \Z_{16}$,
plus an additional one generation of right-handed neutrino $\nu_R$
with the anomaly ${\upnu}=1 \in \Z_{16}$.
This single right-handed neutrino $\nu_R$ can pair with any of the three generation of
left-handed neutrino $\nu_L$ to give a Dirac mass via the Yukawa-Higgs term.
The quantum interference between the three generation of neutrinos (three $\nu_L$ and one $\nu_R$)
in the TQFT background (of ${\upnu}=2$) may give rise to
the neutrino oscillations observed in the phenomenology (see \Sec{sec:neutrino-oscillations}).

\item Appending the
symmetric gapped 4d TQFT with the anomaly ${\upnu}=2 \in \Z_{16}$,
plus an additional 5d iTQFT in an extra dimension
with the anomaly ${\upnu}=1 \in \Z_{16}$.
\end{enumerate}
Either of the above combinations still complete a scenario for the full anomaly matching. 
By the same logic, we can also propose many more other possible full anomaly matching scenarios,
such as the linear combinations of scenarios given in \Sec{sec:matchtheanomaly}.


\subsection{Gapping Neutrinos:  Dirac mass vs Majorana mass vs Topological mass}
\label{sec:HiddenTopologicalSectorsTopologicalmass}

It should be  clear that the 4d TQFT in \Sec{sec:Hidden4dnon-invertiblenon-abelianTQFT}
with an index $\upnu_{\text{even}} \mod 16$ in \Eq{eq:Ngeneration} 
 \emph{cannot} be directly access from gapping the free fermion theory ($\upnu_{\text{even}} \mod 16$ number of 
Weyl fermions or Majorana fermions). 
In order to access the 4d TQFT, we have to:
\begin{enumerate}
\item Go to a higher-energy mother effective field theory (EFT) at a deeper UV (ultraviolet), such that the mother EFT can
land at IR (infrared) with two different types of low energy physics: 4d free fermions on one side, and the 4d TQFT on another side.
\item Access via the dual fermionic vortex zero mode bound state condensation 
and go through a 4d topological quantum phase transition \cite{subirsachdev2011book, XGWen2004Book} 
or ``4d mirror symmetry \cite{Hori2003VafaMirrorsymmetry}'' (i.e., a duality of QFTs in the version of 4d spacetime). We comment more about the 4d duality 
in \Sec{sec:4dduality}. 
\end{enumerate}
Before then, let us clarify the meanings of Mass or Energy Gap.

The masses for {\bf left-handed neutrinos} (thus also right-handed antineutrinos) are experimentally estimated to be very small, nearly million times smaller than the electron mass \cite{PDG2008}
\bea
m_{\nu_e}, m_{\nu_\mu}, m_{\nu_\tau} < 1 e{\rm V} \ll m_{e} \simeq 0.51 {\rm M}e{\rm V}.
\eea
More precisely,
the flavor states (${\nu_e}$, ${\nu_\mu}$, ${\nu_\tau}$) 
are superpositions of the different mass eigenstates (${\nu_1}$, ${\nu_2}$, ${\nu_3}$). So a flavor state should be weighed and averaged over different masses of the different mass eigenstates.
The current experiments show that the sum of the masses of the three neutrinos should also be below about $1e$V. 
These mass bounds hold for neutrinos regardless being Dirac fermion or Majorana fermion particles.
These mass bounds only apply to left-handed neutrinos (thus also right-handed antineutrinos, or its own anti-particle if a neutrino is a Majorana fermion).
But we do not yet know about the mass, or mass bound, or the existence of 
{\bf right-handed neutrinos} (called {\bf sterile neutrinos} because they are in the trivial representation $({\bf 1},{\bf 1},0)_R$
in \Eq{eq:nuR-rep} and do not interact with any of three SM forces).
They also can be regarded as the left-handed $({\bf 1},{\bf 1},0)_L$ with complex conjugation on the representation.

Let us overview the two known mass generation mechanism and propose a new third mechanism (a topological mass or energy gap mechanism to gap the 
neutrinos):\footnote{By a {\bf gap},
we mean giving a {\bf mass gap} or an {\bf energy gap} to the system's energy spectrum,
which is the eigenenergy values of the quantum Hamiltonian of the system. (E.g. Solving a big matrix eigenvalues in the linear algebra.)
The mass gap usually already assumes the free particle descriptions exist. However, in the interacting systems such as
CFT or strongly-correlated many-body quantum matter, we may not always be able to find a suitable free particle description.
In the later case, we can still have an energy gap from the interacting or many-body physics.
The energy gap can be a generalization of the particle mass gap for the interacting or many-body systems.
After all, Einstein had told us long ago the energy is the mass: $\Delta E \sim (\Delta m) c^2$.}

\begin{enumerate}  [leftmargin=.mm, label=\textcolor{blue}{\arabic*}., ref={\arabic*}]
\item \label{Diracmass}
{\bf  Dirac mass} mechanism \cite{Dirac19281, Dirac19282}: Dirac mass and Anderson-Higgs mechanism are believed to give other Standard Model particles their masses. 
%
\item \label{Majoranamass}
{\bf  Majorana mass} mechanism \cite{Majorana1937}:  This requires that the neutrino and antineutrino to be the same particle.
If a neutrino is indeed a Majorana fermion, then lepton-number violating processes such as neutrino-less double beta decay would be allowed. 
The neutrino-less double beta decay is not allowed if neutrinos are Dirac fermions.
\item  \label{Topologicalmass}
{\bf Topological mass} mechanism (or {\bf  Topological Energy Gap} mechanism):  
This describes the gapped excitations of interacting systems 
(typical many-body quantum matter). The underlying low energy theory 
can be either invertible TQFT (invertible or SPT phases), or non-invertible TQFT (topological order).
The energy gap is induced by the interaction mechanism 
\cite{FidkowskifSPT1, FidkowskifSPT2, Wen2013ppa1305.1045, Wang2013ytaJW1307.7480, 
You2014oaaYouBenTovXu1402.4151, YX14124784, BenTov2014eea1412.0154,
BenTov2015graZee1505.04312, Wang2018ugfJW1807.05998, WangWen2018cai1809.11171}. 
The typical quadratic mass term breaks the chiral symmetry; 
but the interaction-induced energy gap can be fully symmetric without any chiral symmetry breaking \cite{RazamatTong2009.05037}.
\end{enumerate}

Is the mass or energy gap necessarily associated with a single particle picture (only defined if there is a free limit)? 
Or can the mass gap associated with the whole quantum system in a strongly correlated interacting limit? 
It turns out that  a free particle mass may be suitable for {\bf  Dirac mass} \ref{Diracmass} and {\bf  Majorana mass} \ref{Majoranamass} mechanisms;
but a free particle mass may not be appropriate for the 
{\bf Topological mass} \ref{Topologicalmass} mechanism.

\subsubsection{Dirac Mass or Majorana Mass}
\label{sec:MajoranaMassorDiracMass}

In more details, we write
Dirac mass (\ref{Diracmass}) and Majorana mass (\ref{Majoranamass}) as follows.
 By the 2-component Weyl spinor notations in 4d,
the undotted indices are for left-handed spinor fields, the
dotted indices are for right-handed spinor fields.
We write the 4-component spinors (such as Dirac or Majorana) in terms of two
2-component Weyl spinors:\footnote{Here we use the 2-component Weyl spinor notation.
Not to confuse the ${``\mu''}$ in muon neutrino index ${\nu_{\mu}}$ with the spacetime index $\mu$.
For two general left-handed Weyl spinors, say $\chi$ and $\chi'$ (where each component is a Grassman number with anti-commutation properties 
$\chi_{\al }  \chi'_{\bt} = -  \chi'_{\bt} \chi_{\al }$), we have
$$
\chi \chi' \equiv \chi^{\al }  \chi'_{\al} \equiv  \epsilon^{\al \bt} \chi_{\bt} \chi'_{\al} =  -\epsilon^{ \bt \al} \chi_{\bt} \chi'_{\al} = 
- (\chi^{\rm T})_{\bt}(\ii \sigma^2)_{\bt \al }\chi'_{\al }
=-  \epsilon^{\al \bt}  \chi'_{\al } \chi_{\bt}= \epsilon^{ \bt \al}  \chi'_{\al} \chi_{\bt} \equiv \chi'  \chi
, \quad\quad
$$
$$\hspace{-14mm}
( \chi  \chi' )^\dagger =(\chi^{\al }  \chi'_{\al})^\dagger=(\chi'_{\al})^\dagger(\chi^{\al } )^\dagger=(\chi'^\dagger_{\dot \al})(\chi^{\dagger\dot \al }) 
\equiv\chi'^\dagger \chi^\dagger 
= (\chi'^{\dagger}_{\dot\al})(\ii \sigma^2)_{\dot \al \dot \bt}\chi^\dagger_{\dot \bt }
= (\chi'^{\dagger}_{\dot\al})(\ii \sigma^2)_{ \dot \al  \dot\bt}\chi^*_{\dot \bt }
=(\chi^\dagger_{\dot \al})(\chi'^{\dagger\dot \al }) \equiv\chi^\dagger \chi'^\dagger  =(   \chi' \chi)^\dagger.
$$
We use the convention $ \epsilon^{\al \bt} =   \epsilon^{\bt \al} =  -\epsilon_{\al \bt}=  \epsilon_{\bt \al} =
\epsilon^{\dal \dbt} =   \epsilon^{\dbt \dal} =  -\epsilon_{\dal \dbt}=  \epsilon_{\dbt \dal} =(\ii \sigma^2)_{\al \bt} =
\begin{pmatrix}
0 & 1 \\
- 1&  0 
\end{pmatrix}_{\al \bt}$,
also
$\chi^{\al }  \equiv  \epsilon^{\al \bt} \chi_{\bt}$ and
$(\chi^{\al } )^\dagger=(\chi^{\dagger\dot \al })=(\epsilon^{\dot \al \bt} \chi_{\bt})^\dagger=(\chi_{\bt})^\dagger \epsilon^{\dot \al\bt}
=(\ii \sigma^2)_{\dot \al  \dot\bt}\chi^\dagger_{\dot \bt }
=(\ii \sigma^2)_{\dot \al  \dot \bt}\chi^*_{\dot \bt }$.
The
Hermitian conjugate switches the two SU(2) Lie algebras in the Lie algebra of the Lorentz group 
$\Spin(3,1)=\rm{SL}(2,\C)$. 
$\Spin(4)=\Spin(3) \times \Spin(3)=\SU(2)_L \times \SU(2)_R$.
The $h.c.$ is Hermitian complex conjugate.
}
 \bea\label{eq:Diracmass}
 &\hspace{-10mm}
\text{Dirac spinor } 
\Psi_{\rm D}\equiv \begin{pmatrix}
\chi_{\al}\\
{\zeta}_{}^{\dagger \dot\al}
\end{pmatrix}. 
\quad
\overline{\Psi}_{\rm D}\equiv({\zeta}_{}^{ \al}, \chi_{\dot \al}^\dagger ).
&
\text{Dirac mass: }
\overline{\Psi}_{\rm D} {\Psi}_{\rm D}= {\zeta}_{}^{ \al}  \chi_{\al} +\chi_{\dot \al}^\dagger {\zeta}_{}^{\dagger \dot\al}
=\zeta  \chi + ( \zeta  \chi )^\dagger.  
 \quad\quad\\
&\label{eq:Majoranamass}
\hspace{-10mm}
\text{Majorana spinor } 
\Psi_{\rm M}\equiv\begin{pmatrix}
\chi_{\al}\\
\chi_{}^{\dagger \dot\al}
\end{pmatrix}. 
\quad
\overline{\Psi}_{\rm M}\equiv({ \chi}_{}^{ \al}, \chi_{\dot \al}^\dagger ).
&\text{Majorana mass: }
\overline{\Psi}_{\rm M} {\Psi}_{\rm M}=
\chi  \chi + h.c.=\chi  \chi + ( \chi  \chi )^\dagger. 
\quad\quad\quad
\eea
If the neutrino mass is generated by Dirac mass (\ref{Diracmass}) like other Standard Model fermions, 
then the mass term needs to be an SU(2) singlet. The left-handed neutrino from the SU(2) doublet ${l_L}_{\nu}$
 would have the Yukawa interactions with 
 the SU(2) doublet Higgs $\phi_H$ and an SU(2) singlet 
 right-handed neutrino $\chi_\nu$.
 
Another mechanism is that neutrino mass can be generated by a Majorana mass (\ref{Majoranamass}), which would require the neutrino and antineutrino to be the same particle (particle as its anti-particle).

The anomaly $\upnu =- N_{\text{generation}}\mod 16$ in \Eq{eq:Ngeneration} dictates that in the free fermion limit,
we need to add $\upnu =N_{\text{generation}} = 3\mod 16$ right-handed spinors to SM to match the anomaly.\footnote{This 
anomaly $\upnu =- N_{\text{generation}}\mod 16$ in \Eq{eq:Ngeneration}
also rules out many BSM introducing more than $N_{\text{generation}} = 3 \mod 16$ of sterile neutrinos or right-handed Weyl spinors.}
Denote the right-handed spinor of sterile neutrinos of 3 generations $(\nu_e, \nu_{\mu}, \nu_{\tau})$ as
$(\chi_{\nu_e},
\chi_{\nu_\mu},
\chi_{\nu_\tau})$, 
we can write down the free
quadratic non-interacting action where  $\bar\sigma \equiv (1, \vec{\sigma})$,
\bea
\chi_{\nu_e}^{\dagger}\ii \bar\sigma^{\mu} {\partial}_{\mu} 
\chi_{\nu_e}
+
\chi_{\nu_\mu}^{\dagger}\ii \bar\sigma^{\mu} {\partial}_{\mu} 
\chi_{\nu_\mu}
+
\chi_{\nu_\tau}^{\dagger}\ii \bar\sigma^{\mu} {\partial}_{\mu} 
\chi_{\nu_\tau}+\frac{1}{2}
\bigg((\chi_{\nu_e},
\chi_{\nu_\mu},
\chi_{\nu_\tau})
\begin{pmatrix}
&&\\
&M_{} &\\
&&
\end{pmatrix}
\begin{pmatrix}
\chi_{\nu_e}\\
\chi_{\nu_\mu}\\
\chi_{\nu_\tau}
\end{pmatrix}
+ h.c.
\bigg).
\eea
Along the rank-3 mass matrix $M$, its diagonalized elements represent the Majorana mass \Eq{eq:Majoranamass}. 

More generally, we can pair 
$\upnu =N_{\text{generation}} = 3$ right-handed Weyl spinors (sterile neutrinos) with each other and with the
3 left-handed Weyl spinors via another
free non-interacting matrix term action: 
\bea
\frac{1}{2}
\bigg(
(
\Big({l_L}_{\nu_e},
{l_L}_{\nu_\mu},
{l_L}_{\nu_\tau}\Big)\frac{\<\phi_H\>}{|{\<\phi_H\>}|},
\chi_{\nu_e}^\dagger,
\chi_{\nu_\mu}^\dagger,
\chi_{\nu_\tau}^\dagger)\quad
\Biggl(\mkern-5mu
\begin{tikzpicture}[baseline=-.65ex]
\matrix[
  matrix of math nodes,
  column sep=1ex,
] (m)
{
\;\;\;\;0_{\text{\;\;\;}} & M_{\text{Dirac}} \\
M_{\text{Dirac}} & M_{\text{S}}\\ 
};
\draw[dashed]
  ([xshift=0.55ex]  m-1-1.north east) -- ([xshift=0.ex]m-2-1.south east);
\draw[dashed]
  ([yshift=-0.2ex]m-1-1.south west) -- ([yshift=-0.19ex] m-1-2.south east);
\node[above,text depth=1pt] at (m-1-1.north) {$\scriptstyle 3$};  
\node[above,text depth=1pt] at (m-1-2.north) {$\scriptstyle 3$};
\node[left,overlay] at ([xshift=-1.8ex]m-1-1.west) {$\scriptstyle 3$};
\node[left,overlay] at ([xshift=-1.2ex]m-2-1.west) {$\scriptstyle 3$};
\end{tikzpicture}\mkern-5mu
\Biggr)
\begin{pmatrix}
{l_L}_{\nu_e} \frac{\<\phi_H\>}{|{\<\phi_H\>}|} \\
{l_L}_{\nu_\mu}\frac{\<\phi_H\>}{|{\<\phi_H\>}|}\\
{l_L}_{\nu_\tau}\frac{\<\phi_H\>}{|{\<\phi_H\>}|}\\
\chi_{\nu_e}^\dagger\\
\chi_{\nu_\mu}^\dagger\\
\chi_{\nu_\tau}^\dagger
\end{pmatrix} + h.c.
\bigg).
\eea
There is a rank-6 mass matrix.
Here the Dirac mass scale, for example, is given by Higgs vev $M_{\text{Dirac}} \sim |{\<\phi_H\>}|$. 
We remind that the flavor states can be superpositions of the different mass eigenstates.
In above, we rewrite the three generations of SU(2) doublet ${\bf 2}$ of left-handed neutrino  
$\Big({l_L}_{\nu_e},
{l_L}_{\nu_\mu},
{l_L}_{\nu_\tau}\Big)$ pair
with the SU(2) doublet ${\bf 2}$ Higgs 
as $\Big({l_L}_{\nu_e},
{l_L}_{\nu_\mu},
{l_L}_{\nu_\tau}\Big)\frac{\<\phi_H\>}{|{\<\phi_H\>}|}$.
The usual seesaw mechanism \cite{Minkowski1977seesaw}
sets the scale of $|M_{\text{S}}| \gg |M_{\text{Dirac}}|$.
So the three mass eigenstates have mass $\simeq \frac{|M_{\text{Dirac}}|^2}{|M_{\text{S}}|} \ll |M_{\text{Dirac}}|$ (for the observed 3 neutrinos much smaller other
Dirac mass of M$e$V or G$e$V scales),
while the other three mass eigenstates have mass $\simeq {|M_{\text{S}}|}$ which can set to be the GUT scale (thus too heavy yet to be detected).

We should emphasize that by having {Dirac mass or Majorana mass} to any of the Weyl fermion spinors,
this quadratic mass would break the $\Z_{4,X}$ symmetry (\Eq{eq:nuRsym} and \Eq{eq:nuLsym} assigned to the complex Weyl fermions).
More precisely, as explained in \Sec{sec:matchtheanomaly}, 
the Dirac mass term preserves the $\Z_{4,X}$ symmetry, but the $\Z_{4,X}$ symmetry is \emph{spontaneously} broken by the electroweak Higgs condensate ${\<\phi_H\>} \neq 0$.
In contrast, the Majorana mass term \emph{explicitly} breaks the $\Z_{4,X}$ symmetry.

Is it necessary to break $\Z_{4,X}$ symmetry in order to saturate the anomaly by a gapped theory?
No, we do not have to break the $\Z_{4,X}$ symmetry if we introduce
a Topological Mass/Energy Gap for Weyl fermions, see
\Sec{sec:TopologicalMassandTopologicalEnergyGap}.

\subsubsection{Topological Mass and Topological Energy Gap}
\label{sec:TopologicalMassandTopologicalEnergyGap}

Now we introduce {\bf Topological mass} and {\bf Topological energy gap} mechanism.
The concept of free particle mass may not be appropriate here, 
so we should digest the mechanism from the interacting theory viewpoint. 
We may colloquially call any of the following mass gaps as the topological energy gap:

\begin{enumerate}[leftmargin=2.5mm, label=\textcolor{blue}{(\arabic*)}., ref={(\arabic*)}]
\item  \label{appr:Interaction}
{\bf Interaction-induced mass or interaction-induced energy gap}:
This idea has been used to classify the topological phases of interacting quantum matter.
Given the symmetry $G$ (including
the ${G_{\text{spacetime} }}$ and ${{G}_{\text{internal}} }$) of the system,
are the two ground states (i.e., two vevs) deformable to each other by preserving the symmetry $G$?
This is the key question for the community studying the classification of Symmetry-Protected Topological state (SPTs).\\[2mm]
$\bullet$  Fidkowski-Kitaev \cite{FidkowskifSPT1, FidkowskifSPT2} had shown
that 1+1d $\Z$ classification \cite{RyuSPT, Kitaevperiod, Wen1111.6341} 
of $T^2=+1$ topological superconductor with $\Z_2^T \times \Z_2^F$ symmetry can be reduced to
$\Z_8$ class. Fidkowski-Kitaev may be the first example of showing the interaction can produce the energy gap
between 8 Majorana fermions in 1+1d without breaking the original symmetry. (See recent discussions along QFT reviewed in \cite{BenTov2015graZee1505.04312, Tong20198Majorana1906.07199}.)\\[2mm]

$\bullet$  Kitaev \cite{Kitaev2015} and Fidkowski-Chen-Vishwanath (FCV) \cite{Fidkowski1305.5851} suggested that
3+1d $\Z$ classification of $T^2=(-1)^F$ topological superconductor (TSC) with $\Z_4^T \supset \Z_2^F$ symmetry can be reduced to
$\Z_{16}$ class. This implies that the 16 number of 2+1d Majorana fermions on the boundary of 3+1d TSC can open up an energy gap
without introducing any free quadratic mass: neither Majorana nor Dirac masses.\\[2mm]
$\bullet$ Wen \cite{Wen2013ppa1305.1045}, and the author and Wen \cite{Wang2013ytaJW1307.7480, Wang2018ugfJW1807.05998, WangWen2018cai1809.11171}, 
suggest that all the $G$-anomaly-free gapless theory can be fully gapped
without breaking $G$-symmetry. 
This idea includes introducing a random disorder new Higgs field \cite{Wen2013ppa1305.1045};
or introducing the proper-designed direct \emph{non-perturbative interactions} \cite{Wang2013ytaJW1307.7480, WangWen2018cai1809.11171}, 
(which are usually irrelevant or marginal operator deformations viewed from IR QFTs)
or introducing a \emph{symmetric gapped boundary} \cite{Wang2018ugfJW1807.05998}. 
Many of such examples are applied to construct chiral fermion or chiral gauge theories on the lattice.
This approach is pursued also by You-BenTov-Xu \cite{You2014oaaYouBenTovXu1402.4151, YX14124784}.
BenTov and Zee \cite{BenTov2014eea1412.0154, BenTov2015graZee1505.04312} names this mechanism as Kitaev-Wen mechanism, 
or \emph{the mass without mass}.
In the lattice gauge theory, gapping the mirror fermions dated back to Eichten-Preskill \cite{Eichten1985ftPreskill1986}; 
a recent work by Kikukawa also suggested the same conclusion that the SO(10) GUT mirror fermion can be gapped \cite{Kikukawa2017ngf1710.11618}.
This deformation of $G$-anomaly-free theory is consistent with Seiberg's conjecture that the 
 deformation classes of QFTs is constrained by symmetry and anomaly \cite{NSeiberg-Strings-2019-talk}.

\item \label{appr:VortexCondensation}
{\bf Vortex condensation}: This is an approach commonly used in condensed matter literature for 2+1d strongly-correlated 
systems. The idea is that the symmetry-breaking defects (such as vortices) may trap the zero modes and which carry nontrivial quantum number.
The question is to find which number of vortex zero modes with what kind of symmetry assignment, 
can the vortices be proliferated to restore the broken global symmetry ---
this would drive the quantum phase transition between the symmetry-breaking phase and the symmetry-restoring phase.
This approach, called the vortex condensation, has been used to construct 2+1d surface topological orders, see the condensed matter review \cite{Senthil1405.4015}.

\item \label{appr:SymmetryExtension}

{\bf Symmetry-extension and symmetry-preserving gapped topological order/TQFT}: 
As mentioned briefly in \Sec{sec:sym-ext}, 
 the symmetry-extension mechanism \cite{Wang2017locWWW1705.06728} 
 and
 higher-symmetry extension generalization \cite{Wan2018djlW2.1812.11955}
 are a rather exotic mechanism. 
 We trivialize the anomaly and introduce a gapped phase, 
 not by breaking $G$ to its subgroup $G_{\text{sub}} \subset G$, but by extending it to a larger group
 $G_{\text{total}}$ which can be regarded as a fibration of the original group $G$ as a quotient group. 
 (See the down-to-earth lattice constructions in any dimension  \cite{Wang2017locWWW1705.06728} and in 1+1d bulk \cite{Prakash2018ugo1804.11236, PrakashJW2011.13921}) 
This mechanism is a useful intermediate step stone, 
to construct a \emph{symmetry-preserving TQFT}, via gauging the extended-symmetry \cite{Wang2017locWWW1705.06728}.
This approach is applicable to bosonic systems \cite{Wang2017locWWW1705.06728, Wang1801.05416} and fermionic systems \cite{GuoJW1812.11959, Kobayashi2019lep1905.05391, PrakashJW2011.13921} in any dimension.
In contrast to the well-known gapped phase saturate the anomaly via {\bf symmetry breaking} (either global symmetry breaking or Anderson-Higgs mechanism for gauge theory), this approach is based on {\bf symmetry extension} (thus beyond symmetry breaking and Anderson-Higgs mechanism).

\end{enumerate}

The {\bf Topological mass} and {\bf Topological energy gap} mechanism (including
\ref{appr:Interaction}, \ref{appr:VortexCondensation} and \ref{appr:SymmetryExtension}), in fact, 
is obviously beyond the familiar Dirac, Majorana mass, and seesaw mechanism \cite{Minkowski1977seesaw}.
In a colloquial sense, we do not have a Higgs field $\phi_H$ breaking the symmetry and gives vev $ \<\phi_H\>  \neq 0$.
In certain case, we can consider a new randomly disordered Higgs field $\phi_h$ such that
\cite{Wen2013ppa1305.1045, WangWen2018cai1809.11171}
\bea
 \<\phi_h\>  = 0, \quad\quad  \<|\phi_h|^2\>  \neq 0.
\eea
So Topological mass/energy gap is a quantum behavior beyond the mean field quadratic semiclassical theory, beyond Anderson-Higgs mechanism, and beyond Landau-Ginzburg symmetry-breaking paradigm.

\subsubsection{4d duality for Weyl fermions and ``mirror symmetry''} 
\label{sec:4dduality}

Before apply the Topological Mass from \Sec{sec:TopologicalMassandTopologicalEnergyGap} to neutrino physics furthur, 
we like to introduce a potential helpful supersymmetry (SUSY) duality in 4d known as Seiberg duality  \cite{Seiberg19949411149} studied in $\CN=1$ theory and supersymmetric quantum chromodynamics (SQCD). Seiberg duality is an $\CN=1$
electric magnetic duality in SUSY non-abelian gauge theories with the weak-strong duality.
On the left-hand side of the duality may have
quarks and gluons; 
on the right-hand side dual theory,
they become the solitons (such as nonabelian magnetic monopoles) of the elementary fields.
When the left-hand side theory is Higgsed by an expectation value of a squark, the  right-hand side dual theory's is confined. 
Massless glueballs, baryons, and magnetic monopoles in the confined strongly coupled description in the left-hand side theory
becomes some weakly coupled elementary quarks in the right-hand side dual Higgs description.

Schematically, there is an IR duality between left-hand side (LHS) and  right-hand side (RHS) under a renormalization group (RG) flow
for $\CN=1$:
\begin{multline}
 \label{eq:4dSeibergdual}
\SU(N_c) \text{ gauge theory with $N_f$ chiral and antichiral multiplets $Q$, $\tilde{Q}$ in color fundamental $N_c$, $\bar{N}_c$} \\
\xleftrightarrow{\text{IR duality}}  
\SU(N_f-N_c)  \text{ gauge theory with $q$ and $\tilde{q}$ in color fundamental $N_f-N_c$, $\overline{N_f-N_c}$ and meson $M$}.
\end{multline}
Include the representations of chiral multiplet/superfields, we have the relations:
\bea
  \begin{tabular}{ccc  }
  \hline
4d $\CN=1$ Seiberg duality   & LHS: {SQCD} & RHS: dual theory \\
  \hline
  \hline
{color gauge group}   & SU($N_c$) & SU($N_f-N_c$) \\
  \hline\\[-4mm]
Same  global internal symmetries  &  
\multicolumn{2}{c}{$\SU(N_f)_L \times \SU(N_f)_R \times \U(1)_B \times \U(1)_R$}\\ 
 \hline
 $\begin{array}{c}
 \text{Chiral multiplet/superfields:}\\	
  \text{Representation {\bf R}}  
   \end{array}$
  & $Q: (N_f, 1, 1, \frac{N_f-N_c}{N_f})$ & $q: ( 1, N_f,  \frac{-1}{N_f-N_c}, \frac{N_c}{N_f})$  \\
    & $\tilde{Q}: ( 1, \bar{N}_f, -1, \frac{N_f-N_c}{N_f})$ & $\tilde{q}: ( \bar{N}_f, 1,  \frac{1}{N_f-N_c}, \frac{N_c}{N_f})$  \\
        & & $M: ( {N}_f, \bar{N}_f, 0,  \frac{2(N_f-N_c)}{N_f})$  \\
  \hline
  \end{tabular}
\eea
We are particularly interested in the case when the RHS flows to free Weyl spinors, which means that
we can choose as simple as $N_c=2$ and $N_f=3$, and $\CN=1$:
\bea \label{eq:15}
\SU(N_c=2) \text{ with $N_f=3$ chiral and anti-chiral multiplets $Q$ and $\tilde{Q}$} 
\xleftrightarrow{\text{IR duality}}  
15 \text{ Weyl spinors}.
\eea
For Weyl spinor counting
we have $N_f=3$ chiral multiplets and $N_f=3$ anti-chiral multiplets,
each is the fundamental or anti fundamental ${\bf 2}$ or $\bar{\bf 2}$ of SU(2),
thus they contribute $2 \cdot 2 \cdot 3 = 12$ Weyl spinors. 
There is also a vector multiplet which sits at the adjoint representation ${\bf 3}$ of SU(2),
this contributes another $3$ Weyl spinors.
So in total we have $2 \cdot 2 \cdot 3 +3 = 15$ Weyl spinors. 
The $N_f=N_c +1=3$ is interesting because it sits at the lower boundary ($3 N_c/2 = N_f$)
of $3 N_c/2 < N_f < 3 N_c$, where the origin of the moduli space is an interacting CFT and
non-abelian Coulomb phase.
Also this case we have $N_f=N_c +1$ thus two moduli spaces are identical but
the interpretations of the singularity at the origin are different --- massless particles can be regarded as, either
strongly coupled mesons and baryons on LHS, or weakly interacting or free quarks on RHS.

This duality helps as $ 15 \mod 16 = -1  \mod 16$ so to cancel the anomaly 
\Eq{eq:nu-1} as $\upnu = -N_{\nu_R}  \mod 16= -1  \mod 16 $ in one generation of SM.
One way to simplify 15 Weyl spinors to
1 Weyl spinor on RHS would be that adding 1 Weyl spinor on both sides in the trivial representation,
and adding \emph{nonperturbative deformations} to gap the RHS completely: 
\begin{multline} \label{eq:4ddual}
\SU(N_c=2) \text{ with $N_f=3$ chiral and anti-chiral multiplets $Q$ and $\tilde{Q}$ +  deformations} \\
\xleftrightarrow{\text{IR duality}}  
(\text{gapping 16 Weyl spinors via nonperturbative interacting deformations}) \\
+ (- 1) \text{ Weyl spinors (in the complex conjugate representation)}.
\end{multline}
We leave details of this construction in an upcoming work \cite{toappear}.
In the following subsections, we can argue several phenomenon of gapping Weyl spinors based on this proposed duality \Eq{eq:4ddual}. 
The hope is that
we can access the 4d TQFT
from the dual fermionic vortex zero mode bound state condensation via the topological quantum phase transition \cite{subirsachdev2011book, XGWen2004Book} 
or ``4d mirror symmetry \cite{Hori2003VafaMirrorsymmetry} description'' (i.e., a duality of QFTs in the version of 4d spacetime).

In a general colloquial sense, this duality is also related to the particle-vortex duality \cite{Peskin19771978, DasguptaHalperin1981}.
The renown supersymmetric version of duality in 3+1d includes, for example, the $\CN=2$ Seiberg-Witten theory \cite{SeibergWitten1994rs9407087},
and the $\CN=1$ Seiberg duality \cite{Seiberg19949411149}, and many other theories (see a review \cite{Intriligator1995auSeiberg9509066}). 
A recent ongoing pursuit, along 
a fermionic non-supersymmetric version of particle-vortex duality in 3+1d, 
includes 
\cite{Anber2018iof1805.12290, Cordova2018acb1806.09592DumitrescuClay, BiSenthil1808.07465, 
Wan2018djlW2.1812.11955, BiLakeSenthil1910.12856, WangYouZheng1910.14664}.\footnote{There were also a fermionic version in 2+1d  of particle-vortex duality proposed in \cite{Son20151502.03446, Wang2015Senthil1505.05141, Metlitski20151505.05142} and formalized in \cite{SeibergWitten20161606.01989, KarchTong20161606.01893, MuruganNastase20161606.01912}. They have achieved great success on understanding condensed matter phenomena in 2+1d. 
They may provide insightful guidelines to the 3+1d, namely 4d duality construction.}

\subsubsection{Gapping 3, 2, or 1 Weyl spinors / sterile right-handed neutrinos + extra sectors} 
\label{sec:Gapping3Weylspinors}

In \Sec{sec:MajoranaMassorDiracMass}, the 4d anomaly $\upnu =- N_{\text{generation}}\mod 16$ in \Eq{eq:Ngeneration} dictates that in the free particle limit,
we need to add $\upnu =N_{\text{generation}} = 3\mod 16$ right-handed spinors to match the anomaly. But as mentioned, in \Sec{sec:Hidden4dnon-invertiblenon-abelianTQFT},
there is an obstruction 
to obtain a Topological Mass: there does \emph{not} exist any 4d noninvertible TQFT to saturate the $\upnu =N_{\text{generation}} = 3\mod 16$ anomaly,
shown by \cite{Hsieh2018ifc1808.02881,Cordova1912.13069}.

{In order to gap 3 Weyl spinors in a topological fashion, we can obtain a 4d $\upnu_{\text{even}}=2$-TQFT,
with a left-over right-handed neutrino (which introduces Dirac/Majorana masses):
\begin{multline} \label{eq:TQCP-1}
\text{4d $(\upnu = 3)$ massless right-handed neutrinos}\\ 
\xrightarrow{\text{4d topological quantum phase transition}}\\
\text{4d $(\upnu_{\text{even}}=2)$-TQFT + } \\
\text{1 right-handed neutrino gives Dirac/Majorana masses} \\
\text{to the other 3 of SM's left-handed neutrinos.}
\end{multline}
We can go through a quantum phase transition 
by using 3 times of \Eq{eq:4ddual}.
We may access the 4d noninvertible TQFT from the LHS.
On the other hand, we do not gain the expectation values for the quadratic mass to two generations of fermions:
\bea \label{eq:interaction-mass}
\newcommand*{\temp}{\multicolumn{1}{c|}{0}}
\begin{pmatrix}
M_{11} & M_{12}& M_{13}\\
M_{21} &M_{22} & M_{23}\\
M_{31} & M_{32}& M_{33}
\end{pmatrix}
\xrightarrow{\text{mean field vev}}  
\text{quadratic mass term vev} \propto
\begin{pmatrix}
0 &\temp &0\\
0 & \temp & 0 \\\cline{1-3}
0 & \temp  & \# \<\phi_H^{\nu}\> \text{ or } M_{\text{Majorana}}
\end{pmatrix}.
\eea
Schematically, on the left-hand side of \eq{eq:interaction-mass}, 
we may include the quadratic fermion pairing term 
with the $M_{ij}$ stands for possible highly-order interaction (or momentum-dependent pairing) terms.
On the right-hand side, we ask what are the quadratic mass at the mean-field level.
Two generations have the zero mean-field mass. But one generation has left with a Dirac mass ($\# \<\phi_H^{\nu}\>$) or Majorana mass ($M_{\text{Majorana}}$).}
If we aim to preserve the $\Z_{4,X}$ symmetry at the Lagrangian level, then we should choose the 
Dirac mass term instead of Majorana mass term.

However, as mentioned in \Sec{sec:matchtheanomaly}, Yukawa-Higgs-Dirac mass term spontaneously breaks the $\Z_{4,X}$,
while the Majorana mass explicitly breaks the $\Z_{4,X}$. Neither scenario involving Dirac or Majorana mass preserves the $\Z_{4,X}$ at the dynamical level.
Nonetheless, we can preserve the $\Z_{4,X}$ at the dynamical level if we trade a single generation of 4d right-handed neutrino
to a 5d $(\upnu = 1)$-iTQFT. So we can consider the $\Z_{4,X}$-symmetric topological quantum phase transition:
\begin{multline} \label{eq:TQCP-2}
\text{4d $(\upnu = 2)$ massless right-handed neutrinos + 5d $(\upnu = 1)$-iTQFT}\\ 
\xrightarrow{\text{4d topological quantum phase transition}}\\
\text{4d $(\upnu_{\text{even}}=2)$-TQFT + }  
\text{5d $(\upnu = 1)$-iTQFT.}
\end{multline}

To complete a story for high-energy phenomenology, 
we can add
``(4d SM with $(N_{\text{generation}} = 3) \times$ 15 Weyl fermions)''
on both sides of topological quantum phase transitions in 
 \eq{eq:TQCP-1} and 
  \eq{eq:TQCP-2}, in order to apply to SM and BSM physics.


\section{Ultra Unification: Grand Unification $+$ Topological Force and Matter} 
\label{sec:UU}

In the previous sections, we had shown that in order to match a nonperturbative $\Z_{16}$ global anomaly (\Sec{sec:SMandGUTdiscrete}) but still preserve\footnote{By
preserving $\Z_{4,{X}}$, we mean the $\Z_{4,{X}}$-symmetry is preserved at some higher energy scale above the electroweak scale. Of course,
the energy scale lower such as the Higgs scale 125 G$e$V, the usual Yukawa-Higgs-Dirac mass terms would spontaneously break the $\Z_{4,{X}}$.}
$\Z_{4,{X}} $ of SM and SU(5) GUT Georgi-Glashow (GG) model with only 15 Weyl fermions per generation,
we can introduce a new hidden gapped sector appending to the SM and GUT: \\[-10mm]
\begin{itemize}
\item 
Topological Mass/Energy Gap (\Sec{sec:Hidden4dnon-invertiblenon-abelianTQFT} and \Sec{sec:HiddenTopologicalSectorsTopologicalmass}) to gap the 16th Weyl fermions (right-handed neutrinos). The outcome low energy gives rise to a 4d non-invertible TQFT.
\item 5d invertible TQFT (iTQFT) with one extra dimension (\Sec{sec:5dSPT}, known as 5d SPTs in a quantum condensed matter analogy).
\end{itemize}
Overall, we consider the combinations of the above two solutions plus additional scenarios enlisted in \Sec{sec:matchtheanomaly}. 
In either cases,
we require new hidden gapped topological sectors beyond SM and GUT. We name the
unification including SM, Grand Unification, plus additional unitary gapped topological sectors with Topological Force and Matter
as {\bf Ultra Unification}.

\subsection{4d-5d Theory: Quantum Gravity and Topological Gravity coupled to TQFT
}
\label{sec:gravity}

\begin{enumerate}[leftmargin=0mm, label=\textcolor{blue}{\Roman*]}:, ref={\Roman*]}]

\item
The combinations of solutions from  
\Sec{sec:matchtheanomaly}, including adding 4d non-invertible TQFT and 5d invertible TQFT, 
mean that we can propose a new schematic partition function / path integral, generalizing \Eq{eq:UU-GUT} to
%
\begin{multline}  \label{eq:UU-GUT-1}
{\bf Z}_{\overset{\text{5d-iTQFT/}}{\text{4d-QFT}}}[\CA_{{\Z_4}}]
=
\exp(\frac{2\pi \ii}{16} \cdot\upnu_{\rm{5d}} \cdot \eta(\text{PD}(\CA_{{\Z_2}})) \rvert_{M^5})  \cdot
\int [{\cal D} {\psi}] [{\cal D}\bar{\psi}][{\cal D} A][{\cal D} \phi_H][{\cal D}  {a}] [{\cal D}   b]\cdots \\
\exp( \ii \left. S_{\text{4d-SM/GUT}}[\psi, \bar{\psi}, A, \phi_H, \dots,  \CA_{\Z_4}] \right  \rvert_{M^4}
+ \ii \left. S_{\text{4d-TQFT}}^{(\upnu_{\rm{4d}})} [a, b, \dots , \CA_{{\Z_2}}] \right\rvert_{M^4}
) 
\bigg\rvert_{\upnu_{\rm{5d}} - \upnu_{\rm{4d}}=-N_{\text{generation}}}.
\quad\quad
\end{multline} 
Partition function depends on the ($\CA_{{\Z_2}} = \CA_{{\Z_4}}$ mod 2) background fields. 
The anomaly \Eq{eq:Ngeneration} is now matched by:
\bea \label{eq:Ngeneration-nu45}
\upnu = \upnu_{\rm{5d}} - \upnu_{\rm{4d}}=-N_{\text{generation}} \mod 16.
\eea
The $S_{\text{4d-SM/GUT}}$ is the 4d SM or GUT action.
The $\psi, \bar{\psi}, A, \phi_H$ are SM and GUT quantum fields,
where $\psi, \bar{\psi}$ are the 15 or 16 Weyl spinor fermion fields,
the $A$ are 12 or 24 gauge bosons given by gauge group Lie algebra generators,
and $\phi_H$ is the electroweak Higgs (we can also add GUT Higgs). 
The $S_{\text{4d-TQFT}}^{(\upnu_{\rm{4d}})}$ is a 4d noninvertible TQFT outlined in \Sec{sec:Hidden4dnon-invertiblenon-abelianTQFT} 
(in particular, $\upnu_{\rm{4d}}= \upnu_{\rm{even}}= \pm 2, \dots$).
The ${a}$ and ${b}$ (and possibly others fields from fermionic invariants such as the ABK invariant)
are TQFT gauge fields (locally differential 1-form $a$ and 2-form $b$, as anti-symmetric tensor gauge connections).

\item Many different perspectives guide to an understanding that at high enough energy scale, every global symmetry should be either 
\emph{dynamically gauged} or 
\emph{explicitly broken}  \cite{Misner1957mtWheeler, Polchinski2003bq0304042, Banks2010zn1011.5120, Harlow2018tngOoguri1810.05338}.
By every global symmetry, we include the internal symmetry ${{G}_{\text{internal}} }$,
the spacetime ${{G}_{\text{spacetime}} }$, the fermion parity $\Z_2^F$, the time reversal symmetry $\Z_2^T$ or $\Z_4^{T} \supset \Z_2^F$, and so on.
Below we discuss the consequences on gauging some of the symmetries on the boundary or in the bulk.
Our notations on the bulk and boundary symmetries associated with the group extension
follow the conventions of \cite{Wang2017locWWW1705.06728, Wang1801.05416}:

\begin{enumerate}[leftmargin=1.5mm]
\item {\bf Gauge the $\Z_2$ on the 4d}:
We can gauge $a$ $\in$ ${\H^1(M, \Z_{2})}$ on the boundary
which sits at the normal subgroup $[\Z_2]$ 
of the symmetry extension 
\eq{eq:Z2-SpinxZ4}:
\bea \label{eq:Z2-SpinxZ4-Z2}
1  \to [\Z_2]_{\text{bdry gauge}} \to \Spin \times {\Z_{4,X}} \to (\Spin \times_{\Z_2^F} {\Z_{4,X}})_{\text{Bulk/bdry sym}} \to 1,
\eea
while the 5d bulk and 4d boundary both keep the global symmetry $ \Spin \times_{\Z_2^F} {\Z_{4,X}}$.
In this case, there is a 4d $\Z_2$ gauge theory living on the boundary of a 5d iTQFT.

\item 
 {\bf Gauge the $\frac{\Z_{4,X}}{\Z_2^F}$ in 5d} and {\bf gauge the $\Z_{4,{X}}$ on the 4d}:
We can also dynamically gauge $\CA_{{\Z_2}}$ $\in$ ${\H^1(M, \Z_{4,X}/\Z_2^F)}$ which sits at the normal subgroup $\Z_2$ 
of the following short exact sequence respect to the bulk:
\bea \label{eq:Z2-Spin}
1  \to [\frac{\Z_{4,X}}{\Z_2^F}]_{\text{Bulk gauge}} \to \Spin \times_{\Z_2^F} {\Z_{4,X}}\to (\Spin)_{\text{Bulk sym}}    \to 1.
\eea
{{If we also gauge the $[\frac{\Z_{4,X}}{\Z_2^F}]$ out of the 5d bulk symmetry $\Spin \times_{\Z_2^F} {\Z_{4,X}}$, 
then the 5d bulk short-range entangled gapped iTQFT can become a
5d bulk long-range entangled gapped $\Z_2$ gauge TQFT with a
fermionic rotational $\Spin(d)$ symmetry definable on manifolds with a $\Spin$ structure.}}
This is due to gauging the normal subgroup of the above group extension \eq{eq:Z2-Spin}.

Once the $[\frac{\Z_{4,X}}{\Z_2^F}]$ is dynamically gauged in 5d,
it also implies that the $[\Z_{4,{X}}]$ should be dynamically gauged on the 4d boundary theory.
 (It is natural that the $\Z_{4,{X}}$ is gauged above some GUT scale,
especially given that the $\Z_{4,{X}}$ is a gauge subgroup $\Z_{4,{X}} = Z(\Spin(10))  \subset \Spin(10)$ of the SO(10) GUT, see more discussions in \cite{JW2008.06499}.)
  By doing so, we need to sum over 
$\CA_{{\Z_4}}$ $\in$ ${\H^1(M, \Z_{4,X})}$ for a given spacetime topology and geometry.
We can gauge the normal subgroup $\Z_{4,X}$ out of the 4d boundary symmetry $\Spin \times {\Z_{4,X}}$ in 
the group extension, 
\bea \label{eq:Z4-Spin}
1  \to [{\Z_{4,X}}]_{\text{bdry gauge}}  \to \Spin \times_{} {\Z_{4,X}} \to (\Spin)_{\text{Bulk/bdry sym}}  \to 1.
\eea


\item 
{\bf Gauge the ${\Z_2^F} \times \frac{\Z_{4,X}}{\Z_2^F}$ in 5d} and {\bf gauge the ${\Z_2^F} \times \Z_{4,{X}}$ on the 4d}:
Other than gauging $\CA_{{\Z_2}}$ $\in$ ${\H^1(M, \Z_{4,X}/\Z_2^F)}$, we can also gauge the fermion parity $\Z_2^F$. 
So we gauge the normal subgroup out of the following short exact sequence respect to the bulk:
\bea \label{eq:Z2-SO}
1  \to [{\Z_2^F} \times \frac{\Z_{4,X}}{\Z_2^F}]_{\text{Bulk gauge}} \to \Spin \times_{\Z_2^F} {\Z_{4,X}}\to (\SO)_{\text{Bulk sym}}    \to 1.
\eea
Then the 5d bulk becomes a 
long-range entangled gapped TQFT with a bosonic
$\SO(d=5)$ symmetry defined on 5d manifolds with a special orthogonal group $\SO$ structure.
For the boundary theory, 
we gauge the normal subgroup out of the following short exact sequence respect to the boundary:
\bea \label{eq:Z4-SO}
1  \to [{\Z_2^F} \times {\Z_{4,X}}]_{\text{bdry gauge}} \to \Spin \times_{} {\Z_{4,X}} \to (\SO)_{\text{Bulk/bdry sym}}  \to 1.
\eea
By gauging the fermion parity $\Z_2^F$, this means the UV completion at a higher energy scale should be a bosonic system, 
presumably emerged from a local tensor product Hilbert space ---
this is analogous to the ``It from Qubit'' and the quantum condensed matter view 
\cite{WangWen2018cai1809.11171}.

\item  {\bf Gauge the $\Z_{4,{X}}$ with the spacetime Spin and SO symmetry}:
We can also sum over $\CA_{{\Z_4}}$ $\in$ ${\H^1(M, \Z_{4,X})}$  together with 
all spacetime ${{M \in \forall\; \text{topology},{\forall \;\text{geometry}}}}$.
Let us consider the case the spacetime is a closed 
 $M=M^5$ without boundary.
But summing over all spacetime $M$ certainly diverges, one needs to make sense of the partition function
by regularization. To deal with such a path integral is a challenging problem similar to topological quantum gravity. 
We are not be able to solve it for now.  In summary, a schematic path integral says:
\begin{multline}  \label{eq:5dQG}
\sum_{{M \in \forall\; \text{topology},{\forall \;\text{geometry}}}}
\sum_{\CA_{{\Z_2}} \in \H^1(M, \Z_{4,X}/\Z_2^F)}
\exp(\frac{2\pi \ii}{16} \cdot\upnu_{\rm{5d}} \cdot \eta(\text{PD}(\CA_{{\Z_2}})) \rvert_{M^5}) \\
\xrightarrow{\;\text{regularize}\;}
\sum_{\underset{\CA_{{\Z_2}} \in \H^1(M, \Z_{4,X}/\Z_2^F)}{{M \in \; \{\text{topo/geo}\},}}}
\exp(\frac{2\pi \ii}{16} \cdot\upnu_{\rm{5d}} \cdot \eta(\text{PD}(\CA_{{\Z_2}})) \rvert_{M^5}). 
\end{multline} 

A few thoughts can help to attack this challenging problem on regularization of summing over spacetimes.
\begin{itemize}
\item
In 2d topological gravity, we can sum over conformal structures.
This can be a finite dimensional integral for a fixed topology.

\item One can simplify the problem to sum over different topologies given by the 5d $\eta(\text{PD}(\CA_{{\Z_2}}))$ invariant
with $M^5$ of ${\Spin \times_{\Z_2} \Z_4}= {\Spin \times_{\Z_2^F} \Z_{4,X}}$ structure

\item 
Recent work by Dijkgraaf and Witten on 2d topological gravity provides a guide to the analogous 2d partition function \cite{Dijkgraaf2018Witten1804.03275} related to the 2d Arf invariant. The theory \Eq{eq:5dQG} may be regarded as a 5d gravity (dynamical, quantum, and topological gravity)
 related to the 5d $\eta(\text{PD}(\CA_{{\Z_2}}))$ invariant.

\end{itemize}

\end{enumerate}


\item Suppose we find a way to regularize the path integral, 
then we can consider the theory \Eq{eq:5dQG} with boundary, where we can place the 4d SM theory. We thus rewrite
\Eq{eq:UU-GUT-1} as the fully gauged version of a schematic path integral:
\begin{multline}  \label{eq:UU-GUT-2-grav}
{\bf Z}_{\overset{\text{5d-TQFT.QG/}}{\text{4d-QFT}}}
=
\sum_{\underset{\CA_{{\Z_2}} \in \H^1(M, \Z_4/\Z_2^F)}{{M \in \{\text{topo/geo}\},}}}
\exp(\frac{2\pi \ii}{16} \cdot\upnu_{\rm{5d}} \cdot \eta(\text{PD}(\CA_{{\Z_2}})) \rvert_{M^5})  \cdot
\int [\cD\CA_{{\Z_4}}]
 [{\cal D} {\psi}] [{\cal D}\bar{\psi}][{\cal D} A][{\cal D} \phi_H][{\cal D}  a ] [{\cal D}  b]\cdots \\
\exp( \ii \left. S_{\text{4d-SM/GUT}}[\psi, \bar{\psi}, A, \phi_H, \dots,  \CA_{\Z_4}] \right  \rvert_{M^4}
+ \ii \left. S_{\text{4d-TQFT}}^{(\upnu_{\rm{4d}})}[ a , b, \dots , \CA_{\Z_2}] \right\rvert_{M^4}
) 
\bigg\rvert_{\upnu_{\rm{5d}} - \upnu_{\rm{4d}}=-N_{\text{generation}}}.
\quad\quad
\end{multline} 
Recall we gauge $\CA_{{\Z_4}}$ as it is the ${\Z_{4,X}}$ gauge field and $\CA_{{\Z_4}} \mod 2 = \CA_{{\Z_2}}$.  
This is a 4d and 5d coupled fully gauged quantum system ---
where the 4d has SM, GUT and noninvertible TQFT, 
and the 5d can allow a certain gravity (dynamical, quantum and topological) coupled to the 5d TQFT.
The 1-form gauge field $\CA_{{\Z_4}}$ (obtained by gauging the 0-form ${\Z_{4,X}}$ symmetry) can mediate between
the 5d bulk, and 4d boundary (the 5d bulk and the 4d SM and BSM gapped topological sectors).

\end{enumerate} 



\subsection{Braiding statistics and link invariants in 4d or 5d: \newline
Quantum communication with gapped hidden sectors from our Standard Model} \label{subsec:braiding}

Follow \Sec{sec:gravity}, above a higher energy scale (way above SM and above SU(5) GUT but around SO(10) GUT), 
the discrete $X=5({\bf B}- {\bf L})-4Y$ symmetry (the ${\Z_{4,X}}$ symmetry) 
becomes dynamically gauged, with a dynamical  gauge vector boson
mediator $X_g$, mathematically associated with a 1-cochain gauge field $\CA_{{\Z_4}}$.\footnote{Again 
this is a discrete $\Z_{4,X}$ or $\U(1)_X$ gauge theory's gauge boson $X_g$, different from the SU(5) GUT's continuous nonabelian Lie group's gauge boson X and Y. 
We have been denoted $X_g$'s 1-cochain gauge field as
$\CA_{{\Z_2}}$ for the case of $\frac{\Z_{4,X}}{\Z_2^F}$ symmetry.
We have been denoted $X_g$'s 1-cochain gauge field as
$\CA_{{\Z_4}}$ for the case of $\Z_{4,X}$ symmetry.}
So the entangled Universe in 4d and 5d may be described by \Eq{eq:UU-GUT-2-grav}.

Once the $\Z_{4,X}$ is dynamically gauged,
the 4d and 5d gauged TQFT sectors become noninvertible TQFTs. 
In fact such a medium can be regarded as some version of {\bf topological quantum computer}  \cite{Kitaev2003, Kitaev2006} with intrinsic topological orders \cite{Wen2016ddy1610.03911}.

There is an immediate question: Do we have any way to communicate or interact with 
objects living in 4d or 5d TQFTs? Could we communicate with 
those objects within SM particles that we human beings and creatures are made of?
Another way to phrase this question is: Can we detect the Topological Force?

For example in the TQFTs of \eq{eq:5dSPT-4dTQFT-explicit}, we have some gauge invariant extended line and surface operators:
\bea
\oint a+\frac{1}{2} \CA_{{\Z_2}},\quad\quad
\oiint b+\frac{1}{2}\CA_{{\Z_2}}^2,
\eea
invariant under the gauge transformation: 
$\CA_{{\Z_2}}\to \CA_{{\Z_2}}+\delta\lambda$, \; 
$a\to a+\lambda\delta\lambda$,
and $b\to b+\lambda\delta\lambda\delta\lambda$, where $\lambda$ is a 0-cochain valued in $\H^1(M,\Z_2)$.
Since the extended ${a}$ line and ${b}$ surface operators do not end on the  SM particles, nor do those $a$ and $b$ extended operators
interact with SM gauge forces, 
it seems that we SM particles cannot communicate with the TQFT
at least n\"aively.
{At least, we SM particles cannot directly interact with the 4d TQFT sector via the SM gauge forces}.  

\begin{figure}[t!] 
  \centering
      (a)  \includegraphics[width=1.24in]{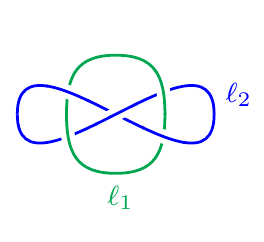}
       (b) \includegraphics[width=4.in]{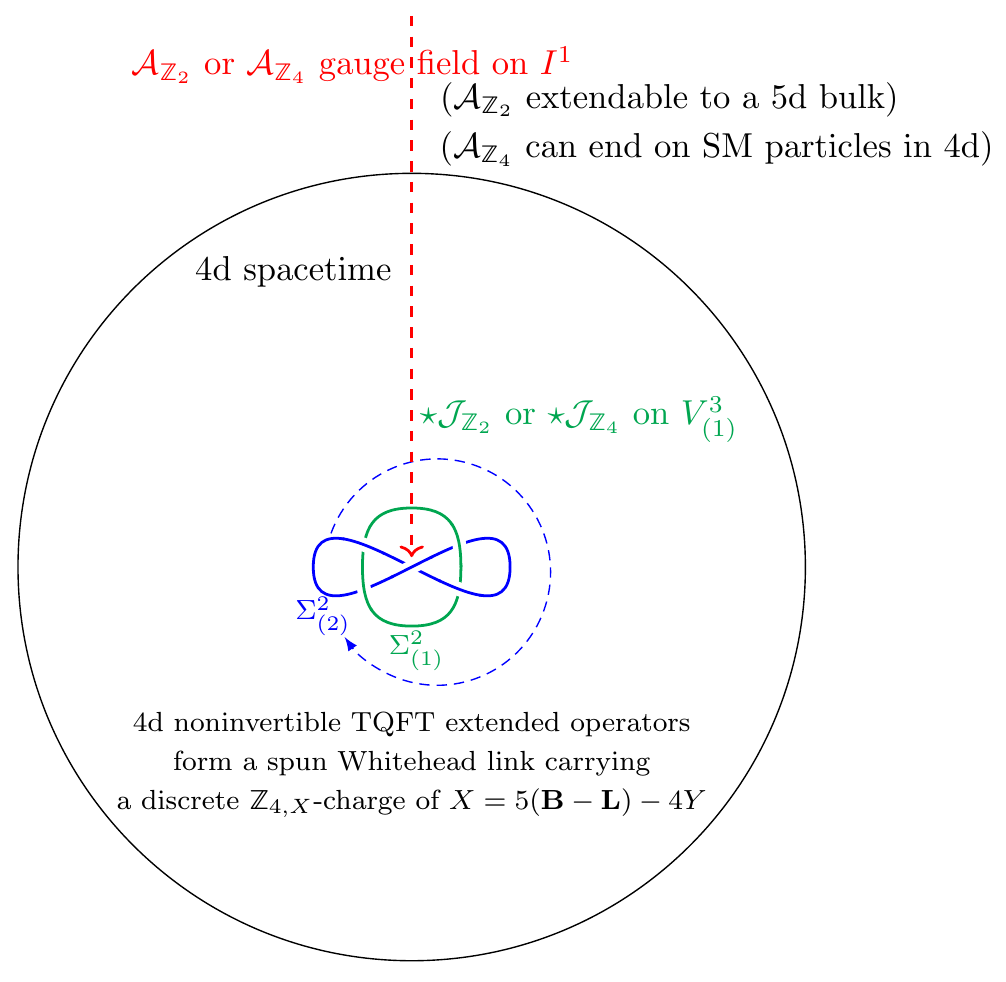} 
  \caption{(a) The Whitehead link formed by a disjoint union of two-component 1-loops, $\ell = \ell_1 \sqcup \ell_2$, is detectable by the Sato-Levine invariant. (See related QFTs explored in  \cite{Putrov2016qdo1612.09298PWY,GuoJW1812.11959}.) 
 (b)
  A schematic nontrivial link configuration in a 4d spacetime (inside the large black circle)
  that can carry an odd $\Z_{4,X}$ charge (also an odd $\frac{\Z_{4,X}}{\Z_2^F}$ charge)
  measured by a codimension 1 operator $\star \CJ_{{\Z_2}}$ (from the 5d bulk perspective) or $\star \CJ_{{\Z_4}}$ (from the 4d boundary perspective).
    The $\Z_{4,X}$ charge is trapped non-locally within the surface link.
  The surface link $L = \Sigma_{(1)}^2 \sqcup \Sigma_{(2)}^2$ contains two surfaces obtained from a spun version of Whitehead link.
  Let a $D^3$ ball contain the Whitehead link $\ell$.
  The $\Sigma_{(1)}^2$ rotates the $\ell_1$ along an ${S^1}'$ of $D^3 \times {S^1}'$ outside the $D^3$. 
  The $\Sigma_{(2)}^2$ rotates the $\ell_2$ along the blue dashed arrow around the $\ell_1$ circle within the $D^3$.
  From a 4d boundary theory perspective, we have the $\CA_{{\Z_4}} $ along a 1-line $I^1$ (drawn in the red dashed)
  Poincar\'e dual (PD) to the $\star \CJ_{{\Z_4}}$.
    From a 5d bulk theory perspective, we have the $\CA_{{\Z_2}} $ along a 1-line $I^1$ PD to the $\star \CJ_{{\Z_2}}$.
  }
  \label{fig:BSMlink1}
\end{figure}

But in fact, the answer is {\bf  YES}, {we SM particles can indeed communicate with the 4d TQFT sector, only via the $\CA_{{\Z_2}}$ gauge field.}
Moreover, {we SM particles can also communicate with the 5d TQFT sector via the $\CA_{{\Z_4}}$ gauge field.}
The idea is that the TQFT path integral \eq{eq:5dSPT-4dTQFT-explicit} schematically
\bea \label{eq:Z4dnabTQFTgaugeZ4}
\int [\cD\CA_{{\Z_2}} ] \int  [\cD b] [\cD a] \;
 \exp{\Big(\ii \pi \int_{ M^4} b (\delta a+\CA_{{\Z_2}}^2) 
+\frac{2\pi\ii}{8}\text{ABK}(c\cup \text{PD}'(\CA_{{\Z_2}} a)) \Big)}.
\eea
contains a term with a schematic coupling:
\bea \label{eq:ABKAa}
 \pi \int (b \smile  \delta a  + \delta \CB \smile  \CA_{{\Z_2}}  + b \smile \CA_{{\Z_2}}  \smile \CA_{{\Z_2}}) + \frac{1}{4} (\text{ABK} \smile \CA_{{\Z_2}} \smile a).
\eea
For the purpose which will become clear soon, 
we introduce a dynamical 2-cochain $\CB$ field which is natural if the $\CA_{{\Z_2}}$ is dynamically gauged, with 
$\CB \smile \delta  \CA_{{\Z_2}} = \delta \CB \smile  \CA_{{\Z_2}}$ on a closed 4-manifold.
Again the schematic cup product term 
$(\text{ABK} \smile \CA_{{\Z_2}} \smile a)$,
coupling between a fermionic invariant ABK and the cohomology class $Aa$, is formally $\text{ABK}(c\cup \text{PD}'(\CA_{{\Z_2}} a))$ defined in \Sec{sec:Z16-nu=2-TQFT}.
We denote $\text{PD}'$ as the Poincar\'e dual on the 4-manifold $M^4= \partial M^5$.
So what are the experimental designs for us to 
  communicate with ``TQFT sectors''?\footnote{In fact,
previous works on braiding statistics and link invariants, 
such as the multi-string 3-loop braiding process 
and other exotic braiding process (\cite{WangLevin1403.7437, Jiang1404.1062, Wang1404.7854} and \cite{CWangMLevin1412.1781, 
Putrov2016qdo1612.09298PWY, 1602.05951WWY, Wang2019diz1901.11537}), 
provide helpful guidelines to these topological link designs, shown in \Fig{fig:BSMlink1}.
}

The action term $(\text{ABK} \smile \CA_{{\Z_2}} \smile a)$ in \Eq{eq:ABKAa}
prompts us to design a configuration in \Fig{fig:BSMlink1}. 

To motivate the discussion,
the fermionic topological term
$(\text{ABK} \smile \CA_{{\Z_2}} \smile a) 
\sim
\text{ABK}(\text{PD}'(\CA_{{\Z_2}} a))$ 
is similar to
the  fermionic topological invariant
$(\text{Arf} \smile a_1 \smile a_2)=
\text{ABK}(\text{PD}'(a_1 \smile a_2) )
=
\text{ABK}(\text{PD}'(a_1) \cap \text{PD}'(a_2) )
$ 
studied in  
\cite{GuoJW1812.11959}.
The $(\text{Arf} \smile a_1 \smile a_2)$ 
corresponds to a $\Z_2$ torsion class cobordism invariant in the subgroup of
$\Omega^{\Spin \times (\Z_4)^2}_{4} = (\Z_4)^2 \times \Z_2$.
The $a_1$ and $a_2$ are the mod 2 classes of 1-cochain of two distinct $\Z_4$ gauge fields.
Only this $\Z_2$ generator is intrinsically fermionic.
\Refe{GuoJW1812.11959} shows that the Sato-Levine invariant \cite{sato1984cobordisms},
$\text{Arf}(V_{(1)} \cup V_{(2)})$, can detect a certain link configuration.
The link configuration $L$ is a disjoint union of two surface links $\Sigma_{(1)}^2$ and $\Sigma_{(2)}^2$,
so $L = \Sigma_{(1)}^2 \sqcup \Sigma_{(2)}^2$. The ``$\sqcup$'' means the disjoint union.
The $L$ is a semi-boundary link \cite{sato1984cobordisms},
so that, by definition, there exists Seifert volumes $V_{(1)}^3$ and $V_{(2)}^3$ for each link component $\partial V_{(j)}^3= \Sigma_{(j)}^2$, 
such that the intersections $V_{(1)}^3 \cap \Sigma_{(2)}^2 = \emptyset$ and
$\Sigma_{(1)}^2 \cap V_{(2)}^3 = \emptyset$ are empty.

To design a detectable link configuration for
the fermionic topological term
$(\text{ABK} \smile \CA_{{\Z_2}} \smile a) 
\sim
\text{ABK}(\text{PD}'(\CA_{{\Z_2}} a))$,
we start from finding appropriate two 2-surfaces $ \Sigma_{(1)}^2$ and $ \Sigma_{(2)}^2$ to be linked, 
see \Fig{fig:BSMlink1}.

We have a 2-cochain $\CB$ field dual to the 1-cochain $\CA_{{\Z_2}}$ which is dynamically gauged.
In this case, the 2-surface $\Sigma_{(1)}^2$ fundamental class can be paired with a 2-surface operator from the cohomology class $\CB$ field,
$
\oiint_{L_1} \CB +\dots.
$
And another 2-surface $\Sigma_{(2)}^2$ fundamental class can be paired with a 2-surface operator from the cohomology class $b$ field,
$
\oiint_{L_2} b +\dots
$
with extra $\dots$ terms to make the 2-surface operator gauge invariant \cite{toappear}.

Let us explain how to design a semi-boundary two-component link $L = \Sigma_{(1)}^2 \sqcup \Sigma_{(2)}^2$  (in \Fig{fig:BSMlink1} (b), 
this link configuration is only an example, there could be other configurations trapping also a $\Z_{4,X}$ charge): 
\begin{enumerate}[leftmargin=2.5mm]
\item
First, we can start from thinking of a Whitehead link $\ell = \ell_1 \sqcup \ell_2$ in a 3-ball $D^3$
shown in \Fig{fig:BSMlink1} (a). 
In \Fig{fig:BSMlink1} (a), the $\ell_1$ and $\ell_2$ are both $S^1$ circles linked in a $D^3$.
\item Then, we make the $\ell_1$ circle spun around another ${S^1}'$ circle along the ${S^1}'$ of $D^3 \times {S^1}'$ outside of the $D^3$ sphere. 
(So the readers can imagine this ${S^1}'$ rotation of $D^3$ going outside of the paper.)
If $M^4$ is a sphere $S^4$, it can be chosen as $S^4= (D^3 \times {S^1}') \cup ( S^2 \times D^2 )$, gluing the $(D^3 \times {S^1}')$ with another $( S^2 \times D^2 )$ 
along their common boundary $( S^2 \times {S^1}')$.
So this spun $\ell_1$ gives us a 2-torus that we call this 2-surface $ \Sigma_{(1)}^2 \cong \ell_1 \times {S^1}'$.
We place a cohomology class $\CB$ field on the $\ell_1$.

\item We also make the $\ell_2$ rotating along the blue dashed arrow, circling around the $\ell_1$ component.
We call this 2-surface $ \Sigma_{(2)}^2 \cong \ell_2 \times S^1$.
We place a cohomology class $b$ field on the $\ell_2$.

\item This semi-boundary two-component 2-surface link $L = \Sigma_{(1)}^2 \sqcup \Sigma_{(2)}^2$ in \Fig{fig:BSMlink1} (b), with $\CB$ and $b$ field inserted, 
is a nontrivial link configuration which detects the fermionic topological term
$(\text{ABK} \smile \CA_{{\Z_2}} \smile a)$.

\item We can also replace the 2-cochain $\CB$ field to its dual 1-cochain $\CA_{{\Z_2}}$ field. 
In the sense, we can interpret the coupling 
$\pi \int  \CA_{{\Z_2}} \smile  \Big(  \delta \CB   +  b \smile \CA_{{\Z_2}} + \frac{1}{4} (\text{ABK}  \smile a) + \dots \Big)$
as the $\CA_{{\Z_2}}$ gauge field coupled to a current $\CJ_{{\Z_2}}$ (with a Hodge dual $\star$), say in a schematic differential form:
\bea \label{eq:AZ2JZ2}
\int \CA_{{\Z_2}} \wedge \star \CJ_{{\Z_2}}  +\dots.
\eea
The $\CB$ sits at the 2-surface $\Sigma_{(1)}^2$,
so the Hodge dual of the current 
$ \star \CJ_{{\Z_2}} \propto  \delta \CB   +\dots $ sits at the Seifert 3-volume $V_{(1)}^3$.
If so, we learn that the 1-cochain gauge field $\CA_{{\Z_2}}$ can sit along a 1-line $I^1$,
Poincar\'e dual (PD) to the Seifert volume $V^3_{(1)}$,
drawn along the red dashed line $I^1$ in \Fig{fig:BSMlink1} (b).

\item The $\CA_{{\Z_2}}$ gauge field in \eq{eq:AZ2JZ2} is actually the dynamical gauge field living in the 5d bulk TQFT perspective; in fact, on the 4d boundary, 
the $\CA_{{\Z_2}}$ gauge field and the $a$ gauge field can be combined together 
to become a $\Z_4$ gauge field: $\CA_{{\Z_4}}$. 
The relation between
the boundary $a$ $\in$ $ {\H^1(M, \Z_{2})}$,
the boundary $\CA_{{\Z_4}}$ $\in$ ${\H^1(M, \Z_{4,X})}$, 
and
the bulk $\CA_{{\Z_2}}$ $\in$ ${\H^1(M, \Z_{4,X}/\Z_2^F)}$ gauge fields 
can be understood
by a group extension (following \eq{eq:Z2-Spin}
and \eq{eq:Z4-Spin}):
\bea
0 \to [{\Z_2}]_{\text{bdry gauge $a$}} \to [{\Z_{4,X}}]_{\text{bdry gauge $\CA_{{\Z_4}}$}} \to  [\frac{\Z_{4,X}}{\Z_2^F}]_{\text{Bulk gauge $\CA_{{\Z_2}}$}} \to 0.
\eea
Recall that $\CA_{{\Z_2}} = \CA_{{\Z_4}} \mod 2$, similarly $\CJ_{{\Z_2}}  = \CJ_{{\Z_4}} \mod 2$.
Thus we have the coupling term \eqn{eq:AZ2JZ2} in 5d, but we have the following term on the 4d boundary:
\bea
\int \CA_{{\Z_4}} \wedge \star \CJ_{{\Z_4}}  +\dots.
\eea
\item Due to \eq{eq:ABKAa}, 
we can view the spun version of Whitehead link as a source of 
an odd $\Z_{4,X}$ charge (also an odd $\frac{\Z_{4,X}}{\Z_2^F}$ charge), wrapped inside a  
current $\star \CJ_{{\Z_4}}$ in a 3-volume (or $\star \CJ_{{\Z_2}}$ in a 4-volume) \cite{GuoJW1812.11959, toappear}.

\end{enumerate}

In fact, the odd $\Z_{4,X}$
charge $q_X$ of  the topological link means that the dynamically gauged Wilson line with a line integral 
$\exp(\ii q_X \int \CA_{{\Z_4}})$ 
can mediate between a topological link \emph{on one end} (${\text{p}_1}$) 
and any other object carrying a nontrivial $\Z_{4,X}$ charges \emph{on the other end} (${\text{p}_2}$).

So we only need to look for all SM and GUT particles carrying the $\Z_{4,X}$ charges (especially the odd $\Z_{4,X}$ charge).
We now look at \Table{table:SMfermion} and \ref{table:SMboson}, all the left-handed Weyl spinors carry $\Z_{4,X}$ charge $+1$ (and the 
right-handed Weyl spinors carry a complex conjugate of $\Z_{4,X}$ charge $-1 = 3 \mod 4$).
The electroweak Higgs $\phi_H$ carries an even $\Z_{4,X}$ charge $2 \mod 4$.
So, yes, in fact, all SM Weyl spinors carry odd $\Z_{4,X}$ charge, thus, all SM Weyl spinors, say $\psi_{\text{SM}}$, can be the other end of the
 line integral $\exp(\ii q_X \int \CA_{{\Z_4}}$) of 
$\Z_{4,X}$. All the odd number of bound states (like proton and neutrons in our body, also electrons) presumably can carry
the odd $\Z_{4,X}$.
In summary, we may have a schematic communication between a SM particle $\psi_{\text{SM}}$ and a topological link via a line operator
\bea
\exp(\ii   \pi \oiint b + \dots ) \bigg\vert_{{\text{a link $L$ formed around }\text{p}_1}}\cdot \exp(\ii q_X \int_{\text{p}_1}^{\text{p}_2} \CA_{{\Z_4}}) \cdot \psi_{\text{SM}}({\text{p}_2}),
\eea
where $\CA_{{\Z_4}}$ is the discrete $\Z_{4,X}$ gauge field.
A candidate link $L$ is shown in \Fig{fig:BSMlink1} (b).
The $\Z_{4,X}$ discrete gauge boson $\CA_{{\Z_4}}$ 
is the mediator of Topological Force. 
Topological Force has effects of long-range remote interactions but only through braiding and fusion statistics, thus
Topological Force is presumably weaker than the GUT forces, strong, electroweak forces.
In summary, if either the 4d or 5d gapped topological sectors exist,
our SM world and the gapped topological sector
can be mediated and communicated by Topological Force.

\subsection{Topological Force as a new force?}

We had mentioned that under some assumptions 
about the discrete $\Z_{4,X}$ symmetry unbroken at high energy and only 15 Weyl fermions per generation,
it is natural to include a nonperturbative topological sector of 4d TQFT, 5d iTQFT, or 5d TQFT, appending to the SM or the GUT.
The Topological Force derived here is not within three known Fundamental Forces or GUT forces.
So we may view the Topological Force as a new force.
Our model in  \Sec{sec:gravity} suggests that perhaps the Topological Force can be a new force in the model \Eq{eq:UU-GUT-1}.
But the Topological Force may mix with the gravity (the Fourth force) in the model  \Eq{eq:UU-GUT-2-grav}, when the $\Z_{4,X}$ is gauged and the
geometry/topology are summed but regularized.
It is not clear how the gravity in  \Sec{sec:gravity} has anything to do with Einstein gravity. 
It may be interesting to explore this direction in the future.

\subsection{Neutrinos}
\label{sec:neutrino-oscillations}

{\bf Neutrino oscillations} may also be explained if we consider the Majorana zero modes of the vortices in the 4d TQFT defects in  \Sec{sec:HiddenTopologicalSectors}. 
The left-handed neutrinos (confirmed in the experiments) are nearly gapless/ massless. When neutrinos traveling through the 4d TQFT defects,
 we may observe nearly gapless left-handed neutrino flavor oscillations interfering with the Majorana zero modes trapped by the vortices in the 4d TQFT defects.

A possible scenario is to arrange an even $\upnu_{\rm{4d}}$ anomaly index for the 4d TQFT  
$$
\upnu_{\rm{4d},\rm{even}}= \pm 2,  \pm 4, \dots \in \Z_{16},
$$
so there is an additional internal rotational symmetry within the 4d TQFT.
For example, given a $\upnu_{\rm{4d},\rm{even}}= 2$ of 4d TQFT,  
there is an additional U(1) internal rotational symmetry.\footnote{We emphasize this additional U(1) is an extra symmetry, 
distinct from the $({\bf B}- {\bf L})$, the $X=5({\bf B}- {\bf L})-4Y$, the SM, or the GUT gauge groups. 
In one lower dimension, for a 3d boundary of a 4d bulk iTQFT given by the 4d APS $\eta$ invariant of $\Omega_4^{\Pin^+}=\Z_{16}$,
\Refe{Metlitski20141406.3032} introduces a similar extra U(1) symmetry when $\upnu =\pm 2 \in \Z_{16}$ of $\Omega_4^{\Pin^+}$.}
We can imagine a mother EFT with a $\upnu_{\rm{4d},\rm{even}}= 2  \in \Z_{16}$ anomaly
as a deeper UV theory which lands to two possible IR theories:\\
$\bullet$ One side has two right-handed neutrinos, a free fermion theory with a $\upnu_{\rm{4d},\rm{even}}= 2  \in \Z_{16}$ anomaly.\\
$\bullet$ Another side has a 4d TQFT also with a $\upnu_{\rm{4d},\rm{even}}= 2  \in \Z_{16}$ anomaly.\\
As a dual description of TQFT on the free fermion side, we have
the internal rotational U(1) or O(2) symmetry realized as an internal flavor rotation between two right-handed neutrinos (as 3+1d Weyl or Majorana fermions).
The TQFT defects are U(1) symmetry breaking vortices (or vortex strings).
The vortices can trap Majorana zero modes below the vortex energy subgap.

\noindent
{\bf SM and GUT Models with right-handed neutrino, 4d TQFT, and 5d iTQFT}:
We may still need to introduce at least a right-handed neutrino in order to give a conventional Dirac mass to
the three left-handed neutrinos.
To match the $\upnu= -N_{\text{generation}}= -3 \in \Z_{16}$ anomaly,
we had considered the scenario in \eq{eq:TQCP-1}: 
\bea \label{eq:system-1}
\left\{
\begin{array}{l}
\text{$\bullet$ 4d SM or GUT with $\upnu = 3 \times 15$ Weyl fermions.}\\
\text{$\bullet$ 4d $(\upnu_{\text{even}}=2)$-TQFT.}  \\
\text{$\bullet$ 4d $\upnu=1$ right-handed neutrino gives Dirac/Majorana masses} \\
\quad  
\text{to the other 3 of SM's left-handed neutrinos.}
\end{array}
\right.
\eea
If we want to introduce also a 5d iTQFT but still keep a 4d TQFT and a right-handed neutrino (in order to give a conventional Dirac mass to
left-handed neutrinos), we can also propose another scenario:
\bea \label{eq:system-2}
\left\{
\begin{array}{l}
\text{$\bullet$ 4d SM or GUT with $\upnu = 3 \times 15$ Weyl fermions.}\\
\text{$\bullet$ 4d $(\upnu_{\text{even}}=4)$-TQFT or $(\upnu_{\text{even}}=2+k)$-TQFT.}  \\
\text{$\bullet$ 5d $(\upnu = -2)$-iTQFT or $(\upnu = -k)$-iTQFT.}\\
\text{$\bullet$ 4d $\upnu=1$ right-handed neutrino gives Dirac/Majorana masses} \\
\quad  \text{to the other 3 of SM's left-handed neutrinos.}
\end{array}
\right.
\eea
We have a generic $k \in \Z_{\text{even}}$. The above models, \eq{eq:system-1} and \eq{eq:system-2}, all have a 0 mod 16 index, thus free from the $\Z_{16}$ anomaly.

\noindent
{\bf Right-handed sterile neutrinos are not sterile to $\Z_{4,X}$ gauge field}:
We should emphasize again that although the
right-handed neutrinos are sterile to SM forces,  
they are \emph{not} sterile to the $\Z_{4,X}$ gauge field  $\CA_{{\Z_4}}$, because they carry the odd $\Z_{4,X}$ charge.

\subsection{Dark Matter}

{\bf Dark Matter}: In the heavy sector of TQFTs that we described in \Sec{sec:HiddenTopologicalSectors},
the surface $b$ is the worldsheet of heavy string excitations. 
Those are new heavy objects whose mass is in the scale of a topological order energy gap, $\Delta_{\text{TQFT}}$,
possibly around the GUT scales (for example, suggested in \cite{JW2008.06499}, the $\Delta_{\text{TQFT}}$ can be in between the SU(5) GUT and the SO(10) GUT scales).
So these new heavy higher-dimensional extended objects may be a suitable candidate account for the abundant Dark Matter in the Unvierse.

\noindent
{\bf Dark Energy}: We have very little to say about the Dark Energy. Except that in the gravity version (sum over geometry and topology; dynamical, quantum, and topological gravity)
of partition function in \Eq{eq:UU-GUT-2-grav}, it may be worthwhile to investigate the underlying energy stored in this partition function.
It seems that 4d and 5d topological sectors can be very heavy and store a much huge amount of energy than what we knew of from our Standard Model world.


\section{Acknowledgements} 

JW is grateful to his previous collaborators for fruitful past researches  
as helpful precursors for the present work.
JW appreciates Email correspondences with Robert Gompf, Miguel Montero, Kantaro Ohmori, Pavel Putrov, Ryan Thorngren,  Zheyan Wan, and Yunqin Zheng;
and the mental support 
from Shing-Tung Yau.\footnote{Instead of writing or drawing an image of the author's mental feelings, a piece of  
Johannes Brahms's  music ``The Variations and Fugue on a Theme by Handel, Op. 24 (1861)'' may illuminate this well.}
JW thanks the participants of Quantum Matter in Mathematics and Physics program at
Harvard University CMSA for the enlightening atmosphere.
Part of this work is presented by JW in the workshop Lattice for Beyond the Standard Model physics 2019 at Syracuse University (May 2-3, 2019) 
\cite{JW-BSM-2019-talk}, also at Higher Structures and Field Theory at Erwin Schr\"odinger Institute in Wien (August 4, 2020) \cite{ESIJWUltraUnification},
and at Harvard University Particle Physics Lunch (November 30, 2020) and String Lunch (December 4, 2020).
%
JW was supported by
NSF Grant PHY-1606531. 
This work is also supported by 
NSF Grant DMS-1607871 ``Analysis, Geometry and Mathematical Physics'' 
and Center for Mathematical Sciences and Applications at Harvard University.

\section{Bibliography}
\bibliographystyle{Yang-Mills}
\bibliography{BSM-SU3SU2U1-cobordism.bib}

\end{document}